\documentclass{aa}  

\usepackage[normalem]{ulem}
\usepackage{listings}

\usepackage{graphicx}

\usepackage{amsthm}
\usepackage{xcolor}
\usepackage{array,booktabs}
\usepackage{multirow}
\usepackage{txfonts}
\begin{document}

   \title{J-PLUS: Uncovering a large population of extreme [OIII] emitters in the local Universe}

\author{A. Lumbreras-Calle
        \inst{\ref{CEFCA}}\fnmsep\thanks{\email{alumbrerascalle@gmail.com}}
        \and C.~L\'opez-Sanjuan\inst{\ref{CEFCA}}
        \and D.~Sobral\inst{\ref{Lancaster}}
        \and J.~A.~Fern\'andez-Ontiveros\inst{\ref{CEFCA}}
        \and J.~M.~V\'{\i}lchez\inst{\ref{IAA}}
        \and A.~Hern\'an-Caballero\inst{\ref{CEFCA}}
        \and M.~Akhlaghi\inst{\ref{CEFCA}}
        \and L.~A.~D\'{\i}az-Garc\'{\i}a\inst{\ref{IAA}}
        \and J.~Alcaniz\inst{\ref{ON}}
        \and R.~E.~Angulo\inst{\ref{DIPC},\ref{ikerbasque}}
        \and A.~J.~Cenarro\inst{\ref{CEFCA}}
        \and D.~Crist\'obal-Hornillos\inst{\ref{CEFCA}}
        \and R.~A.~Dupke\inst{\ref{ON},\ref{MU},\ref{Alabama}}
        \and A.~Ederoclite\inst{\ref{USP}}
        \and C.~Hern\'andez-Monteagudo\inst{\ref{IAC},\ref{ULL}}
        \and A.~Mar\'{\i}n-Franch\inst{\ref{CEFCA}}
        \and M.~Moles\inst{\ref{CEFCA}}
        \and L.~Sodr\'e Jr.\inst{\ref{USP}}
        \and H.~V\'azquez Rami\'o\inst{\ref{CEFCA}}
        \and J.~Varela\inst{\ref{CEFCA}}
}
\institute{Centro de Estudios de F\'{\i}sica del Cosmos de Arag\'on (CEFCA), Unidad Asociada al CSIC, Plaza San Juan 1, 44001 Teruel, Spain\label{CEFCA}
        \and
        Department of Physics, Lancaster University, Lancaster, LA1 4YB, UK\label{Lancaster}
        \and
        Instituto de Astrof\'isica de Andaluc\'ia (IAA-CSIC), P.O.~Box 3004, 18080 Granada, Spain\label{IAA}
        \and
        Observat\'orio Nacional - MCTI (ON), Rua Gal. Jos\'e Cristino 77, S\~ao Crist\'ov\~ao, 20921-400 Rio de Janeiro, Brazil\label{ON}
        \and
        Donostia International Physics Centre (DIPC), Paseo Manuel de Lardizabal 4, 20018 Donostia-San Sebastián, Spain\label{DIPC}
           \and
        IKERBASQUE, Basque Foundation for Science, 48013, Bilbao, Spain\label{ikerbasque}
        \and
        University of Michigan, Department of Astronomy, 1085 South University Ave., Ann Arbor, MI 48109, USA\label{MU}
        \and
        University of Alabama, Department of Physics and Astronomy, Gallalee Hall, Tuscaloosa, AL 35401, USA\label{Alabama}
        \and
        Instituto de Astronomia, Geof\'{\i}sica e Ci\^encias Atmosf\'ericas, Universidade de S\~ao Paulo, 05508-090 S\~ao Paulo, Brazil\label{USP}
        \and
        Instituto de Astrof\'{\i}sica de Canarias, La Laguna, 38205, Tenerife, Spain\label{IAC}
        \and
        Departamento de Astrof\'{\i}sica, Universidad de La Laguna, 38206, Tenerife, Spain\label{ULL}
}

   \date{}

  \abstract
{Over the past decades, several studies have discovered a population of galaxies undergoing very strong star formation events, called extreme emission line galaxies (EELGs).}
 {In this work, we exploit the capabilities of the Javalambre Photometric Local Universe Survey (J-PLUS), a wide field multifilter survey, with 2000 square degrees of the northern sky already observed. We use it to identify EELGs at low redshift by their [OIII]5007 emission line. We intend to provide with a more complete, deep, and less biased sample of local EELGs.}
 {We select objects with an excess of flux in the J-PLUS mediumband $J0515$ filter, which covers the [OIII] line at z$<$0.06. We remove contaminants (stars and higher redshift systems) using J-PLUS and WISE infrared photometry, with SDSS spectra as a benchmark. We perform spectral energy distribution fitting to estimate the physical properties of the galaxies: line fluxes, equivalent widths (EWs), masses, stellar population ages, etc.}
 {We identify 466 EELGs at ${\rm z} < 0.06$ with [OIII] EW over 300 \text{\AA} and $r$-band magnitude below 20, of which 411 were previously unknown. Most show compact morphologies, low stellar masses ($\log (M_{\star}/M_{\odot}) \sim {8.13}^{+0.61}_{-0.58}$), low dust extinction ($E(B-V)\sim{0.1}^{+0.2}_{-0.1}$), and very young bursts of star formation (${3.0}^{+2.7}_{-2.0}$ Myr). Our method is up to $\sim$ 20 times more efficient detecting EELGs per Mpc$^3$ than broadband surveys, and as complete as magnitude-limited spectroscopic surveys (while reaching fainter objects). The sample is not directly biased against strong H$\alpha$ emitters, in contrast with works using broadband surveys.}
 {We demonstrate the capability of J-PLUS to identify, following a clear selection process, a large sample of previously unknown EELGs showing unique properties. A fraction of them are likely similar to the first galaxies in the Universe, but at a much lower redshift, which makes them ideal targets for follow-up studies.}


   \maketitle

\section{Introduction}
 
Galaxies undergoing strong events of star formation and presenting compact morphologies have been identified and analyzed since the middle of the XXth century, with pioneering works by \cite{Haro56}, \cite{Zwicky66}, and \cite{Markarian67}. One of the most fruitful developments in the detection of this kind of galaxies was the use of an objective prism to identify strong emission lines \citep{Markarian67}. Early analysis of this class of galaxies noticed their usually compact morphologies and blue colors, with their spectra resembling those of galactic HII regions, thus the label "HII galaxies" \citep{Sargent70,Melnick85}. Low metallicities were also identified, as well as recent enhancement of star formation \citep{Searle72}.

With the advent of modern CCD detectors and wide field surveys, new windows into the analysis of these objects were opened. An observational alternative for the identification of emission line galaxies (ELGs) is the use of multi-band surveys, with some examples including CADIS \citep{Meisenheimer98}, COMBO-17 \citep{Wolf01a,Wolf03}, HiZELS \citep{Geach08,Sobral15}, ALHAMBRA \citep{Moles08}, MUSYC \citep{Cardamone11},  SHARDS \citep{Perez-Gonzalez13}, COSMOS mediumband \citep{Taniguchi15}, PAU \citep{Benitez09}, SC4K \citep{Sobral18}, J-PLUS \citep{Cenarro19}, S-PLUS \citep{MendesdeOliveira19} and J-PAS \citep{Benitez14}. Using a set of several narrowband or mediumband filters, sometimes including also broadband ones, these surveys detect easily emission lines by identifying an excess of flux in one of the bands \citep[e.g.][]{Hippelein03,Maier03,Gronwall07,Sobral13,Sobral15,Cava15,Lumbreras-Calle19a,Spinoso20,Vilella-Rojo21}. While usually providing smaller fields of view and less wavelength coverage than broadband or objective prism surveys, they can reach deeper magnitudes than the former and higher spectral resolution than the latter. These surveys have yielded large samples of ELGs, sometimes identifying simultaneously different emission lines and covering a wide redshift range. 

\begin{figure*}
   \centering
   \includegraphics[width=0.33\textwidth,keepaspectratio]{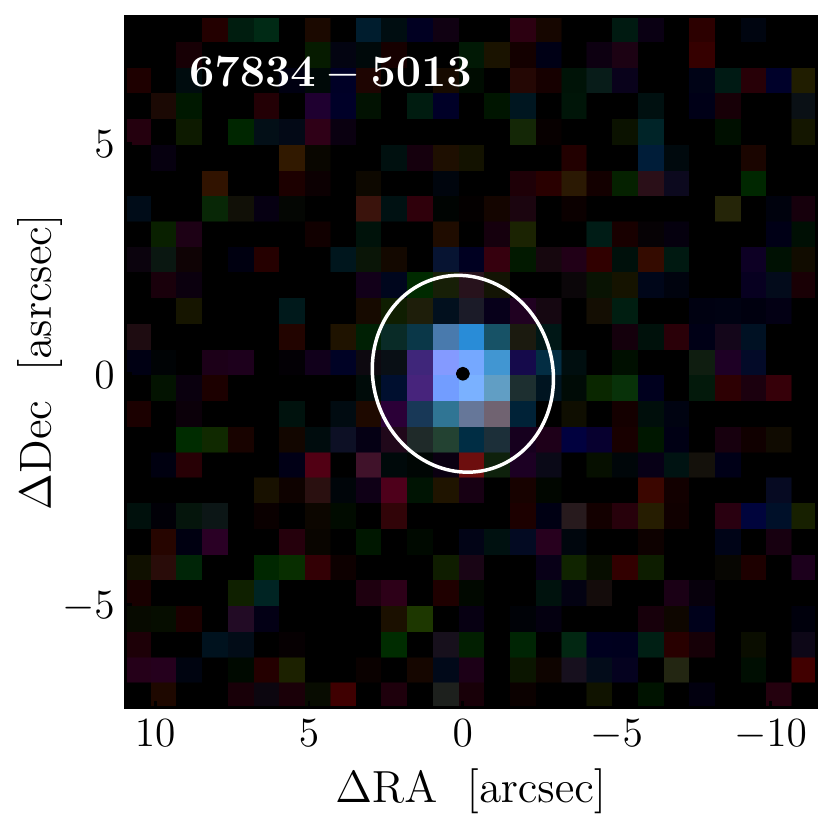}
   \includegraphics[width=0.66\textwidth,keepaspectratio]{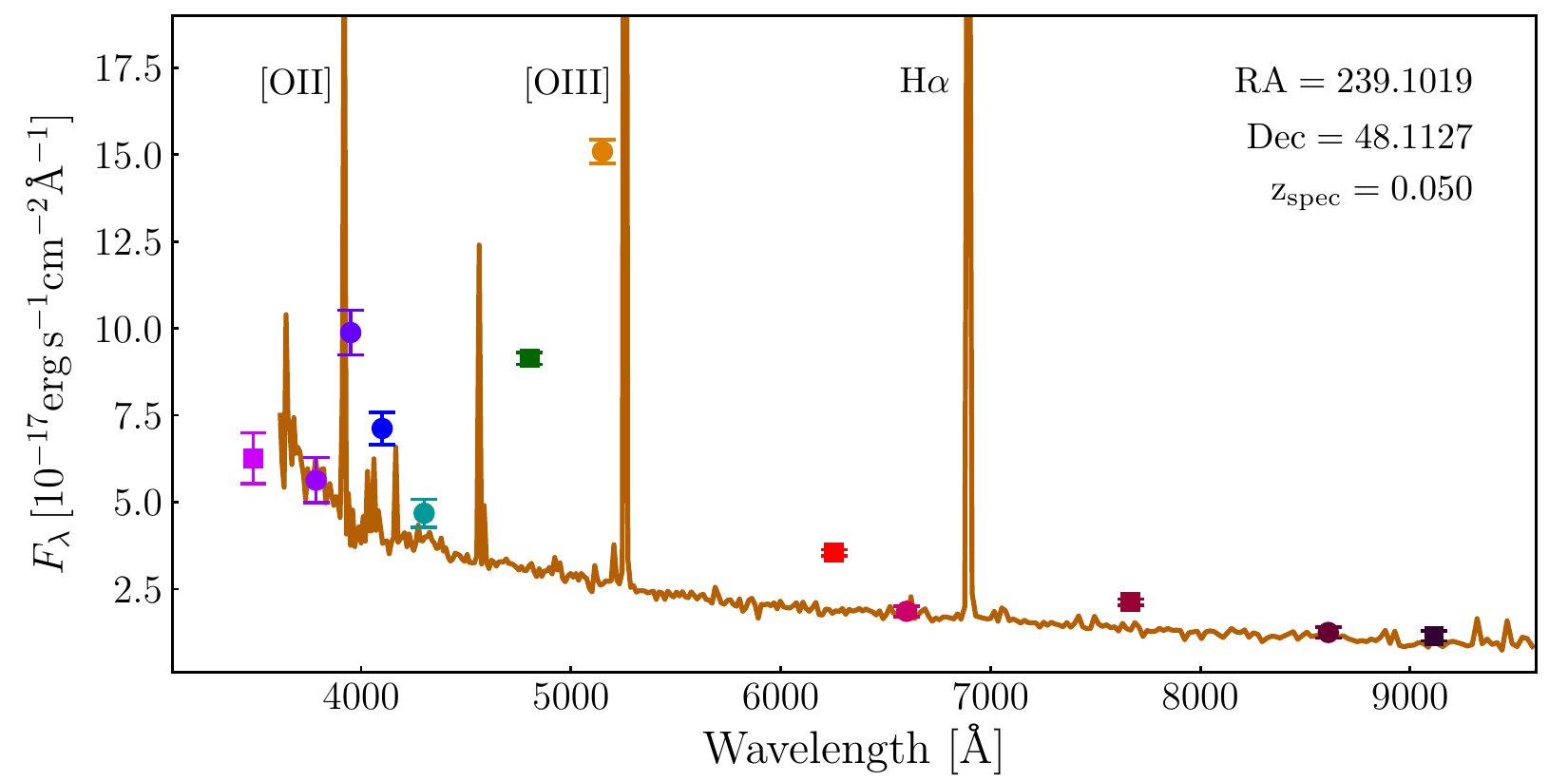}
      \caption{
      Illustrative example of an extreme [OIII] emitter, the J-PLUS source 67834-5013 (${\rm RA} = 239.1019$, ${\rm Dec}= 48.1127$, ${\rm z}_{\rm spec} = 0.050$). {\it Left panel:} Color composite of the galaxy, obtained from the $gri$ J-PLUS images. The sky location of the source is shown as a black dot and the white ellipse indicate the three effective radius contour for the source. {\it Right panel:} J-PLUS twelve-band PSFCOR photometry of the galaxy. The squares show the five SDSS-like filters ($ugriz$), and circles the seven mediumband filters ($J0378$, $J0395$, $J0410$, $J0430$, $J0515$, $J0660$, and $J0861$). The solid line shows the spectra from SDSS with a downgraded resolution of $R \sim 180$ and normalized to the flux in the filter $J0660$. The location of the most prominent emission lines is marked: [OIII] (traced by $J0515$ and $g$), H$\alpha$ (traced by $r$), and [OII] (traced by $J0395$).
      }
     \label{fig:example}
\end{figure*}

Focusing on higher redshifts, the search for ELGs has been performed using deep, broadband surveys, specially those with data from the \textit{Hubble} Space Telescope (HST), such as GOODS, COSMOS, or CANDELS \citep{Kakazu07,vanderWel11,Maseda18}. They found an increase in the density of ELGs at higher redshifts, in coherence with the evolution of the cosmic star formation history \citep[SFH,][]{Madau14}. In fact, at very high redshift (z$>6$), typical low mass galaxies are expected to be extreme emission line galaxies (EELGs), playing a crucial role in the reionization of the Universe \citep{Ouchi09,Bouwens15}. From an observational point of view, space-based grism spectroscopy in some of the HST deep fields provides an even greater insight into the nature of EELGs \citep{Pirzkal13}. This approach has only been feasible in small fields, which is enough for higher redshifts but insufficient for local Universe studies. Upcoming survey telescopes such as Euclid, the Nancy Grace Roman Telescope and the Chinese Space Station Telescope will extend this grism analysis to fields covering thousands of square degrees. It is therefore vital to gather samples of EELGs that can be used as a reference for these kind of surveys, testing their limitations and providing a readily available reference sample.

The precise definition of EELGs varies in different works in the literature. It is very often described as a threshold in the rest-frame equivalent width (EW) of the [OIII]5007 line, but values range from 100 \text{\AA} in \cite{Amorin15} and \cite{Perez-Montero20}, to 300 \text{\AA} in \cite{Jiang19}, with several surveys providing most objects above 500 \text{\AA} \citep{vanderWel11,Yang17,Maseda18}.

Recently, several projects have aimed at detecting samples of EELGs at low redshift using photometric surveys, such as \cite{Cardamone09}, \cite{Yang17} and \cite{Senchyna19}. They have demonstrated the power of broadband surveys to identify this kind of galaxies with relatively high purity, while covering much wider areas than previous surveys, allowing the study of rare populations. Studies using narrowband filters have also proven useful in the detection of ELGs over relatively modest fields \citep{Kellar12,Salzer20} or brighter galaxies in very wide fields \citep{Cook19}.

In addition to photometric surveys, the extensive database provided by the SDSS spectra has been thoroughly explored in the search of EELGs \citep{Izotov11} or extremely metal poor galaxies \citep{SanchezAlmeida16}. The physical information for individual galaxies accessible through this database is unparalleled, showing for instance how these local samples can serve as analogs for the very high redshift systems responsible of the reionization of the Universe \citep{Izotov21}. Nevertheless, a drawback of this data set is its poorly defined selection function, with galaxies often observed in surveys targeting other type of sources (such as quasars). As a result, no meaningful study regarding e.g. number densities can be performed, except for the very bright end, where completeness is high \citep{Strauss02}.  A very deep spectroscopic survey (with exposure times of $\sim$ 10 h)  that does not suffer from these selection biases is the MUSE \textit{Hubble} ultra deep field \citep{Bacon17}. Using the MUSE integral field unit they detected a large sample of ELGs, some extremely faint and undetected in the HST data. Nevertheless, this work covered a small area ($\sim$ 10 arcmin$^2$), insufficient for studies of uncommon sources at low redshift.

In this context, the Javalambre Photometric Local Universe Survey (J-PLUS, \citealt{Cenarro19}) combines multiband observation with a very wide field, making it a unique tool for the identification of very rare emission line objects. In addition, as an imaging survey, it allows us to analyze the emission in the whole galaxy, not only the aperture where the spectra is extracted, a limitation found in fiber-fed spectroscopy surveys (such as SDSS, GAMA, or DESI). This photometric survey has already produced relevant results in the detection and analysis of extragalactic systems presenting emission lines in their spectra. In the local Universe, previous works by \cite{Logrono-Garcia19} and \cite{Vilella-Rojo21} have analyzed H$\alpha$ emitters and measured accurately the local star formation main sequence. In addition, \cite{Spinoso20} performed a search for high-redshift quasars with Lyman $\alpha$ emission in J-PLUS, finding an unprecedented number of very high luminosity sources.

In the present work, we select low redshift EELGs in J-PLUS by identifying objects showing intense emission in the $J0515$ filter, as illustrated in Fig.~\ref{fig:example}. We cover a much wider area than previous multiband surveys, reaching deeper magnitudes than wide field spectroscopic surveys, and are able to identify emission lines with much more precision than broadband surveys. We aim at providing a new sample that will uncover many unclassified EELGs in the Local Univers, that can complement previous photometric and spectroscopic studies and serve as targets for follow-up observations.

\begin{table}[t] 
\caption{J-PLUS filter system. The filters $g$, $r$, $i$, and $z$ match their namesakes in SDSS.}
\label{tab:JPLUS_filters}
\centering 
	\begin{tabular}{c c c } 
	\hline\hline 
               & Central        &   		 \\ 
       Filter  & Wavelength     & FWHM       \\ 
	       	   & [\AA]          & [\AA]      \\ 
	\hline
	$u$	 	    &3485 	&508		\\ 
	$J0378$ 	&3785 	&168 	    \\ 
	$J0395$ 	&3950	&100		\\ 
	$J0410$ 	&4100	&200		\\ 
	$J0430$	    &4300 	&200	    \\ 
	$g$		    &4803 	&1409	    \\ 
	$J0515$ 	&5150	&200		\\ 
	$r$ 		&6254	&1388	    \\ 
	$J0660$ 	&6600 	&138		\\ 
	$i$		    &7668 	&1535	    \\ 
	$J0861$	    &8610 	&400		\\ 
	$z$ 		&9114	&1409	    \\ 
	\hline 
\end{tabular}
\end{table}

The paper is structured as follows: In Sect. \ref{sec:databas} we describe the J-PLUS survey and database, as well as our procedure for selecting the candidate sample and removing contaminants. In Sect. \ref{sec:results} we described the spectral energy distribution (SED) fitting analysis and its results, along with the final definition of the EELG sample and their visual morphologies. Finally, we compare these results with the available spectra for the sample. We discuss the results in Sect, \ref{sec:discussion}, testing the sample selection, the number densities of EELGs we find, and reviewing the main physical properties of the EELG sample. Finally, in Sect, \ref{sec:conclusions} we summarize our work and present the conclusions.

Throughout this paper, all quoted magnitudes are in the AB system, and all logarithms used are in base 10, and coordinates refer to the 2000 equinox. All mentions to the [OIII] line, unless specified, refer to the [OIII]5007+4959 doublet. The median is used to show the typical value of a magnitude, while the upper and lower limits presented refer to the 16th and 84th percentiles of the distributions. We assumed a $\Lambda$CDM cosmology with with $\Omega_{\Lambda}=0.7$, $\Omega_{\rm M}=0.3$, and $H_{0}=70$ km s$^{-1}$ Mpc$^{-1}$. In the SED fitting we use the \cite{Salpeter55} initial mass function (IMF). 

\section{Database and sample selection}

\label{sec:databas}

\subsection{J-PLUS second data release}
J-PLUS\footnote{\url{www.j-plus.es}} is being conducted at the Observatorio Astrof\'{\i}sico de Javalambre (OAJ; \citealt{oaj}) using the 83\,cm Javalambre Auxiliary Survey Telescope (JAST80) and T80Cam, a panoramic camera of 9.2k $\times$ 9.2k pixels that provides a $2\deg^2$ field of view, with a pixel scale of 0.55 arsec pix$^{-1}$ \citep{t80cam}. The J-PLUS filter system is composed of twelve passbands (Table~\ref{tab:JPLUS_filters}) spanning the full optical range ($3\,500 - 10\,000$ \text{\AA}), with seven mediumband and five broadband filters. The J-PLUS observational strategy, image reduction, and main scientific goals are presented in \citet{Cenarro19}. The J-PLUS photometric calibration is described in \citet{clsj19jcal} and \citet{clsj21zsl}.

This work is based on the second data release (DR2) of J-PLUS. It covers $2\,176$ square degrees ($1\,941$ deg$^2$ after masking), with $1\,088$ individual images of 2 deg$^2$. The survey reaches limiting magnitudes ranging from 21.8 mag in $r$ to 20.5 mag in $z$, with 21.0 mag in the $J0515$ filter (considering 5$\sigma$ detection).

In order to create the galaxy sample we will describe in this work, we run queries using the Astronomical Data Query Language (ADQL) interface in the J-PLUS database. In the queries we use the PSFCOR photometry of the catalog, which is designed to better capture the colors of the objects rather than their total flux \citep{Molino19}. Briefly, PSFCOR photometry is measured considering elliptical apertures with semi-major axis equal to the Kron radius \citep{Kron80} in the reference band ($r$), half the size of the AUTO aperture. Then, the photometry in the rest of the bands is measured, correcting for the different point spread functions (PSFs), to produce accurate colors (see \citealt{Molino19} and \citealt{Hernan-Caballero21} for more details). This photometry, while optimal for comparing fluxes in different bands, underestimates the total flux of galaxies with respect to the AUTO photometry by 0.5 mag in average \citep{GonzalezDelgado21}. Therefore, in the rest of the analysis of this work (unless specified), we re-scale the PSFCOR photometry to the AUTO one using the ratio between AUTO and PSFCOR fluxes in the $r$ band. In addition, we apply a galactic extinction correction, as provided in the J-PLUS database.

\subsection{Sample selection}
\subsubsection{Selection of intense $J0515$ emitters}
   
The first step in the process is to query the J-PLUS database for objects that show a large excess of emission in the $J0515$ filter, since all extreme [OIII]4959+5007 emitters between redshift 0.007 and 0.06 should show strong emission in this band (Fig.~\ref{fig:example}). The precise definition of "strong emission" will be discussed in Section \ref{sec:cigale}, but given the width of the $J0515$ filter, in order to secure clear detections we will aim at selecting objects with EW(J0515)$>$200 \text{\AA}. First, in order to compute the excess of emission corresponding to the line flux, we need to estimate the continuum underneath it. To perform a simple query, we assumed that the continuum flux (in erg s$^{-1}$ cm$^{-2}$) at the wavelength of $J0515$ is roughly equivalent to the flux of the $r$ band. This obviously depends on the slope of the SED and relative intensity of the H$\alpha$ line, but it is enough to draw a first sample of emitters, of which we will later select the most extreme ones. A possible bias against extreme H$\alpha$ emitters introduced by this assumptions will be addressed in Sect. \ref{sec:selecEELG}. Specifically, we imposed the following condition:
\begin{equation}
\label{eq:selec}
\frac{F(J0515) - F(r)}{F(r)} > 1,
\end{equation} 
which, if we assume that $r$ traces the continuum, would imply that the emission lines that lie within $J0515$ have EW $\gtrsim$ 200 \text{\AA}. A precise estimation of the EW will be performed in Section \ref{sec:sedresults}, when accurate estimations of the continuum have been computed. In addition to this cut, we imposed other constrains to ensure the quality of the sample. We selected objects with $r < 22$ mag and signal-to-noise ratio (SNR) larger than $3$ in the $r$, $g$, and $J0515$ bands. We also remove objects with a FLAGS parameter indicating bad photometric quality for those bands (saturation, proximity to a bright star or the edge of an image, etc.), except when the flag indicates the existence of close companions or deblending. We prefer to draw a broad sample in order to only later apply further cuts, motivated by the properties and limitations of the sample. The full ADQL query is reproduced in Appendix \ref{appendix:query}.

Nevertheless, in order to analyze the selected objects, we separate them into four groups, considering their \texttt{CLASS\_STAR} value ($>$=0.5 and $<$0.5) and their deblending flag (either no deblending or some deblending). This allows us to better identify different types of spurious detections that only appear in some of the groups, and to develop techniques to address them specifically. This first rough selection yields a sample of $30\,336$ objects with flux excess in the $J0515$ filter.

\subsubsection{Cross-match with SDSS spectra}
Multiple types of sources can show excess in the $J0515$ filter, such as ELGs, quasi-stellar objects (QSOs), foreground stars, or any object with spurious photometric measurements. In addition, ELGs and QSOs at different redshifts can show emission in this filter, but due to different emission lines. Since we are only interested in low redshift ELGs (${\rm z} < 0.06$, where it is the [OIII] line that causes the excess emission), we need to purge the sample from all other sources. To test the accuracy of the methods for removing contaminants, we cross matched the excess-flux sample with the 16th data release of the SDSS spectroscopic database \citep{Ahumada20}. This survey covers a large fraction ($\sim$ 80\%) of the J-PLUS DR2 footprint (more than any other spectroscopic survey) and provides a spectral classification for the sources. We consider a 3 arcseconds maximum separation in the cross-match, identifying $2\,561$ objects. 

\subsubsection{Removal of contaminants via J-PLUS photometry}
When performing a visual inspection of the samples with blending flag, we identified several objects that are classified as galaxies but in fact were stars measured with large apertures. These stars are included in our sample because of a spurious deficit in $r$ emission, which implies a large $J0515$ over $r$ ratio. Therefore, removing candidates with low $r$-band flux compared to redder filters (such as $J0861$ and $z$) allows us to eliminate these detections and keep all other candidates. Additionally, this removes several QSOs with red $(r-z)$ and $(r-J0861)$ colors. Specifically, we rejected the $4\,491$ objects that fulfill $(r-z) < 1$ and $(r-J0861) < 1$. The selection was confirmed by checking that none of the $148$ objects with SDSS spectra removed in this step is a low redshift galaxy (as well as inspecting a subsample of their images and SEDs).

Another subsample of contaminants are QSOs at ${\rm z} \sim 1.7$ with a prominent C III]1909 emission line, which at that redshift creates an excess of flux in the $J0515$ filter, mimicking a low redshift ELG. In those cases, the object often shows also excess in the $J0430$ filter, corresponding to the strong C IV 1549 line. In fact, these emission lines have been used to detect AGNs in narrowband surveys, for example in \cite{Stroe17a,Stroe17b}. We therefore removed the $255$ objects with $(J0515-J0430)>0$. Out of those, $46$ candidates have SDSS spectra, without any low redshift galaxy among them.
\begin{figure}
   \centering
   \includegraphics[width=0.48\textwidth,keepaspectratio]{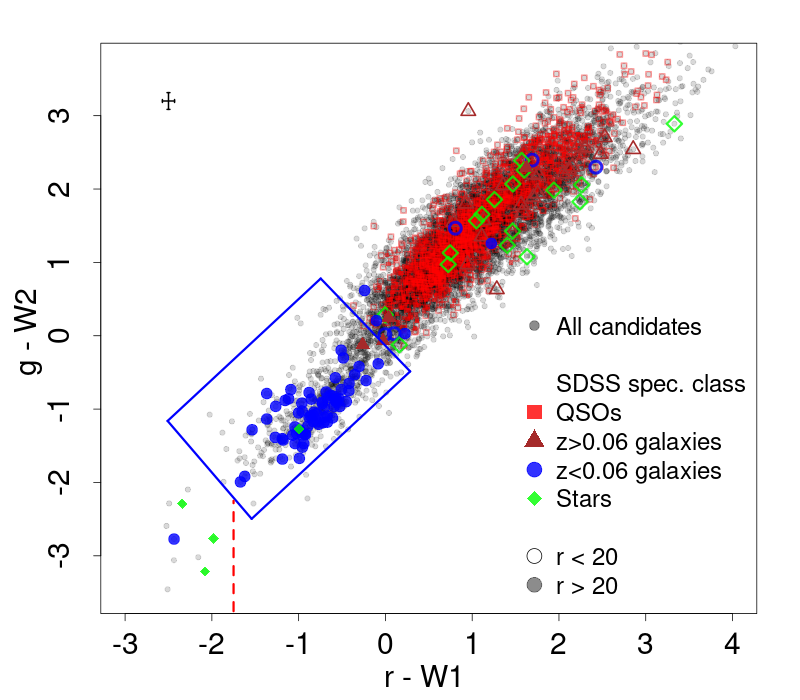}
      \caption{Color-color diagram used to separate low redshift ELGs from other types of sources. The $g$ and $r$ data are taken from J-PLUS DR2, and the $W1$ and $W2$ from the $unWISE$ catalog. All candidates from the sample with $J0515$ excess flux are shown in dark grey dots. Sources with SDSS spectra are shown in colors according to their nature (red squares for QSOs, brown triangles for galaxies at z$>$0.06, green diamonds for stars), and in filled symbols or outlines only if they are brighter or fainter than $r$= 20 mag. The objects selected as candidates to be EELGs are enclosed by the blue lines, while those to the left of the dashed red line are visually inspected to select only the extended ones as EELGs. In the top left corner the typical errors (for sources with $r$<20 mag) are shown.}
         \label{fig:grw1w2}
   \end{figure}
      \begin{figure}
   \centering
   \includegraphics[width=0.48\textwidth,keepaspectratio]{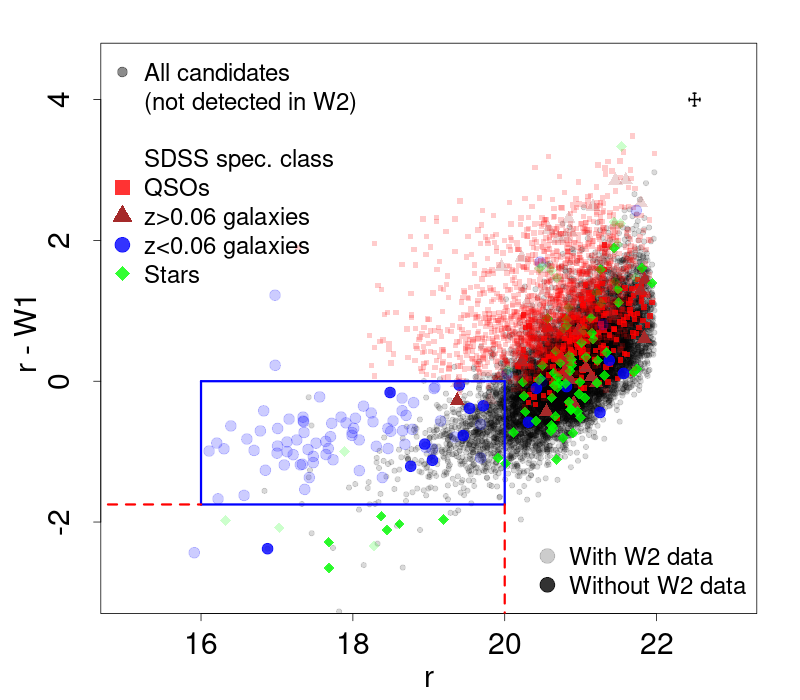}
      \caption{Color-magnitude diagram used to separate low redshift ELGs from other types of sources, for those objects where the $W2$ flux was not available (grey dots). All sources with SDSS spectra are shown in colors. The color code for those is the same as the one used in Fig. \ref{fig:grw1w2}, but in this case objects with $W2$ data are shown in a lighter shade. The objects selected as candidates to be EELGs are enclosed by the blue lines, while those to the left of the dashed red line are visually inspected to select only the extended ones as EELGs. In the top right corner the typical errors (for sources with $r$<20 mag) are shown.}
         \label{fig:rw1_r}
   \end{figure}
   
      \begin{figure}
   \centering
   \includegraphics[width=0.48\textwidth,keepaspectratio]{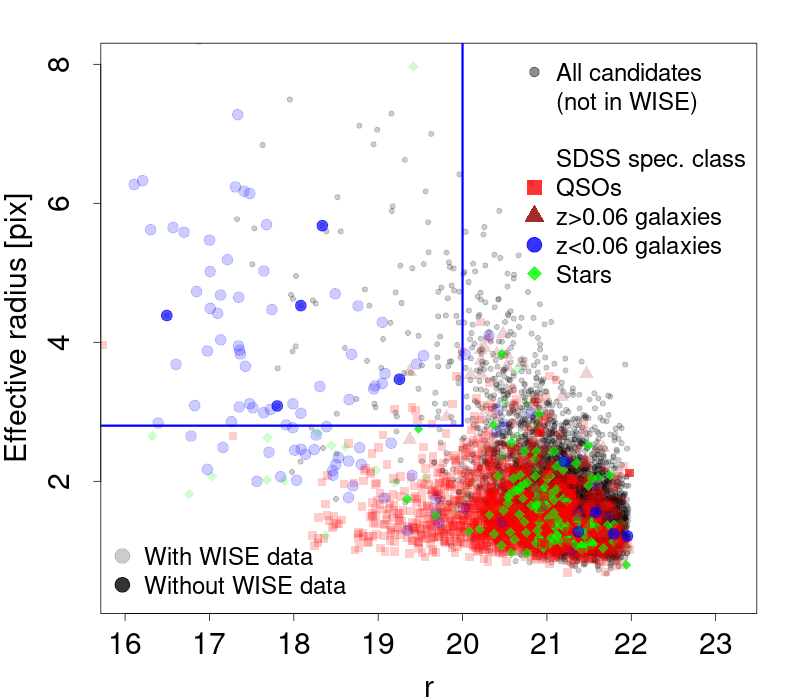}
      \caption{Diagram showing the effective radius as a function of $r$ magnitude, used to select candidates to be EELGs for those objects with excess flux in $J0515$ and no data in the $unWISE$ catalog (grey dots). All sources with SDSS spectra are shown in colors. The color code for those is the same as the one used in Fig. \ref{fig:grw1w2}, but in this case objects with $W2$ data are shown in a lighter shade.}
         \label{fig:sinunwise}
   \end{figure}
   
The sample with $J0515$ excess after this cleaning steps includes $25\,590$ sources.

\subsubsection{Removal of contaminants: WISE photometry and object sizes}

\begin{figure*}
   \centering
   \includegraphics[width=0.95\textwidth,keepaspectratio]{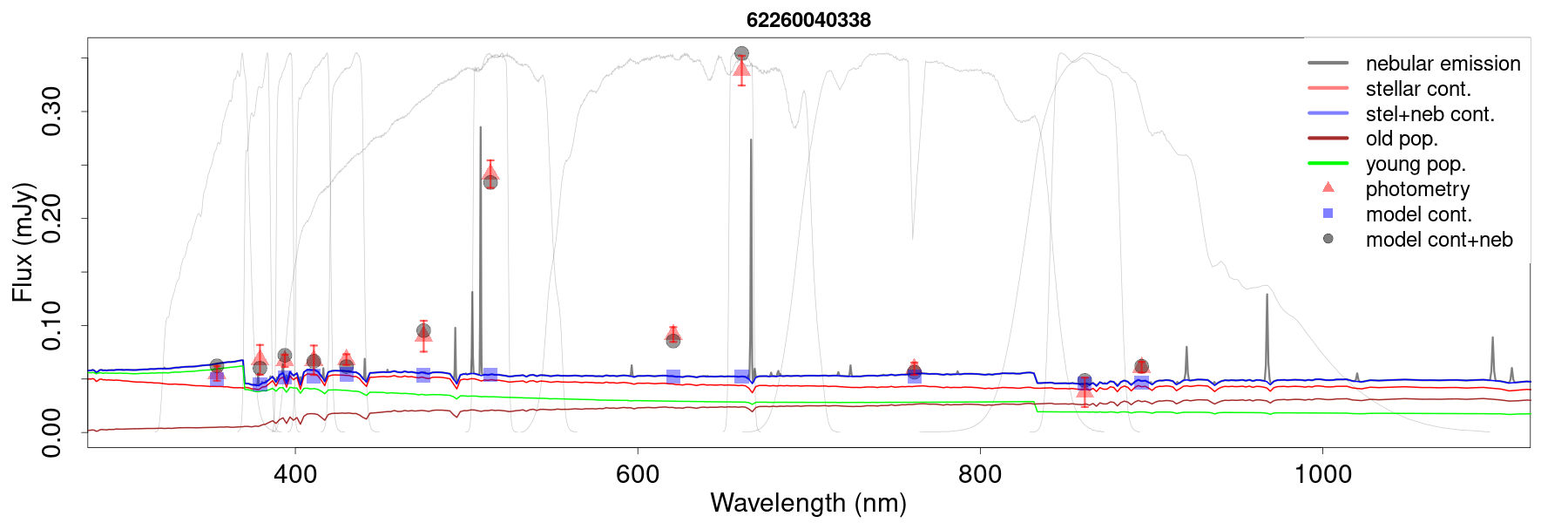}
      \caption{Example of the J-PLUS SED and \texttt{CIGALE} fit of an EELG. Red triangles correspond to the observed photometry, grey dots to the \texttt{CIGALE} synthetic photometry, and blue squares to the \texttt{CIGALE} synthetic photometry only considering the stellar and nebular continua (not considering the emission lines). The thick lines represent the \texttt{CIGALE} synthetic spectrum, considering different components: the grey line takes into account the full model, while the blue does not take into account the emission lines. The red represents the stellar continuum, while the brown and green only consider the old and young stellar populations, respectively.}
         \label{fig:ejemCIG}
   \end{figure*}

In an early analysis, it became clear that J-PLUS photometry alone was not enough to distinguish precisely between different types of sources, specially at fainter magnitudes. Following the method performed by \cite{Spinoso20}, we cross-matched our excess flux selection with the infrared (IR) data from the Wide-Field Infrared Survey Explorer (WISE) satellite. This mission performed a very wide survey in four IR bands, from $3.4$ to $22$ $\mu$m. We use the $unWISE$ catalog \citep{Schlafly19}, which only has data for the bluest bands ($W1$ and $W2$), but provides a higher depth and spatial resolution. Out of the $25\,590$ sources with excess flux at this stage, $19\,922$ are present in the $unWISE$ catalog. We remove from that sample 71 stars that have been identified using Gaia DR2 data available in the J-PLUS tables.

After testing several alternatives, we found that the most precise way of separating the low redshift [OIII] emitters from the rest of the types of objects using $unWISE$ data was the $g-W2$ versus $r-W1$ diagram (Fig.~\ref{fig:grw1w2}). The separation in this diagram between low-redshift galaxies and quasars can be understood regarding the different components that dominate the infrared light emitted by these sources. In galaxies, the $W1$ and $W2$ filters cover a local minimum in their SEDs, at longer wavelength than the peak of emission from low-mass stars but at shorter wavelength than the peak of dust emission \citep[see e.g.][]{Boquien19}. This results in blue optical-$W1$/$W2$ colors. In low redshift quasars, the $W1$/$W2$ bands are dominated by dust emission, much brighter than their optical emission (dominated by the accretion disk), resulting in very red optical-$W1$/$W2$ colors. Even for higher redshift quasars, where the $W1$/$W2$ range is affected by a combination of disk and dust emission, the optical-$W1$/$W2$ colors remain slightly red or close to zero \citep[see e.g.][]{Hernan-Caballero16}, thus different from low-redshift galaxies. We plot all sources with available $W1$ and $W2$ photometry in that diagram, indicating in different colors the type of source, for those that have available SDSS spectroscopy. It becomes clear that the low-z [OIII] emitters are clustered in a very specific region of Fig.~\ref{fig:grw1w2}, while the vast majority of stars, QSOs and higher redshift galaxies are excluded. We define a set of limits in Fig.~\ref{fig:grw1w2} to select a sample of candidates to be EELGs, preserving a very high purity and completeness (a full account of those will be provided in Sec. \ref{sec:summarysel}). The selected objects are those located in the area that fulfills the following set of equations (blue lines in Fig. \ref{fig:grw1w2}):

  \begin{align}
 g-W2 &> 1.1 \cdot (r-W1) - 0.8\nonumber\\
 g-W2 &< -1.3 \cdot (r-W1) - 0.19 \label{eq:colores}\\
 g-W2 &< 1.1 \cdot (r-W1) +1.6\nonumber\\
 g-W2 &> -1.3 \cdot (r-W1) - 4.5 \nonumber
\end{align}

Nevertheless, there remains a large fraction of objects without $W2$ photometry. From Fig. \ref{fig:grw1w2} it can be seen that the $r-W1$ color alone acts as a good discriminant on its own. When plotting $r-W1$ as a function of $r$ magnitude (Fig. \ref{fig:rw1_r}), we see that, for high values of $r$, the separation between AGNs and galaxies becomes less clear. In addition, there is a lack of spectroscopic data at $r > 20$ and $(r - W1) < 0$ which casts doubts about the nature of the objects with these characteristics.

Once again, we select the area in Fig. \ref{fig:rw1_r} where the low redshift galaxies are clustered, avoiding as many contaminants as possible. This results in the following conditions, that objects without $W2$ detection must fulfil:
\begin{align}
 (r-W1) &< 0\nonumber\\
(r-W1) &> -1.75 \label{eq:colores2}\\
r &< 20 \nonumber\\
r &> 16\nonumber
\end{align}

After applying both sets of criteria (from eq. \ref{eq:colores} and \ref{eq:colores2}), we are left with $1\,447$ candidates to be EELGs. If we applied to all sources with $unWISE$ data the selection in Eq. \ref{eq:colores2}, the final sample would be slightly different: it would be missing 16 objects and including 78 new ones. Despite the small change, given the very clear separation and gap between spectroscopically confirmed QSOs and low redshift galaxies in Fig. \ref{fig:grw1w2}, we consider that including the $g-W2$ color adds valuable information. Considering the uncertainties in the color and magnitude values, a small fraction of the selected galaxies could fall outside the selection areas (and vice versa). This effect is small in the color cuts, with $\sim$ 10 \% of selected galaxies less than 1$\sigma$ away from the dividing line between low-z and quasar dominated regions in Fig. \ref{fig:grw1w2} . Only 4\% of the selected galaxies fulfil that condition for the $r-W1=0$ line in Fig. \ref{fig:rw1_r}. The share of galaxies that could swap from the selected/rejected areas is higher for those with magnitudes close to $r=20$ in Fig. \ref{fig:rw1_r}, reaching values close to 20\% of the selected sample. Nevertheless, since we do not expect a swift physical change at $r=20$ mag, we consider that the contamination induced by this uncertainties is relatively small.

We also observe in both Figures \ref{fig:grw1w2} and \ref{fig:rw1_r} that there is an area with a small population of sources with very negative $r-W1$ color (marked in the figures with a dashed red line at ($r-W1$)$<$1.75, and in Fig. \ref{fig:rw1_r} also with a dashed line at $r$=20 mag). Some of the objects are spectroscopically confirmed to be low redshift galaxies, yet most are stars. In order to avoid missing a small but extreme population, we inspect the morphology and SEDs of these galaxies (10 selected in Fig. \ref{fig:grw1w2} and 19 selected in Fig. \ref{fig:rw1_r}). This subsample is heavily contaminated by stars, but still shows some promising candidates, so we add to the selected sample the 5 objects that show extended morphology (reaching $1\,452$ candidates up to this point).

There are nevertheless $5\,668$ objects that remain in the parent sample and do not have any counterpart in the $unWISE$ catalog. We removed 60 stars identified with a cross-match to Gaia DR2. For this subsample without $WISE$ data, we used a method of removing contaminants considering only the $r$ band magnitude and the apparent effective radius of the object ($R_{\rm eff}$), as measured with the \texttt{R\_EFF} parameter in the J-PLUS database. This value is defined as the radius that encloses half of the total flux in the object in the $r$ band, considering the run of the SExtractor software \citep{Bertin96}, performed in the J-PLUS data reduction pipeline \citep{Cenarro19}. As shown in Fig. \ref{fig:sinunwise}, at $R_{\rm eff}>2.8$ arcsec and $r<20$ mag, the vast majority of sources with spectra are low redshift ELGs: all $5$ sources with spectra but without $unWISE$ counterpart fulfilling those conditions are in fact low redshift ELGs. Using $R_{\rm eff}$ to separate contaminants is less efficient than using $WISE$ photometry (which is why we do not use it in the rest of the sample), but we still recover $92$ extra candidates. At this stage, we have $1\,544$ candidates to be EELGs.

While this final selection biases our sample towards less compact EELGs, the effect is very small. Only 45 objects in the parent sample have no WISE data, $r$<20 and  $R_{\rm eff}<2.8$. Considering the selection of targets with WISE data, we would only expect around half of those objects to be low redshift galaxies. Therefore, missing those galaxies would mean a very small reduction in the sample of candidates ($\sim$ 1.4\%), while the contamination rate would increase by $\sim$ 30 \%. Therefore no attempt is made to include objects without WISE detection and $R_{\rm eff}<2.8$.

\subsubsection{Cosmic ray removal}
In some cases, one of the (usually three) individual frames that are combined to form the final $J0515$ image of an object is affected by a cosmic ray, increasing significantly its flux. In a few of those instances, the data reduction process fails to remove the contaminated frame from the co-added frames, and the object appears in our selection since it apparently shows a strong excess in $J0515$. In order to remove these contaminants, we downloaded all frames that contribute to the $J0515$ image of each candidate, measured the object flux in each one, and checked if any individual frame deviates more than $20$\% from the median flux. This way we removed $21$ objects, one of which is a spectroscopically confirmed star.

\subsubsection{Photometric correction for deblended objects}
\label{sec:blended}
In our sample, in order to reach the highest possible completeness, we have chosen to include objects with a blending flag. This allowed us to accurately analyze objects with relatively close companions without being contaminated by their emission, since the masks for each object have been separated. Nevertheless, this implies that for some extended galaxies, we only selected a small star-forming region, and the rest of the galaxy was considered as a separate object. This would bias our selection, identifying as "galaxies" what in fact are simply star-forming regions. In order to avoid this, we re-computed the photometry of all $375$ selected objects that have a deblending flag. We used SExtractor in dual mode, with detection in the $r$ band, tayloring the parameters of the code to adequate them to the sample of galaxies we were analyzing. In this case we did not perform any correction like the one applied to create the PSFCOR photometry, but given that the galaxies we are dealing with in this step are extended, such a correction would play a very minor role. After a visual inspection, $171$ galaxies were confirmed to be better represented by our SExtractor run rather than the original J-PLUS photometry, and we kept these data during the rest of the analysis. During this stage, $30$ galaxies were removed from the sample: 20 because they were spurious objects (almost all, spikes from bright stars), 1 was a repeated object, and in 9 cases neither the original photometry nor the new SExtractor run were able to properly measure them. For more details, see Appendix \ref{appendix:rephot}.

\subsection{Summary of the sample selection}
\label{sec:summarysel}
With all these considerations, we have built a sample of 1493 sources that are candidates to be extreme [OIII] emitters at ${\rm z}<0.06$, by selecting objects with high $J0515$ over $r$ ratio. We have removed the vast majority of contaminants from other types of sources or higher redshifts galaxies, using J-PLUS and WISE photometry, and SDSS spectra. Out of $85$ objects with available SDSS spectra in the candidate sample only three are not galaxies at ${\rm z}<0.06$: one is a star (likely selected due to an error in the photometric measurement) and two are galaxies at ${\rm z}=0.072$ and ${\rm z}=0.123$, with relatively high uncertainty in the $J0515$ flux. With those three interlopers, the purity our candidate selection is $\sim$96\%. Out of the $2560$ objects with spectra in the excess flux sample, $89$ are galaxies at ${0.006<\rm z}<0.056$ with $r<20$ mag. $82$ of them were selected in the candidates sample, which translates into a completeness of $\sim92$\% in our selection, compared to the excess flux sample.

\section{Results}
\label{sec:results}
To further confirm and characterize the nature of the sample of 1493 candidates to be EELGs at ${\rm z}<0.06$, we performed SED fitting using the \texttt{CIGALE} code \citep{Noll09,Boquien16}, which yields physical properties for the galaxies. We used the synthetic spectra to select only those systems with the highest EW values. Additionally, the galaxy images were inspected to classify their morphologies and the line fluxes estimated with \texttt{CIGALE} were compared to the available SDSS spectra.

\subsection{SED fitting}

To analyze in detail the physical properties of the sample of galaxies candidates to be EELGs, we performed SED fitting with \texttt{CIGALE}. It is a fast and flexible software implementing theoretical models for the different galaxy components. \texttt{CIGALE} creates a large grid of composite stellar populations based on single stellar population models and a variety of SFHs. It includes as well models for dust extinction (both for gas and stars), dust emission, and most importantly, nebular emission (both lines and continuum). The resulting grid of models are fitted to the photometric data, and the galaxy properties are estimated analyzing the posterior likelihood distribution, producing a best-fit model, and a bayesian estimate for each parameter.

\subsubsection{CIGALE parameters}
\label{sec:cigale}
We use a simple model of two stellar populations with an exponentially declining SFH: an old population selected to represent the underlying galaxy, and a young one to reproduce the strong starburst causing the extreme emission lines. This type of modeling has been used in the literature, specially when dealing with galaxies with recent events of star formation \citep[e.g.][]{Nilsson11,Catalan-Torrecilla15,Lopez-Sanjuan17,Lumbreras-Calle19a,ArrabalHaro20}. The very high EW of the emission lines detected in the galaxies in our sample implies a very short timescale for the event, therefore a very young population is more appropriate than a continuous star-formation model. The existence of an old, underlying stellar population has been demonstrated in systems similar to ours, such as blue compact dwarf galaxies \citep{Amorin07,Amorin09}, and it has to be included to account for the majority of the stellar mass in the galaxies. In order to simplify the modelling, we fix the $\tau$ value of the exponential to 50 Myr for the old population and 1 Myr for the young one.

We use \cite{BC03} stellar population models, while the nebular emission \texttt{CIGALE} uses is predicted based on the photo-ionization models by \cite{Inoue11}. We assume the \cite{Calzetti00} extinction law for dust extinction. 

In Table \ref{tab:cigale} we summarize the main free parameters used to fit the galaxy SEDs, and present a representative example in Fig.~\ref{fig:ejemCIG}. For a more detailed account of the SED fitting process, including the values used for all the different parameters, see Appendix \ref{appendix:cigale}.

\begin{table}
\caption{CIGALE parameters}             
\label{tab:cigale}     
\centering                
\begin{tabular}{c c c c}      
\hline\hline                
Parameter & Values  \\    
\hline                        
Redshift & 0 - 0.06 in 0.0025 intervals \\
Age young pop. [Myr]   &  1, 2, 3, 4, 6, 7, 9, 12  \\   
Age old pop. [Myr]    & 200, 500, 1000, 2000, 5000  \\  
Burst ratio & 0.0005, 0.0025, 0.005, 0.01 \\
 &0.03, 0.05, 0.075, 0.15, 0.3 \\
Metallicity [Z] & 0.0001,0.0004, 0.004, 0. 008, 0.02 \\
Escape fraction    &  0, 0.2 \\   
log (U) & -4.0, -3.5, -3.0, -2.5, -2., -1.5\\
E(B-V)$_{\rm young}$  & 0, 0.1, 0.2, 0.3, 0.5 \\   
E(B-V)$_{\rm old}$/(B-V)$_{\rm young}$ & 0.44, 0.9 \\

\hline                                   
\end{tabular}
\end{table}

\subsubsection{SED Results}
\label{sec:sedresults}
    \begin{figure*}
   \centering
   \includegraphics[width=0.95\textwidth,keepaspectratio]{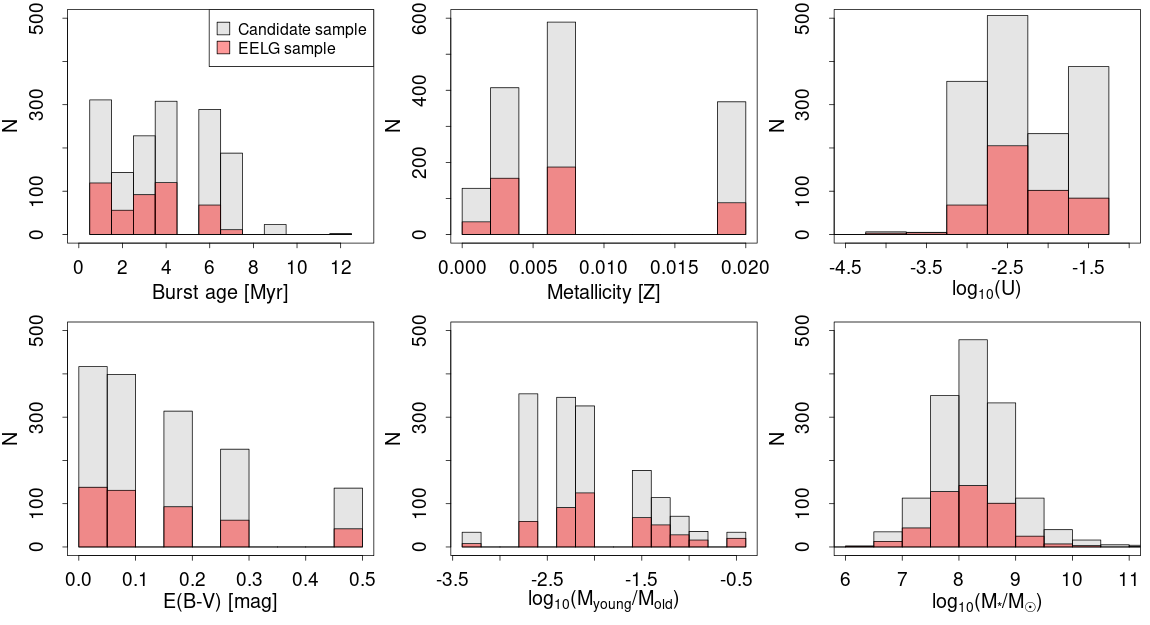}
      \caption{Histograms showing the best fitting values for the J-PLUS galaxies, derived using the SED fitting software \texttt{CIGALE}. We show in grey the results for the 1493 galaxies in the candidate sample and in red the 466 in the EELG sample (EW[OIII]$>$300 \text{\AA}). From left to right and top to bottom, we show the age of the young population, the stellar metallicity, the ionization parameter, the colour excess, the mass ratio between young and old populations, and the total stellar mass. The gaps observed in some of the histograms are due to the sampling in the parameters (see Table \ref{tab:cigale} and Appendix \ref{appendix:cigale}).}
         \label{fig:CIGparam}
   \end{figure*}
In Fig. \ref{fig:CIGparam} and Table \ref{tab:maincig} we show the distribution of the main stellar and nebular parameters that \texttt{CIGALE} delivers for our sample of 1493 EELG candidates. We have chosen to use the best-fit parameters instead of the Bayesian estimates in the \texttt{CIGALE} output for consistency, since \texttt{CIGALE} does not provide Bayesian-estimated for individual components in the synthetic spectra, and we rely on them to compute EWs. In addition, the Bayesian-estimated photometric redshifts are less accurate than the best fitting ones (see section \ref{sec:comparspec}). 

It is important to note the limits on the accuracy of these SED fits. The main driver of them is to obtain accurate estimations of the flux of the most intense emission lines, as well as the underlying continuum flux. Additionally, we rely on the stellar mass estimates for comparison purposes, and in other parameters (burst age, extinction, metallicity) for broad properties of the sample, given the assumptions in the \texttt{CIGALE} run. The results of the fits cannot be analyzed in extreme detail, specially in some sparsely sampled, highly degenerated parameters like metallicity, log(U), escape fraction, etc. For an additional discussion on this topic, see appendix \ref{appendix:cigale}.

Considering the best-fit parameters, the age of the star formation burst is very low, with almost no galaxies showing values higher than 8 Myr. This is consistent with the extreme EW values measured, since that parameter decreases very rapidly in the first Myrs of a star-formation burst \citep[see e.g.][]{Leitherer99}.

The metallicity and ionisation parameter of the sample are low. This is typical for low mass galaxies with strong bursts of star formation, specially at high redshift  \citep{Khostovan15,Tang20,Matthee21}. Nevertheless, the selection process for our sample, imposing a very high EW of [OIII], may prevent the selection of some extremely metal poor galaxies (see Sect. \ref{sec:o3hbeta} for more details).

The typical extinction values that \texttt{CIGALE} derives for the main sample are low, which is consistent with previous results obtained for samples of low-mass star-forming galaxies \citep{Garn10,DuartePuertas17,Lumbreras-Calle19a}.

Even considering the strong burst of star formation that these galaxies are enduring, the mass ratio between old and young population is typically low (log$(M_{\rm you}/M_{\rm old)} =-2.00\substack{+0.70\\ -0.60}$). This is however not surprising, since the mass-luminosity ratios vary strongly between old and very young populations. Therefore, while the old population dominates the total mass of the galaxy, a relatively small (in mass) young population can have strong effects in the integrated photometry of a low-mass galaxy. The total stellar masses of the galaxies in the sample are not constrained by input parameters, and they span most of the range of what is usually considered dwarf galaxies, with log$(M_{\star}/M_{\odot}) \sim 6.5 - 9.5$.
    
\begin{figure}
   \centering
   \includegraphics[width=0.45\textwidth,keepaspectratio]{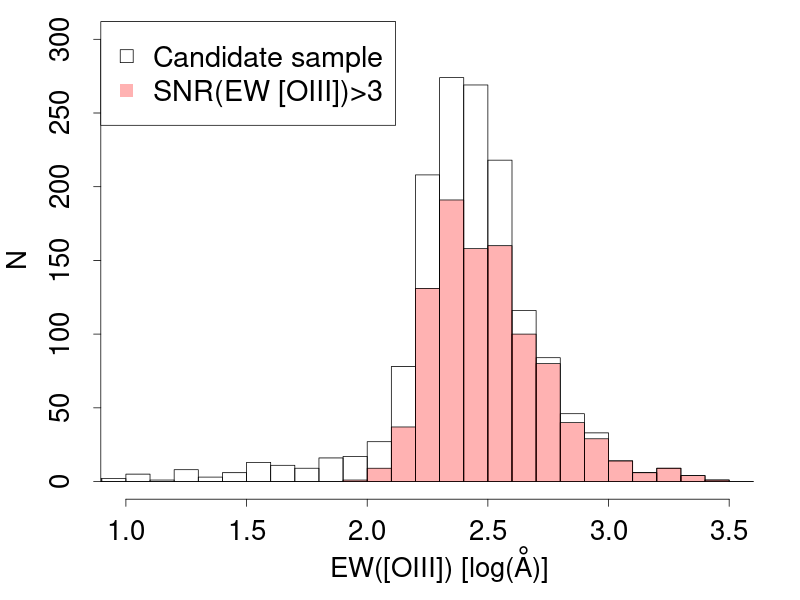}
      \caption{Histogram of EW([OIII]), measured using the J-PLUS $J0515$ photometry and the \texttt{CIGALE} fits for the continuum. The white filled histogram represents the whole sample of candidates, while the red filled one shows only the galaxies with SNR$>$3 in the EW([OIII]) value.}
         \label{fig:EWO3hist}
   \end{figure}

Two of the main parameters in the analysis of ELGs are the flux and EW of the emission lines (in our case specially the [OIII] line, used for the sample selection). We focus in the $J0515$ filter, where most of the line flux comes from the [OIII]4959+[OIII]5007 lines. To comput them, we first convolve the synthetic stellar and nebular continuum derived by \texttt{CIGALE} (blue line in Fig. \ref{fig:ejemCIG}) through the transmission curve of the $J0515$ filter, to obtain an estimation of the continuum in that filter ($F_{\rm cont.}$, blue squares in Fig. \ref{fig:ejemCIG}). Then, using the measured flux in that filter ($F_{\rm J0515}$, red triangles in Fig. \ref{fig:ejemCIG}) and the classical formula (assuming a flat continuum within the filter, and considering fluxes in units of erg/s/cm$^2$/\text{\AA}), we obtained the EW as
\begin{equation}
    {\rm EW [\text{\AA}]}=\frac{F_{J0515}-F_{\rm cont.}}{F_{\rm cont.}}\Delta=\frac{F_{\rm em}}{F_{\rm cont.}}\Delta,
\end{equation}
where $\Delta$ is the width of the filter, 200 \text{\AA}, and $\Delta F_{\rm em}$ is the line flux. In Fig. \ref{fig:EWO3hist} we show the histogram of EW for the $J0515$ filter, separating the galaxies with ${\rm SNR} > 3$ in the $J0515$ EW.


\begin{table*}
\caption{Main \texttt{CIGALE}-derived properties of the J-PLUS selected EELGs. The complete table is available online; only the first row is shown here as guidance.}             
\label{tab:maincig}      
\centering                          
\begin{tabular}{c c c c c c c c c c}        
\hline\hline                 
ID & Redshift & Mass & Mass ratio & Age young & Metallicity & log(U) & E(B-V) & F$_{\rm [OIII]5007}$ & F$_{\rm H\alpha}$\\    
 & & [$log(M_{\odot})$]& [$log(\frac{M_{\rm you}}{M_{\rm old}})$]  & [Myr] & [Z] & & mag &  \multicolumn{2}{c}{$log(erg/s/cm^2)$}\\

\hline                        
63414026530 & 0.0075 & 6.96 & -1.3 & 4.0 & 0.004 & -2.0 & 0.10 & -12.88 & -13.16 \\
\end{tabular}
\end{table*}

\begin{table}
\caption{Main photometric properties of the J-PLUS selected EELGs. The complete table is available online; only the first row is shown here as guidance.}             
\label{tab:main}      
\centering                          
\begin{tabular}{c c c c c c c c c c}        
\hline\hline                 
ID & $r$ & EW [OIII] & EW H$\alpha$ \\    
 & mag & \text{\AA} & \text{\AA} \\
\hline            63414026530 & 17.80 $\pm$ 0.04 & 505  $\pm$ 12 & 480 $\pm$ 100 \\
  
\end{tabular}
\end{table}

\subsubsection{Comparison between \texttt{CIGALE} and J-PLUS fluxes}

\begin{figure}
   \centering
   \includegraphics[width=0.48\textwidth,keepaspectratio]{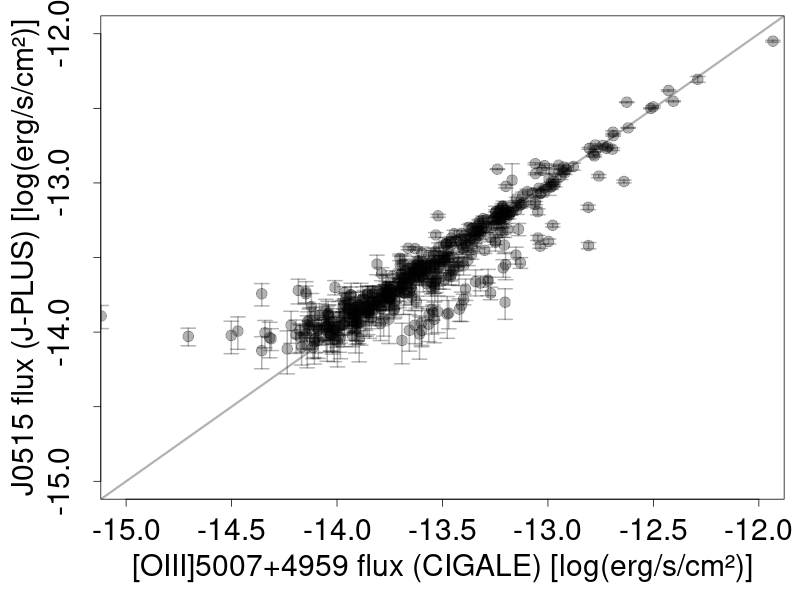}
      \caption{Comparison between the emission line flux measured in the $J0515$ filter and the [OIII]5007+4959 flux estimated by the \texttt{CIGALE} SED fit for the EELG sample.  The black line represents the one-to-one relation.}
         \label{fig:fluxJ515_OIII}
   \end{figure}

\begin{figure}
   \centering
   \includegraphics[width=0.48\textwidth,keepaspectratio]{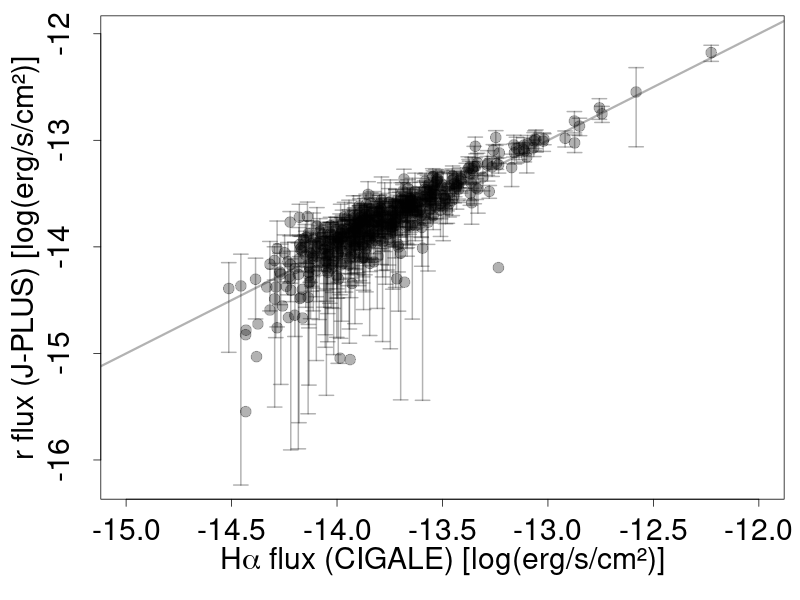}
      \caption{Comparison between the emission line flux measured in the $r$ filter and the H$\alpha$ flux estimated by the \texttt{CIGALE} SED fit for the EELG sample. The black line represents the one-to-one relation.}
         \label{fig:fluxr_Ha}
   \end{figure}

Given the redshift range covered, $0.0075 < {\rm z} < 0.06$, for the majority of the sample the $J0515$ filter is affected by both the [OIII]5007 and [OIII]4959 emission lines. According to the photometric redshifts obtained from \texttt{CIGALE}, for a small fraction of galaxies ($\sim$ 18 \%, at the higher redshift end), the [OIII]5007 line lies in a low transmission region of the $J0515$ filter (or even entirely out of it). The H$\beta$ line enters in the filter wavelength range in almost half of the sample, while at the very low redshift end, some galaxies ($\sim$ 5 \%) lack the contribution of the [OIII]4959 line. Nevertheless, the limited photometric redshift precision (see Sect. \ref{sec:comparspec}) prevents us from providing a detailed account of this distribution of emission lines, galaxy by galaxy. We can still study the whole sample comparing the emission line flux in the whole filter, estimated using the $J0515$ photometry and the \texttt{CIGALE}-derived continuum value, with the fluxes computed by \texttt{CIGALE} for each emission line. The same can be done for the $r$ broadband filter, which is significantly contaminated by the H$\alpha$ emission line. In Figs. \ref{fig:fluxJ515_OIII} and \ref{fig:fluxr_Ha} we show the logarithm of line fluxes derived directly from the J-PLUS photometry, as a function of the \texttt{CIGALE}-estimated individual line fluxes. The integrated fluxes are clearly dominated by the brightest emission lines in several filters. In the $r$ filter (Fig. \ref{fig:fluxr_Ha}), only taking into account the H$\alpha$ line is enough to recover a very good one-to-one relationship, with a small offset ($\sim$ -0.04 dex) and low scatter (1$\sigma\sim$0.13). If we include in the analysis the [NII]6584 flux, the offset is greatly reduced (down lo -0.008 dex), with a similarly low scatter (1$\sigma\sim$0.12 dex). Nevertheless, given the small impact of [NII]6584, we consider that the line emission in the $r$ filter is mostly dominated by H$\alpha$. For the $J0515$ filter (Fig. \ref{fig:fluxJ515_OIII}), taking into account the [OIII]4959 and [OIII]5007 lines for all galaxies provides a reasonably good fit, with a negligible offset ($\sim$ -0.007 dex) and and very low scatter ($\sim$ 0.096 dex). Nevertheless, in the highest redshift range of our sample (z$\geq$0.05), the [OIII]5007 line falls in a very low transmission region of the $J0515$ filter. This translates into the parallel subsample, $\sim$ 0.35 dex below the one-to-one relationship, clearly noticeable in Fig. \ref{fig:fluxJ515_OIII}. Only $\sim$ 7.5\% of the sample suffers this effect, and given the redshift uncertainties, we cannot exclude these galaxies from further analysis. In conclusion, we consider that the $J0515$ EW and line fluxes correspond to the sum of the [OIII]4959 and [OIII]5007 emission lines, while the $r$ line flux and EW correspond to H$\alpha$. For the $18$ galaxies with photometric redshift z$_{\rm phot}<0.017$, the H$\alpha$ line falls within the $J0660$ filter wavelength range, and the precision of the H$\alpha$ measurement is much higher, with an offset of $\sim0.002$ dex and a scatter of $\sim0.02$ dex. 

\subsection{Selection of the EELG sample}
\label{sec:selecEELG}
\begin{figure}
   \centering
   \includegraphics[width=0.48\textwidth,keepaspectratio]{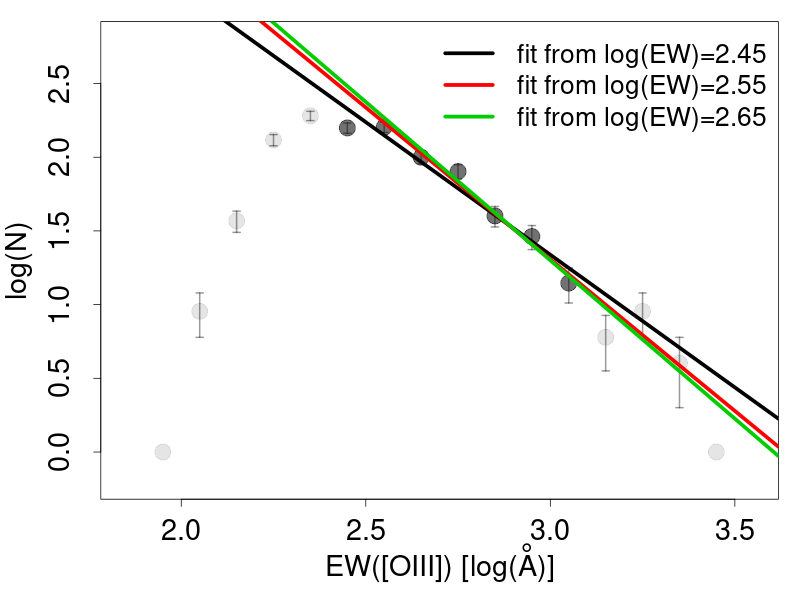}
      \caption{Diagram used to asses the completeness of the EELG sample. We plot the number of galaxies per log(EW[OIII]) bin as a function of log(EW[OIII]). We overplot the linear fits to the data considering three different limits in the lower end 2.45, 2.55, and 2.65 log(\text{\AA}).}
         \label{fig:completeness_ew}
   \end{figure}

Our original selection of candidates was performed using a very rough estimation of the stellar continuum below the [OIII] line (simply the $r$ band flux), and this results in an extended [OIII] EW distribution (see Fig. \ref{fig:EWO3hist}). We decided to perform an additional cut to obtain a clearly defined EELG sample, with a strict limit in EW. We considered (only for galaxies with SNR$>3$ in [OIII] EW) how the logarithm of the number of galaxies in each EW bin varies with EW, given the expected decreasing exponential relation (see Fig. \ref{fig:completeness_ew}). This exponential relation is clear (with R$^2$=0.998), for example, when selecting galaxies with EW([OIII])$>$10 \text{\AA} and SNR$>$3 in the SDSS spectroscopic database (the GalSpecLine table from DR8, \citealt{Aihara11}). In our case, for a given lower limit of EW, the number of galaxies will not increase as fast as an exponential: that would be our completeness limit. In order to compute it accurately, we perform linear fits with EW bins of 0.1 dex in log(\text{\AA}), limiting the analysis in the high EW range to the bins showing more than 10 galaxies, which translates into log(EW) $<$ 3.05 log(\text{\AA}). For the low EW limit, we perform three separate linear fits, covering the bins from 2.45, 2.55, and 2.65 log(EW), respectively. For log(EW) values lower than 2.5 (EW $\sim$ 300 \text{\AA}), the measured value drops below that of a linear fit to the bins above it (Fig. \ref{fig:completeness_ew}), with the fits limited to 2.55 and 2.65 showing comparable parameters. Therefore we place our threshold at EW = 300 \text{\AA} in [OIII] to keep a complete sample, obtaining a total of 466 EELGs. In the higher EW limit, completeness is limited by the $r$<20 threshold defined in Sect. \ref{sec:databas}, given that fainter (and lower mass) galaxies tend to show higher EWs (see Sect. \ref{sec:ewmass}). This magnitude cut is nevertheless necessary in order to provide with a clearly defined sample, and not to include significant contamination of high-z AGNs (as seen in Fig. \ref{fig:rw1_r}).

The value we selected is higher than some limits found in the literature for starburst galaxies or EELGs (100 \text{\AA} in \citealt{Amorin15}, 80 \text{\AA} in \citealt{Hinojosa-Goni16}), but lower than most of the galaxies in other surveys \citep{Cardamone09,Yang17}, especially those at high redshift \citep{vanderWel11,Atek11}. We consider that the intermediate value of 300 \text{\AA} is appropriate, given the width of the filter used to select the sample. It provides a large sample, showing undoubtedly strong emitters, allowing both a statistical analysis and low contamination. A higher threshold would have removed interesting objects and restricted the sample, while a lower one could provide a high rate of contaminants unless we removed the lower brightness objects, which would be large fraction or the sample. Comparing with the SDSS spectroscopic database (using again the GalSpecLine table from DR8), we see that with a 300 \text{\AA} threshold we keep only the top 2\% of the EW([OIII]) distribution (considering only galaxies with EW([OIII])>10\text{\AA}), which further confirms that the emission properties of our sample are extreme.

We consider as well the effect of the initial selection (Eq. \ref{eq:selec}) in the H$\alpha$ flux distribution of the final sample. Since the H$\alpha$ lines falls within the wavelength range of the $r$ filter and Eq. \ref{eq:selec} selects objects with high $J0515$ to $r$ ratio, we could be biasing the sample, removing objects with relatively high [OIII] EW yet low [OIII]5007/H$\alpha$ ratio. To test if this effect was present, we artificially increased the \texttt{CIGALE}-estimated H$\alpha$ fluxes in the selected galaxies (those with EW([OIII])$>$300 \text{\AA}) and check if they still fulfilled Eq. \ref{eq:selec}. A 25\% increase in the H$\alpha$ flux was used, since that produced a significant offset in the [OIII]5007/H$\alpha$ ratio as a function of EW([OIII]), unobserved in the spectroscopic data. Using that 25\% increase, only 16 objects (3.2\%) were rejected, therefore we are confident that no significant bias is added to our sample against galaxies with low [OIII]5007/H$\alpha$ ratio due to Eq. \ref{eq:selec}.

This sample selection results in 466 EELGs. In order to determine how novel the J-PLUS database and the present work are, we investigate the amount of objects in our sample that have not been previously identified as EELGs. To do so, we first query the NASA Extragalactic Database (NED) to select catalogued objects near the positions of the galaxies in our sample. Only 53 have a reported spectroscopic redshift, and we consider that those galaxies have been already classified as EELGs. In addition, 14 galaxies without spectroscopic redshift in NED have been referenced in at least one publication, with 2 of them described as EELGs or a similar category. Finally, we check that all objects with spectroscopic confirmation in this EELG sample are indeed low redshift galaxies, which translates into a theoretical purity of 100\%. In conclusion, the present work identifies as EELGs 411 galaxies (88\% of the sample) that were previously unknown to belong to this class. This highlights how the J-PLUS survey is well suited to find new extreme emitters, reaching fainter magnitudes than spectroscopic surveys covering wide areas of the sky.

\subsection{Morphology of the EELG sample}

      \begin{figure}
   \centering
   \includegraphics[width=0.48\textwidth,keepaspectratio]{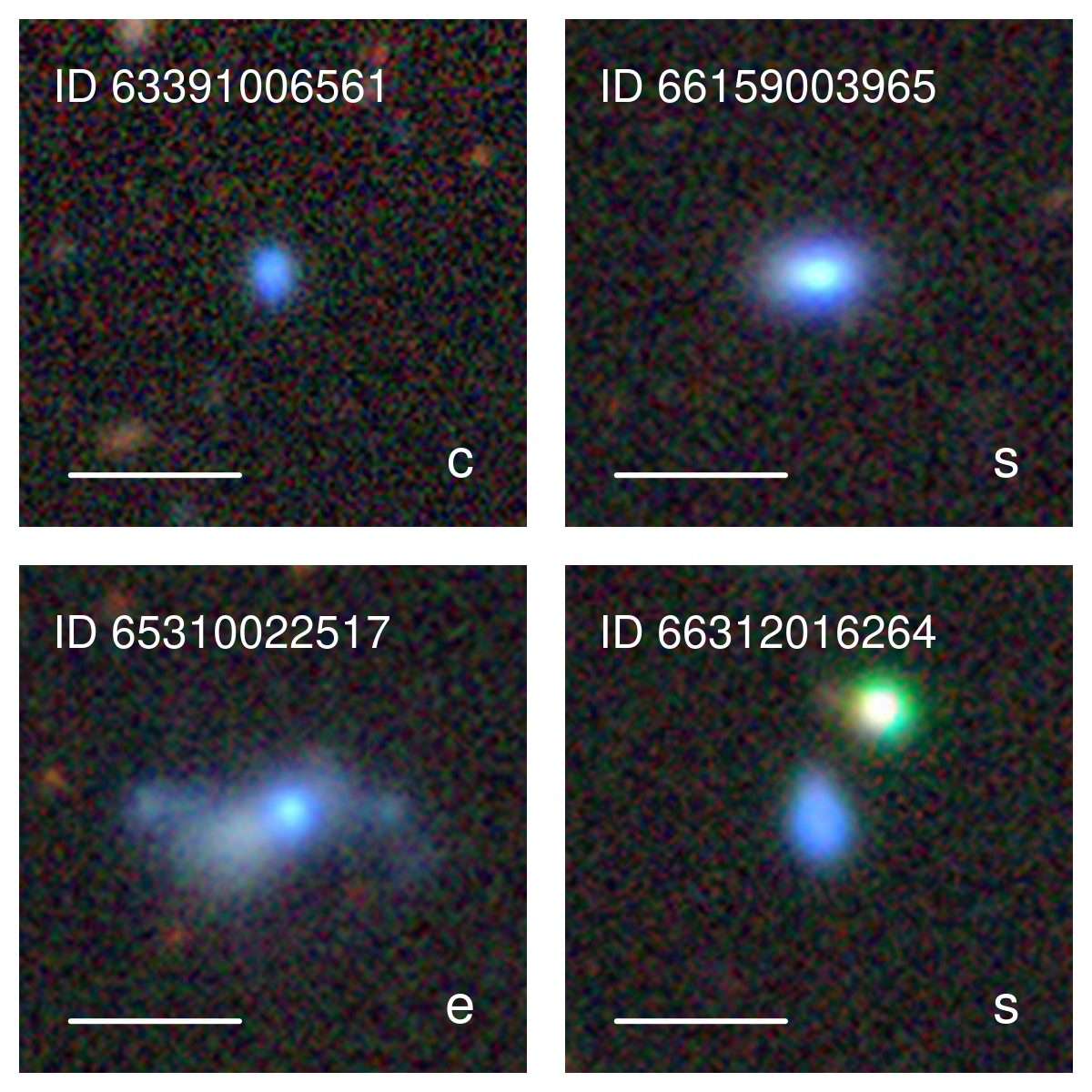}
      \caption{RGB composites (made using images taken with $z$, $r$, and $g$ filters) of four galaxies in the EELG sample. In the bottom right of each panel a letter indicates the type of morphology: compact (c), semi-compact (s), or extended (e). All images have been obtained using the Legacy Survey viewer, with data from the BASS and MzLS surveys \citep{Zou19, Dey19}.}
         \label{fig:exam_morpho}
   \end{figure}

\label{sec:morpho}
We have performed a simple visual classification of the morphology of the extreme [OIII] emitters. The EELGs in the sample have been classified into three categories: compact (c), semi-compact (s), and extended (e). Compact those galaxies show a circular shape, with no further structure discernible. We classify them in the semi-compact category when their shape is dominated by a circular and bright clump, but some fainter structure is revealed: a tail, a halo, etc. We consider extended galaxies those showing a complex morphology, those with no bright clump, or those where the clump or clumps does not dominate the light profile in the galaxy. As an example, we show the RGB post stamps of four galaxies in Fig. \ref{fig:exam_morpho}. 

For this analysis we took the images from the Legacy Survey database\footnote{\url{https://www.legacysurvey.org/viewer}} that had been obtained in the framework of the BASS and MzLS surveys \citep{Zou19, Dey19}. They provide images in the $g$, $r$, and $z$ bands, with median depths of 23.65, 23.08, and 22.60 mag respectively (considering 5$\sigma$ detections of point sources), and typical FWHM values of their PSFs of 1.61", 1.47", and 1.01" \citep{Dey19}. These images reach low surface brightness detection limits in $r$ of 27.9 mag arcsec$^2$ for 3$\sigma$ detection of a 100 arcsec$^2$ feature, fainter than the SDSS and Panstarrs PS1 surveys, and comparable to SDSS Stripe-82 \citep{Hood18}. Therefore, even if fainter features than those visually noticeable in the RGB composite images may exist (and change the morphological classification), we consider the present analysis provides with results accurate enough for this work.

The compact class is the most common with $43$\% of the EELG sample, followed by the semi-compact one ($38$\%), the rest being classified as extended. This is consistent with previous studies in this topic (e.g. \citealt{Izotov11,Yang17}), focused in the compact or semi-compact galaxies. Nevertheless, we show that a significant amount (19\%) of the EELGs in our sample are more properly classified as extended systems, suggesting possible extensions of previous EELG searches.

\subsection{Comparison with spectra}
\label{sec:comparspec}
We assessed the accuracy of the properties derived using J-PLUS photometry and the \texttt{CIGALE} SED analysis by disucssing the results obtained for the 82 sources in the candidate sample with SDSS spectra. For this analysis, we removed the misclassified objects (one star and two high-redshift galaxy).
      \begin{figure}
   \centering
   \includegraphics[width=0.48\textwidth,keepaspectratio]{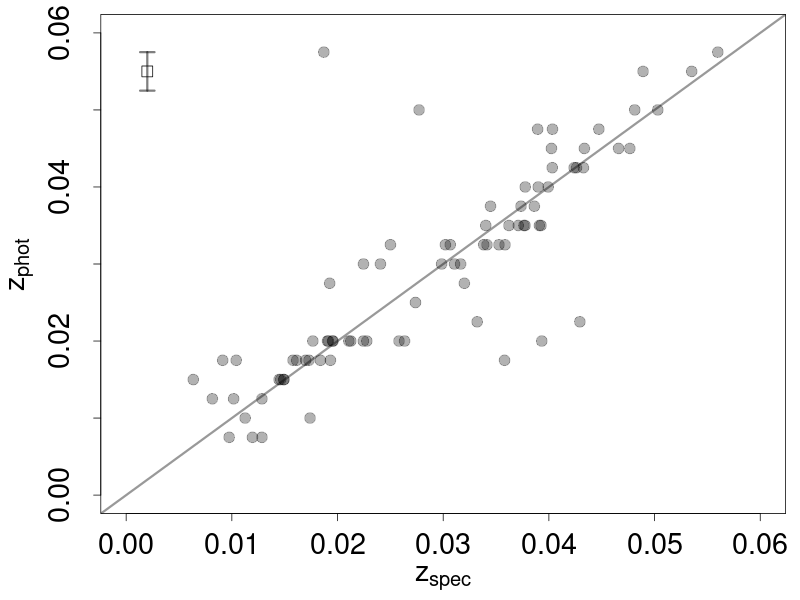}
      \caption{\texttt{CIGALE} photometric redshift as a function of SDSS-derived spectroscopic redshift, for the galaxies in our sample of candidates that have available spectra (excluding the three misclassified objects). The line represent the one to one relationship. On the top left of the figure we show as error bars the minimum step in photometric redshift considered with \texttt{CIGALE} (0.0025).}
         \label{fig:zspeczphot}
   \end{figure}
First, we measured the accuracy of the photometric redshift determined with \texttt{CIGALE}, which are compared in Fig. \ref{fig:zspeczphot} with the spectroscopic redshifts, showing very good agreement. We computed a parameter to quantify the quality of the agreement, the $\sigma$NMAD, defined as
\begin{equation}
\sigma_{\rm{NMAD}} = 1.48 \times \textrm{median}\big{\vert}\Delta z - \textrm{median}(\Delta z)\big{\vert},
\end{equation}
where $\Delta z = (z_{\rm best} - z_{\rm spec}$)/(1+$z_{\rm{spec}}$). We obtained a value of  $\sigma{\rm NMAD}\sim0.003$, similar to the mini-JPAS results, with $56$ narrow band filters \citep{Hernan-Caballero21}. This high accuracy is likely due in part to the limited redshift range covered, to the extreme intensity of the emission lines, but most importantly, to the fact that those lines ([OII]3727, [OIII], and H$\alpha$) lie inside or near the wavelength range of narrow or mediumband filters, which produces very clear photometric features. In fact, for more than half of this subsample (56\%), the difference between photometric and spectroscopic redshift is smaller than the minimum step used in the photometric redshift determination (0.0025). This high accuracy diminishes the possible uncertainties in our determinations of absolute parameters, such as luminosities and stellar masses. We used the best fitting \texttt{CIGALE} models instead of the Bayesian estimates, in part because the agreement with spectroscopic redshift values drops when using the latter. In consequence, the values discussed for the SED fitting (ages, masses, etc.) correspond to the best-fit models.

      \begin{figure}
   \centering
   \includegraphics[width=0.48\textwidth,keepaspectratio]{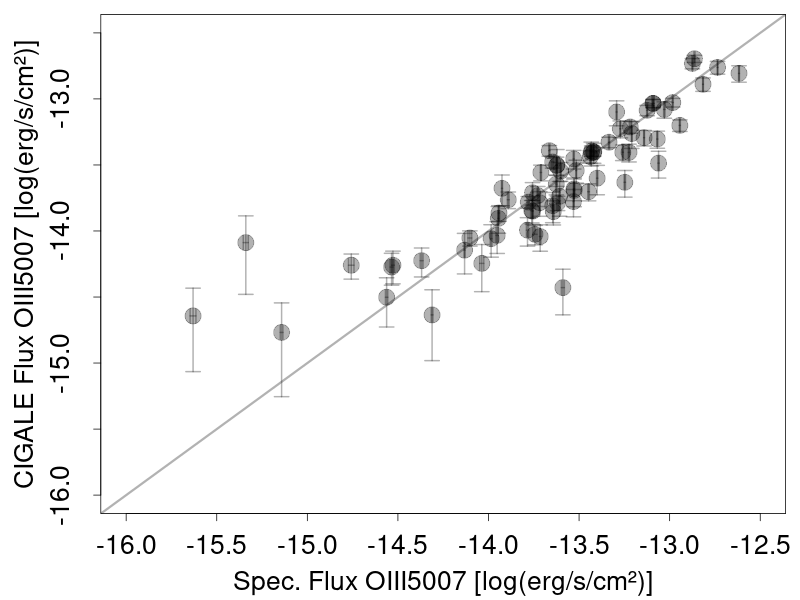}
      \caption{[OIII]5007 flux estimated using \texttt{CIGALE} SED fitting on J-PLUS data as a function of [OIII] flux measured in SDSS spectra. The black line represents the one-to-one relation.}
         \label{fig:fluxo3_compar}
   \end{figure}

      \begin{figure}
   \centering
   \includegraphics[width=0.48\textwidth,keepaspectratio]{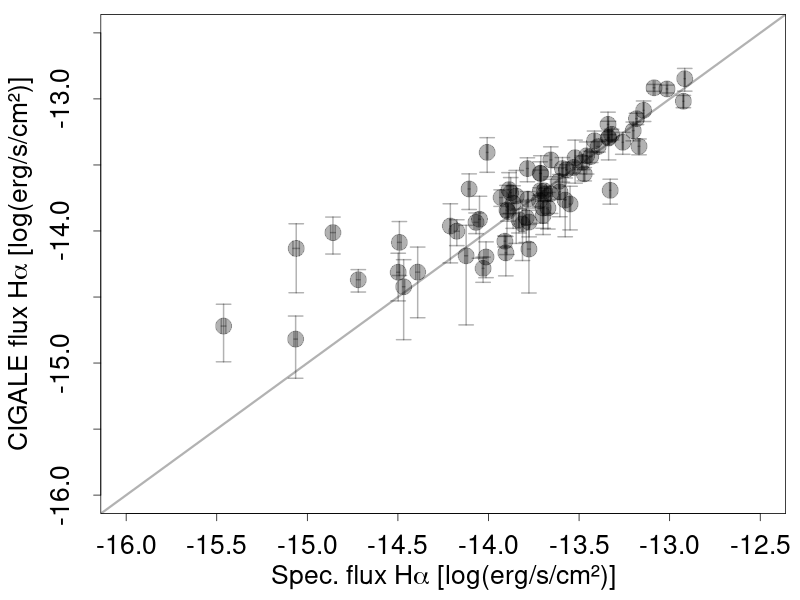}
      \caption{H$\alpha$ flux estimated using \texttt{CIGALE} SED fitting on J-PLUS data as a function of H$\alpha$ flux measured in SDSS spectra.The black line represents the one-to-one relation.}
         \label{fig:fluxha_compar}
   \end{figure}

We also compare the fluxes and EWs estimated in the J-PLUS photometry with those measured using the SDSS spectra. We use our own code to measure the emission line fluxes and EWs in the spectra, a revised version of the procedure used in \cite{Lumbreras-Calle19a}. Briefly, we compute the continuum underneath each emission line by masking the line and performing a linear fit to the remaining spectrum in a 200 \text{\AA} aperture around the line. We estimate the values and uncertainties of the linear fits by performing Bootstrap simulations. We compute the flux of each emission line simply adding the measured flux over the spectral window considered and subtracting the continuum value (and then dividing by the continuum value to get the EW).

It is important to remind the reader that the photometry used in this work has been the PSFCOR J-PLUS photometry, re-scaled to the AUTO aperture value in $r$. In order to perform an accurate comparison, we need to use the 3ARCSEC aperture in the J-PLUS catalog, which should match more closely the SDSS spectroscopic measurement (which is performed with a 3 arcsecond wide fiber in most of the sample). Nevertheless, some sources of discrepancy still exist: different seeing conditions, small differences between the sky position of the objects, or a possible offset between absolute calibrations. We compared the 3ARCSEC $r$ fluxes in J-PLUS with the synthetic $r$ fluxes computed from the spectra, and found very good agreement (a 1$\sigma$=0.12 dex scatter) with a small but noticeable offset (0.08 dex). Therefore, we re-scaled the 3ARCSEC photometry to the SDSS flux using this $r$ band offset.

For a precise comparison, line by line, we need to use the \texttt{CIGALE} model output, which provides us with fluxes for several emission lines ([NII]6584, H$\alpha$, H$\beta$, [OIII]5007, [OIII]4959, and [OII]3727). In Figures \ref{fig:fluxo3_compar} and \ref{fig:fluxha_compar} we show the comparison between spectra and SED models for the [OIII]5007 and H$\alpha$ lines respectively. For [OIII]5007 the agreement between the measurements is very good, with almost no bias (a median difference of -0.001 dex) and very low scatter, a 1$\sigma$ value of $\sim$ 0.18. We can also compare our result with others in the literature. In the Census of the Local Univers (CLU) preliminary fields, \cite{Cook19} show the comparison between photometric and spectroscopic H$\alpha$ fluxes in their figure 10. Considering only their 5$\sigma$ detections, we estimate a 1$\sigma$ scatter of $\sim$ 0.25. This value is higher than ours, which is striking considering that their filters are notably narrower than ours (from 76 to 92 \text{\AA} compared to 200 \text{\AA}). For a more similar comparison, using J-PLUS data, we look at the results in \cite{Logrono-Garcia19}. Considering H$\alpha$ fluxes measured in the $J0660$ filter, these authors reach more accurate values than ours, with a 1 $\sigma$ scatter of $\sim$ 0.11. This result is expected, given that the $J0660$ is narrower than $J0515$ ($\sim$ 140 vs. $\sim$ 200 \text{\AA}), and that they took extreme care in matching the apertures of the photometric and spectroscopic measurements, which included integral field unit data. 

      \begin{figure}
   \centering
   \includegraphics[width=0.48\textwidth,keepaspectratio]{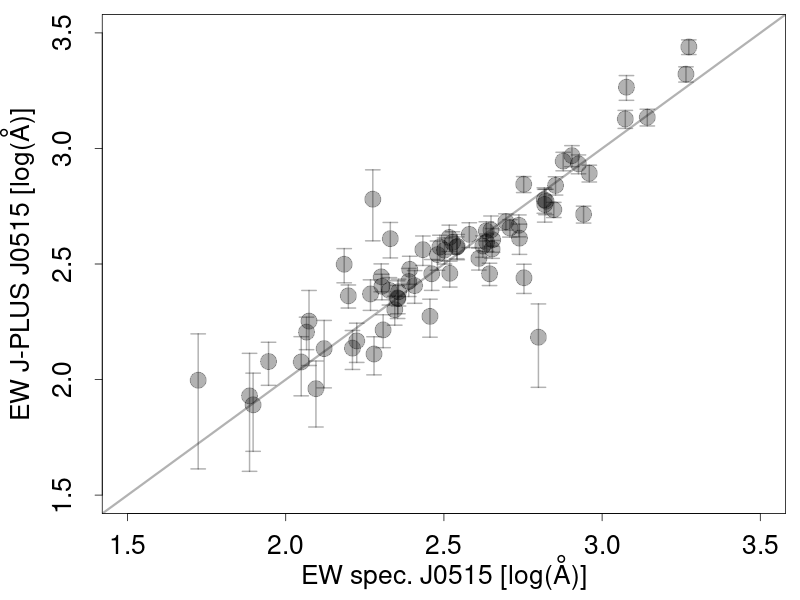}
      \caption{J0515 EW directly measured on J-PLUS data as a function of $J0515$ EW estimated convolving SDSS spectra with the filter transmission. }
         \label{fig:ewJ0515compar}
   \end{figure}

      \begin{figure}
   \centering
   \includegraphics[width=0.48\textwidth,keepaspectratio]{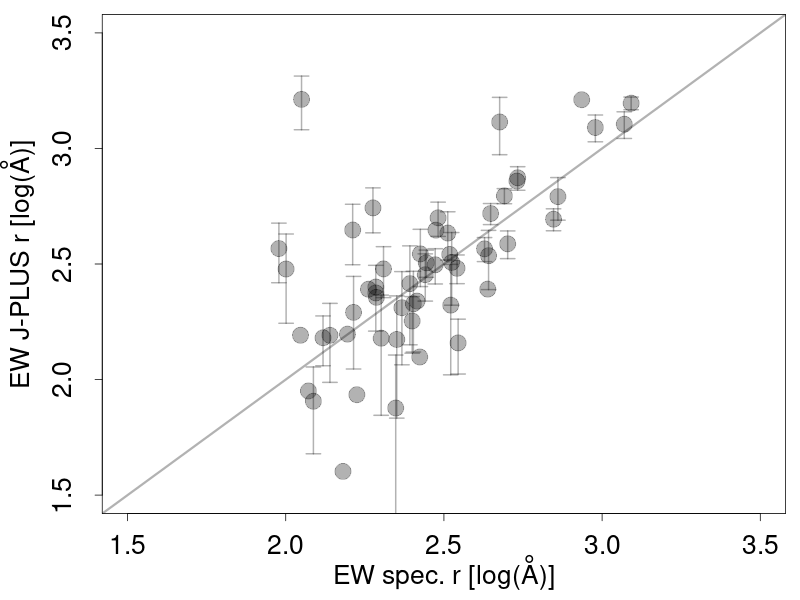}
      \caption{$r$ EW directly measured on J-PLUS data as a function of $r$ EW estimated convolving SDSS spectra with the filter transmission.}
         \label{fig:ew$r$compar}
   \end{figure}

The comparison for the H$\alpha$ line is also very successful, with  a similar offset ($\sim$  dex) and scatter ($\sim$ 0.18 dec.) even if the line flux is estimated using the broadband $r$ filter. This value is likely low thanks to the data in the J0660 filter, that very accurately traces the continuum near the H$\alpha$ line for galaxies at z$>$0.017. For the few galaxies in our sample with available spectra and z$<$0.017, \texttt{CIGALE} uses data from the J0660 filter to estimate the H$\alpha$ flux, obtaining an even better agreement (with an offset of $\sim$-0.01 and a scatter of $\sim$ 0.12). Other emission lines show a worse agreement between spectra and \texttt{CIGALE} models, some simply because they are less intense  (for example, the [NII]6584 line, which in addition is very close to the much more intense H$\alpha$ line). Some others are not only faint, but also located in regions were the spectroscopic and photometric analysis is more complex. For example, the [OII]3727 line lies outside the wavelength range of most of the spectra (and on the edge of some), and it can fall in the gap between two blue narrowband J-PLUS filters ($J0378$ and $J0395$) which are also harder to calibrate than the redder ones. Nevertheless, it is worth noting that making direct measurements on the J-PLUS data, the [OIII]/[OII] ratio of the sample of EELGs reaches extremely high values compared to typical star-forming galaxies. This indicates very hard ionizing radiation, and is similar to very high redshift galaxies. We will explore this result in depth in an upcoming paper analyzing the spectra of some of the galaxies in the current sample. 

The previous comparison between line fluxes, even if very successful for [OIII]5007 and H$\alpha$, relies heavily on the SED models. For a less model-dependant version, we look into the comparison between the EW measured directly in the J-PLUS photometry and the synthetic EW values for each filter, obtained convolving the emission lines in the SDSS spectra with the filter transmission curves. We should note that there is still some modeling implied, since we used the \texttt{CIGALE} fits in order to estimate the continuum in each filter. We show the results for $J0515$ and $r$ in Figures \ref{fig:ewJ0515compar} and \ref{fig:ew$r$compar}, respectively. In this case, the agreement is even better, with similarly small offsets (0.02 for $J0515$ and 0.04 for $r$) but lower scatter values (0.12 and 0.17, respectively). This result strengthens our confidence in the accuracy of J-PLUS photometry and our SED and EW analysis.

\section{Discussion}
\label{sec:discussion}
In this section we compare our sample with different works in the literature, regarding their broadband colors, number density values, and the ratio between [OIII] and the H$\alpha$ and H$\beta$ emission lines. In addition, we discuss the physical properties derived for our EELG sample, considering EW, star formation rate (SFR), mass, and other properties in the context of the literature at different redshifts.

Given the good agreement between the photometric and spectroscopic redshifts, as well as between the photometric and spectroscopic line fluxes and EWs, we are able to discuss the results of this work without performing follow-up spectroscopic observations on the whole sample of galaxies. That step has been however necessary in previous works that used only broadband selection \citep{Yang17,Senchyna19,Kojima20} or narrow band selection like H$\alpha$ dots \citep{Kellar12,Salzer20} and the CLU survey \citep{Cook19}. Even if the physical information we can obtain is limited compared to what can be derived from spectroscopic observations, it is still significant, considering the data spans $2\,176$ sq. deg. down to $r=20$. In addition, this photometric analysis allows for a more efficient spectroscopic follow-up, targeting galaxies with specific physical properties. This characteristics will improve significantly in the upcoming J-PAS survey, with deeper observations and higher spectral resolution.

\subsection{Testing the sample selection}

\subsubsection{Comparison with broadband color-color selection}

\begin{figure}
   \centering
   \includegraphics[width=0.48\textwidth,keepaspectratio]{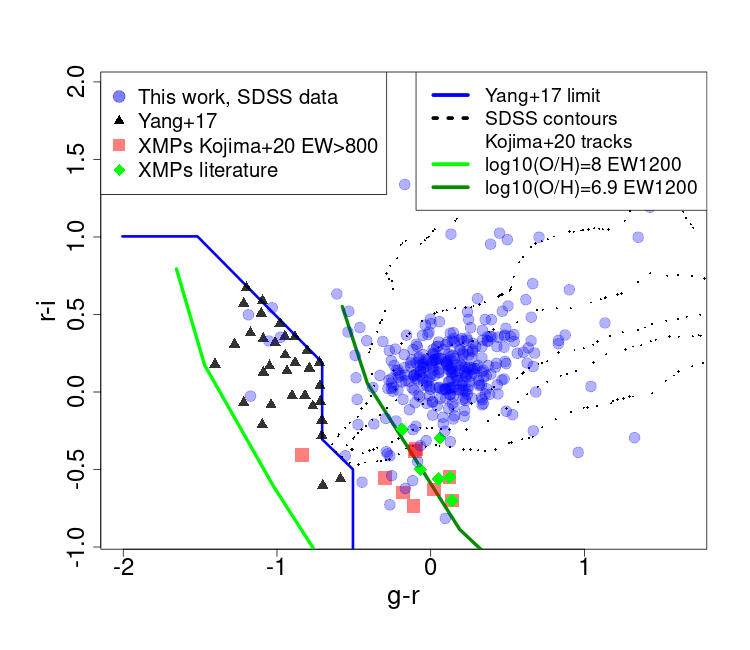}
      \caption{Broadband color-color diagram, created using SDSS data. We plot our EELG sample in blue dots, alongside the blueberry galaxies in \cite{Yang17} and extremely metal poor galaxies (XMP) from \cite{Kojima20} and the literature. The blue line represent the \cite{Yang17} sample selection limits, while the green lines follow the evolutionary tracks of models in \cite{Kojima20}. The small black dots represent the density contours of typical SDSS galaxies \citep{Kojima20}.}
         \label{fig:broadcol}
   \end{figure}

Several works over the past years have used broadband colors from large photometric surveys (most notably SDSS) in order to select galaxies with strong emission lines and/or extremely metal poor gas. \citep{Cardamone09,Yang17,Senchyna19,Kojima20}. For the selection of extreme [OIII] emitters with broadband data, it is necessary to define regions of extreme $g-r$ color (or $r-i$ at higher redshift) in order to avoid selecting typical galaxies. Line ratios and redshift affect these colors, preventing some galaxies with very high [OIII] emission from showing extreme colors. This is presented in Figure \ref{fig:broadcol}, where many extreme [OIII] emitters in our sample are indistinguishable from the main SDSS galaxy population using only the $g$, $r$ and $i$ bands. Our sample of EELGs covers as expected the region of the \cite{Yang17} blueberry galaxies (since they present extreme [OIII] emission at our redshift range), but it also covers the same color space as extremely metal-poor galaxies both from observations and models \citep{Kojima20}. This effect is mainly due to the different H$\alpha$/[OIII] line ratios, given how dominated the broadband fluxes are by the line emission in this kind of galaxies. If this ratio is high, the $g$-$r$ color will not be extreme, while if the ratio is low, the $g$-$r$ color will be very negative. This ratio, even if high, does not prevent us from identifying these extreme emitters, since we look at the flux in the $J0515$ mediumband filter, proving again the added value of the mediumband filters in the J-PLUS survey.

\subsubsection{Number density of EELGs}

\begin{table*}
\caption{Number density of EELGs in this work, and comparison with several literature samples.}             
\label{tab:densities2}      
\centering                          
\begin{tabular}{c c c c c c c c}        
\hline\hline                 
Sample & Selection & N & Density & Redshift & EW limit& Mag. & Mass limit\\    
 & method &  & [Mpc$^{-3}$] &  range & [\text{\AA}]$^a$ & limit$^b$ & $\log\,(M_{\star}/M_{\odot})$\\    
\midrule    
\toprule
J-PLUS & Mediumband & 394 &$(2.45\pm0.27)\cdot 10^{-4}$ & 0.016 - 0.048 & 300 & 20.0 & - \\
\midrule                        
\toprule
  GAMA  & Spectra  & 13 & $(8.77\pm3.40)\cdot 10^{-5}$  & \multirow{2}{*}{0.016 - 0.048} & \multirow{2}{*}{300} & \multirow{2}{*}{19.0}  & \multirow{2}{*}{-} \\            
    \multicolumn{2}{c}{J-PLUS comparison} &  113 & $(6.96\pm0.96)\cdot 10^{-5}$  & &  &   &  \\            
   \midrule
         SDSS & Spectra & 90 & $(1.37\pm0.23)\cdot 10^{-5}$&  \multirow{2}{*}{0.016 - 0.048} & \multirow{2}{*}{300} & \multirow{2}{*}{17.7} & \multirow{2}{*}{-} \\        
   \multicolumn{2}{c}{J-PLUS comparison} &  16 & $(0.99\pm 0.27)\cdot 10^{-5}$  &  &  &  &  \\            
   \midrule  
   \cite{Yang17} & Broadband & 7 & $(0.56\pm0.22)\cdot 10^{-6}$&  \multirow{2}{*}{0.016 - 0.048} & \multirow{2}{*}{1600} & \multirow{2}{*}{20.0} & \multirow{2}{*}{-} \\        
   \multicolumn{2}{c}{J-PLUS comparison} &  14 $^c$ & $(8.38\pm2.39)\cdot 10^{-6}$  &  &  &  &  \\            
   \midrule   
 \cite{Cardamone09}   & Broadband & 27 & $(0.020\pm0.004)\cdot 10^{-6}$ & 0.18 - 0.32 & \multirow{2}{*}{400} & \multirow{2}{*}{-} & \multirow{2}{*}{9.52} \\        
 \multicolumn{2}{c}{J-PLUS comparison}  &  2 $^c$ & $(1.240\pm0.890)\cdot 10^{-6}$ & 0.016 - 0.048 &  &  &  \\      
\hline                                   
  H$\alpha$ dots & Narrowband & 6 & $(1.60\pm1.50)\cdot 10^{-3}$&  0.001 - 0.024 & \multirow{2}{*}{300} & \multirow{2}{*}{20.0} & \multirow{2}{*}{-} \\        
   \multicolumn{2}{c}{J-PLUS comparison} &  104 $^c$ & $(0.70\pm0.10)\cdot 10^{-3}$  & 0.016 - 0.024  &  &  &  \\            
   \hline  
   
   \multicolumn{8}{l}{\footnotesize{$^a$ This limit refers to the [OIII]4959+5007 EW.}}\\
   \multicolumn{8}{l}{\footnotesize{$^b$ The magnitude limits are placed considering $r$ magnitude in all samples except for the H$\alpha$ dots, where $R$ magnitude was used instead. }}\\

   \multicolumn{8}{l}{\footnotesize{   In the GAMA and SDSS samples and comparisons, petrosian magnitude was used. For the \cite{Yang17} sample (where they used }}\\
   \multicolumn{8}{l}{\footnotesize{    \texttt{cmodelmag} magnitudes) and for the H$\alpha$ dots sample (circular apertures), we used AUTO apertures in the J-PLUS comparisons.}}\\

   \multicolumn{8}{l}{\footnotesize{$^c$ In the \cite{Yang17}, \cite{Cardamone09}, and H$\alpha$ dots comparisons, we restrict the J-PLUS sample to compact or semi-compact}}\\
    \multicolumn{8}{l}{\footnotesize{    objects, to reproduce their selection criteria.}}\\
\end{tabular}
\end{table*}

The detection of EELGs presented in this work follows clear and reproducible procedures, without pre-selection of targets (other than a magnitude limit). We can therefore use our sample to estimate the number density of this class of galaxies in the redshift range probed, and compare it with other works in the literature. In order to roughly estimate uncertainties, we consider two sources of errors: Poisson noise (associated with the discreet nature of galaxies) and cosmic variance. We approximate the Poisson noise by $\sqrt{N}$, where $N$ is the number of galaxies, and estimate the error associated with cosmic variance using the code presented in \cite{Driver10}. We add in quadrature both errors to obtain an uncertainty estimation. The results of this analysis are shown in Table \ref{tab:densities2}.

First, the number density of EELGs in our J-PLUS sample is computed. We limit the analysis of our sample to the redshift range where both [OIII] lines lie within the $J0515$ filter, $0.016<{\rm z}<0.048$, in order to ensure completeness. We find 394 EELGs in this range, which translates into a number density of (2.45 $\pm$ 0.27) $\cdot 10^{-4}$ Mpc$^{-3}$. This is about 1 EELG every 4000 Mpc$^{3}$.

In order to compare the number density of EELGs we measure with other works, we impose limitations in magnitude (or mass), [OIII] EW, and redshift range. This is done both for the literature samples and ours, in order to match them appropriately. We start the comparisons with the GAMA survey, which provides with precise spectroscopic data, while being still relatively deep and complete. They obtained fiber spectra of essentially all sources brighter than a certain magnitude threshold in the targeted fields. In order to perform a density comparison we limit ourselves to three of the four fields with available data (excluding G02 due to the use of a different input catalog). To further homogenize the selection, we limit the comparison to galaxies brighter than petrosian magnitude 19 in the $r$ band, where both GAMA and our sample are complete. Using the data in the GaussFitSimple table within the SpecLineSFR data management unit, we select galaxies with EW([OIII]5007)$>$225 \text{\AA} in GAMA, which would correspond to $\sim$ 300 \text{\AA} in our J-PLUS analysis ([OIII]4959+5007). This yields 13 GAMA galaxies over the $0.016 < {\rm z} < 0.048$ redshift range. A similar cut in petrosian magnitude and EW in our sample yields 113 objects. When the relative areas and the volume of the Universe between those redshifts are taken into account, the density of GAMA sources is (8.77 $\pm$ 3.40) $\cdot 10^{-5}$ Mpc$^{-3}$ and (6.96 $\pm$0.96) $\cdot 10^{-5}$ Mpc$^{-3}$ for J-PLUS.

We also compare to the SDSS database, limiting our analysis in this Section to the main legacy survey and the MPA-JHU catalog \citep{Kauffmann03,Brinchmann04,Tremonti04,Salim07}, presented in the GalSpecLine table from DR8 \citep{Aihara11}. The clear selection of this sample (essentially petrosian magnitude brighter than 17.7 in $r$, \citealt{Strauss02}) allows for an accurate comparison. We choose galaxies in the same redshift and EW ranges as with the GAMA survey. Their density in this case is (1.37 $\pm$ 0.23) $\cdot 10^{-5}$ Mpc$^{-3}$, compatible with the (0.99 $\pm$ 0.27) $\cdot 10^{-5}$ Mpc$^{-3}$ density in our J-PLUS data for the same cuts.

The results from the previous two comparisons show that the density of EELGs computed in this work using J-PLUS is compatible with the ones derived from magnitude-limited spectroscopic surveys, which can be considered as "ground truth". This result shows the strength of our contribution, which is able to select essentially all EELGs at this redshift range down to a certain magnitude threshold. Moreover, our work has some advantages over spectroscopic analyses: we cover much wider areas than pencil beam surveys such as GAMA, and we are complete down to deeper magnitudes than a wide-field spectroscopic survey like SDSS.

We can also compare our density values with broadband-selected samples, like the Blueberry Galaxies \citep{Yang17} and Green Peas \citep{Cardamone09}. In this case, we restrict our sample to only the compact and semi-compact categories, to more closely match their sample selection. In both works the authors select only galaxies with very high EW, but without a clear threshold value. Therefore we limit our comparison to the galaxies with the highest EW values of these samples, in order to approach completeness. For \cite{Yang17}, we have placed the EW limit roughly where the number of galaxies per EW bin decreases with increasing EW of the [OIII]5007 emission line (around 1200 \text{\AA}, corresponding to 1600 \text{\AA} in [OIII]5007+4959). In addition, we cut the comparison sample at $r<20$ to simulate our limit in brightness, given the similar redshift range covered in both works. Estimating these limits for \cite{Cardamone09} is more complex, given their EW distribution and higher redshift range (which results in larger masses). As an exercise, we place the limits in EW$>$300 \text{\AA} and $\log\,(M_{\star}/M_{\odot})>9.5$, obtaining only two galaxies in our sample, against 27 in their case. These thresholds result in a density of ($0.56 \pm 0.22$) $\cdot 10^{-6}$ Mpc$^{-3}$ for \cite{Yang17}, and more than ten times higher in J-PLUS, ($8.38 \pm 2.39$) $\cdot 10^{-6}$ Mpc$^{-3}$. The comparison between the \cite{Cardamone09} sample and our work is even more striking, with values of ($0.020 \pm 0.004$) $\cdot 10^{-6}$ Mpc$^{-3}$ vs. ($1.24 \pm 0.890$) $\cdot 10^{-6}$ Mpc$^{-3}$, but more uncertain.

Other surveys use narrowband imaging to identify emission line galaxies. In principle, they should be more sensitive to low EW emission lines than J-PLUS, since their filters are narrower than our $J0515$ filter, and therefore the contrast in brightness between narrow and broadband should be higher. Nevertheless, given that we limit our analysis only to the most extreme events of star formation, the strong contrast is enough to avoid missing a significant amount of EELGs. The sky area and redshift range that narrowband surveys cover is smaller than ours, which limits their ability to identify extreme, rare objects. To our knowledge, there has not been any narrowband survey targeting the [OIII] emission line in the local Universe, so we cannot perform any direct comparison. A similar example are the H$\alpha$ dots identified in the ALFALFA H$\alpha$ survey \citep{Kellar12,Salzer20}. While they select galaxies based on H$\alpha$ emission at our redshift range, they provide spectroscopic follow-up and thus [OIII] EW. We compute the density of H$\alpha$ dots showing [OIII] EW higher than 300 \text{\AA} and brighter than $r=20$ located at ${\rm z}<0.024$ (beyond that the transmission of their reddest H$\alpha$ filter drops). The value we obtain for their number density (1.60$\pm$1.50$\cdot 10^{-3}$ Mpc$^{-3}$) is higher than the corresponding value for our sample (0.70$\pm$0.10$\cdot 10^{-3}$ Mpc$^{-3}$), considering again only compact or semi-compact galaxies in our J-PLUS comparison. Nevertheless, given the high uncertainty in the density value derived for the H$\alpha$ dots survey ($\sim$ 80\%, driven mostly by cosmic variance), the density values are compatible within one standard deviation.

\subsubsection{[OIII]5007/H$\alpha$ and [OIII]5007/H$\beta$ ratios}
\label{sec:o3hbeta}
The present work intends to provide a sample of EELGs selected purely on magnitude and EW([OIII]), with no direct bias on line ratios, since it is only based on the detection of flux excess in a mediumband filter. In contrast with this, other works based on broadband photometry detection (such as \citealt{Yang17} and \citealt{Cardamone09}) are biased against low [OIII]5007/H$\alpha$ systems (and therefore also against low [OIII]5007/H$\beta$ ones). This is because they demand that the flux of the filter where the [OIII] doublet falls ($g$ in \citealt{Yang17}, $r$ in \citealt{Cardamone09}) must be significantly stronger than the filter where the H$\alpha$ line flux contribution lies ($r$ in \citealt{Yang17}, $i$ or $z$ in \citealt{Cardamone09}). According to stellar population synthesis and nebular photoionization models (such as \citealt{Inoue11}), the [OIII]5007/H$\alpha$ ratio reaches a maximum value for gas metallicities around 12+log(O/H) $\sim$ 8.0, which is further confirmed by the typical and metallicities of \cite{Yang17} and \cite{Cardamone09}. Therefore, selecting galaxies with high [OIII]5007/H$\alpha$ ratios biases the samples against very low metallicities \citep{Senchyna19,Kojima20}.

      \begin{figure}
   \centering
   \includegraphics[width=0.48\textwidth,keepaspectratio]{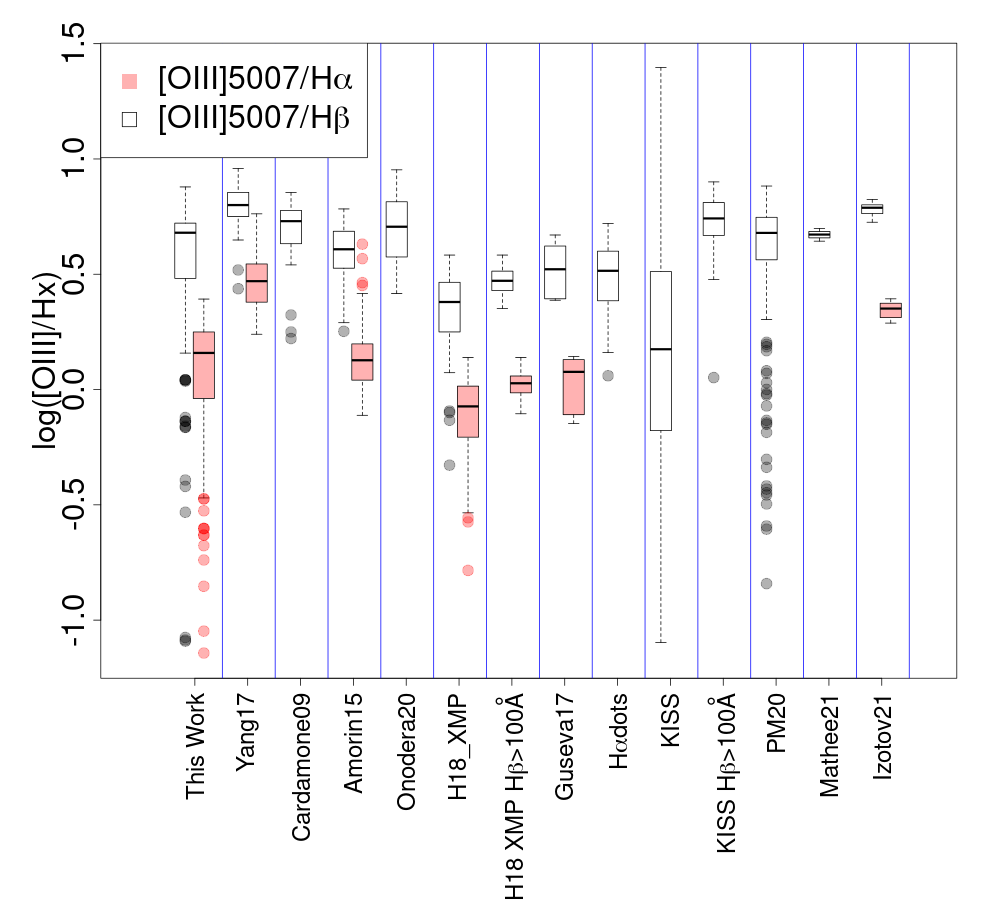}
      \caption{Distribution of the ratios [OIII]5007/H$\alpha$ (red-filled rectangles and dots) and [OIII]5007/H$\beta$ (white-filled rectangles and grey dots). The thick black line represent the median value, while the box ranges from the first quartile to the third one. The error bars represent the maximum and minimum value without outliers, which are plotted as dots (and defined as the values that lie beyond 1.5 times the inter quartilic range). We plot in this figure results from the present work and several samples from the literature \citep{Yang17,Cardamone09,Amorin15,Onodera20,Hsyu18,Guseva17,Perez-Montero20,Matthee21,Izotov21}, as well as the KISS survey \citep{Wegner03,Gronwall04,Jangren05,Salzer05} and the H$\alpha$ dots class \citep{Kellar12,Salzer20}. We have selected the subsample with extremely low metallicity in \cite{Hsyu18} (H18\_XMP), as well as those galaxies with EW(H$\beta$)>100 \text{\AA} in both \cite{Hsyu18} XMPs (H18\_XMP H$\beta$>100 \text{\AA}) and the KISS survey (KISS H$\beta$>100 \text{\AA}). Different samples are separated by vertical blue lines.}
         \label{fig:EWHbHbrat}
   \end{figure}

Since we only select extreme [OIII] emitters, we are certainly missing systems that have [OIII] EW below our threshold, but above it we should recover all galaxies, regardless of their [OIII]5007/H$\alpha$ ratio. In Fig. \ref{fig:EWHbHbrat} we compare the distribution of the [OIII]5007/H$\alpha$ and [OIII]5007/H$\beta$ ratios for several samples of ELGs. We consider more reliable the [OIII]5007/H$\beta$ ones, since the emission lines are much closer in wavelength, minimizing the effect of dust extinction correction and/or flux calibration issues. We see that our work shows lower median values for these ratios than broadband selections such as \cite{Yang17} or \cite{Cardamone09}, and we reach much lower values. Our selection shows slightly higher values than a pure spectroscopic selection \citep{Amorin15}, specially in the [OIII]5007/H$\alpha$ ratio. A stronger difference is present when comparing with H$\alpha$ selected samples (the H$\alpha$ dots, \citealt{Kellar12,Salzer20}) and extremely metal poor galaxies \citep{Hsyu18,Guseva17}, specially if we include systems with low H$\beta$ EW (a proxy for low sSFR). Nevertheless, we can see that the range of values that our survey covers overlaps with most of the range of the extremely metal poor and strongly star forming samples, in contrast with \cite{Yang17} and \cite{Cardamone09}.
In addition, we plot as well the typical values for other types of systems, some with strong ionizing spectra: HeII emitters \citep{Perez-Montero20} and Lyman $\alpha$ emitters \citep{Matthee21,Izotov21}. In these cases, their ratios are more similar to ours.

Finally, it is also important to keep in mind that several properties in the samples affect the [OIII]5007/H$\alpha$ ratios, apart from selection biases and metallicity. The SFR is also positively correlated with the ratio (as seen for example when restricting the \cite{Hsyu18} sample only to objects with high H$\beta$ EW). Stellar mass also plays a role (with higher mass galaxies showing lower ratios), but it is less relevant than the other factors: the \cite{Cardamone09} sample has a typical mass $\sim$ 100 times higher than those of \cite{Hsyu18} or \cite{Guseva17}, yet their ratios are higher. While these factors play a role in the results shown in this section, our selection method is clearly open to the selection of low [OIII]/H$\alpha$ and H$\beta$ galaxies, in contrast with others.

\subsection{Physical properties of the EELG sample}

\subsubsection{[OIII] EW and stellar mass}
\label{sec:ewmass}
\begin{figure}
   \centering
   \includegraphics[width=0.48\textwidth,keepaspectratio]{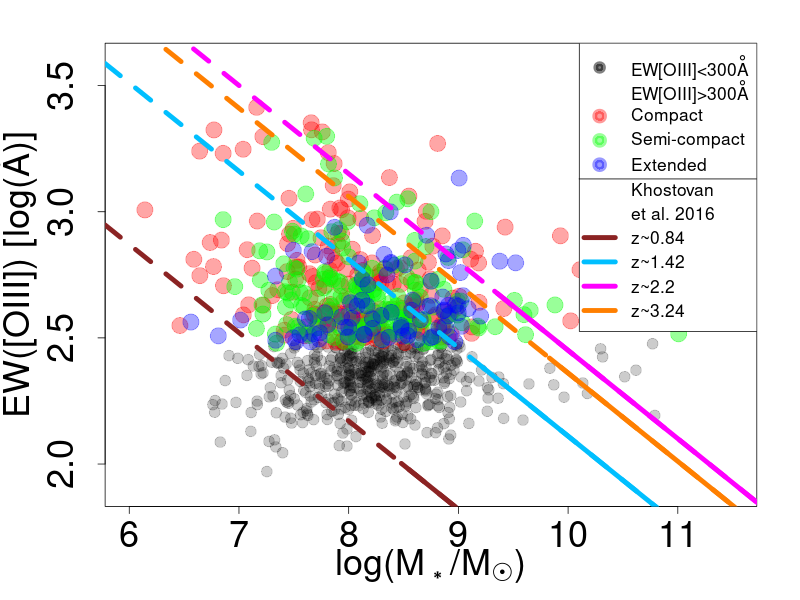}
      \caption{[OIII] EW of the EELG (color dots) and candidate (gray dots) samples as a function of stellar mass. We plot in different colors the galaxies in the EELG sample according to their morphology: red corresponds to compact objects, green to semi-compact, and blue to extended. In addition, we overplot linear relations from \cite{Khostovan16}. }
         \label{fig:EWmass}
   \end{figure}

A clear negative correlation between emission line EW and stellar mass has been found in the literature (e.g. \citealt{Fumagalli12,Sobral14,Khostovan16,Reddy18}; Lumbreras-Calle et al. 2022 \textit{in prep.}). This indicates that galaxies with lower masses tend to have stronger recent star-formation events relative to their mass (higher sSFR). The relationship between the two variables has been fitted with a linear model (in logarithmic units), and depends on the emission line studied (both intercept and slope) and the redshift of the galaxies (mostly the intercept), although discrepancies are still present (i.e. between \citealt{Khostovan16} and \citealt{Reddy18}, but see \citealt{Khostovan21}). It is beyond the scope of the present paper to shed light into the values of that linear relation, given that our selection process focuses only in a region of the EW - mass diagram, making a linear fit to our data meaningless in this context. Nevertheless, we can compare the values obtained to those available in the literature for the [OIII] line (see Fig. \ref{fig:EWmass}). 

The most clear result is that, even if our galaxies reside in the very low redshift Universe, they populate almost exclusively the regions of the diagram defined by the linear fits to galaxies at intermediate redshift (${\rm z}>0.84$), with EW values similar to those typically seen at ${\rm z} \sim 1.4$ \citep{Khostovan16}. We have as well in our sample some galaxies with EW similar to those at medium-high redshift: ${\rm z}\sim 2.2 - 3.4$ for \cite{Khostovan16}, ${\rm z} \sim 3.4$ for \cite{Reddy18}, and even ${\rm z}\sim 8$ \citep{deBarros19}. The very high EW values, low [OII]/[OIII] ratios, and typically compact morphologies are similar to those found in very high redshift galaxies \citep[e.g.][]{Onodera20}. According to detailed spectroscopic studies in the low redshift universe \citep{Izotov21}, this type of galaxies share physical properties with those forming stars at very high redshift, likely leaking Lyman continuum radiation, and playing an important role in the reionization of the Universe. In consequence, we consider the present sample may be useful in improving the understanding of the very early Universe. 

We split the sample in the three morphological categories described in Sect.~\ref{sec:morpho}, and show them in different colors in Fig.~\ref{fig:EWmass}. Compact and semi-compact galaxies have slightly lower masses, with typical values of $8.06\substack{+0.65\\ -0.54}$ and $8.15\substack{+0.58\\ -0.59}$ respectively, compared to $8.38\substack{+0.56\\ -0.67}$ for extended galaxies. Driven mostly by this difference and the EW - mass anticorrelation, compact galaxies show lower typical EW values. Nevertheless, for the lower mass range, compact galaxies show clearly higher EW values than extended ones even at similar masses. This  suggests that there is an additional physical effect that favors higher EW values in compact galaxies over extended ones.

The linear relation shown in \cite{Khostovan15} and \cite{Reddy18} between EW([OIII]) and stellar mass appears to have a limit at EW([OIII]) $\sim 3\,000$ \text{\AA} \citep{Reddy18}. This limit, reached for galaxies dominated by the star-forming burst, is likely to be physical in nature rather than purely observational, given that for galaxies with a given mass, there is no bias against selecting higher EW systems. Understanding the reason behind this limit could provide insights into the star formation cycle in the most extreme environment, in galaxies at the highest redshifts and with the lowest masses and metallicities. Our sample provides several previously unknown nearby candidates around that limit, which can be targeted with follow-up spectroscopic observations to extract relevant conclusions on this topic.

\subsubsection{SFR and Main Sequence}
      \begin{figure}
   \centering
   \includegraphics[width=0.48\textwidth,keepaspectratio]{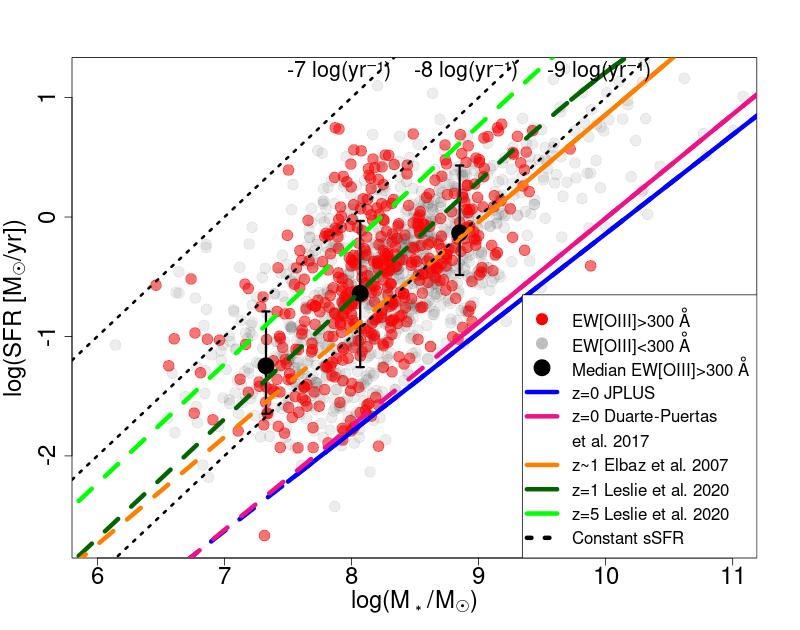}
      \caption{SFR as a function of stellar mass for the galaxies in the EELG sample (red dots) and the rest of the candidate sample (grey dots). Large black dots represent the median values for the EELG sample grouped in mass bins. Colored lines represent main sequence fits from the literature, with black dotted lines represent constant sSFR. }
         \label{fig:mainseq}
   \end{figure}
   
One of the most important properties of star-forming galaxies is their SFR. We will use one of the most prominent methods to estimate this parameter: the flux in the H$\alpha$ emission line. In particular, we will use the flux estimated by the \texttt{CIGALE} SED fit, with the prescription in \cite{Kennicutt98} to transform H$\alpha$ flux into SFR. We correct for extinction using the E(B-V) value from the SED fit and the \cite{Calzetti00} extinction law with R$_V$=4.05 for consistency with the \texttt{CIGALE} implementation (using R$_V$=3.1 would only create an offset with a median value of $\sim$ -0.04 dex in SFR). We use the H$\alpha$ flux since it is the most direct SFR tracer that we can access with the J-PLUS data, and using other sources would entail other corrections beyond the scope of this paper. The H$\alpha$ flux value, even if calculated using \texttt{CIGALE} (and thus not entirely independant from the stellar mass value we use), is in very good agreement with both photometric and spectroscopic estimations (see Figs. \ref{fig:fluxr_Ha} and \ref{fig:fluxha_compar}). Taking directly the SFR estimation by \texttt{CIGALE} would be much less independent and make it harder to compare with other works in the literature. In any case, we only use the SFR values to illustrate the extreme nature of our sample, not to provide any definitive physical conclusions.

 A tight correlation has been proven to exist between the stellar mass of galaxies and their SFR often called ``star formation main sequence'' \citep[e.g.][]{Brinchmann04,Elbaz07,Noeske07,Speagle14}. This relation has been used to define both starburst galaxies (above the relation) and quiescent galaxies (below it). In the present work, our selection method is intentionally oriented to identify galaxies with very high SFR, and thus our data cannot be used to compute the main sequence parameters. In Fig. \ref{fig:mainseq}, we plot the SFR of the galaxies in our sample as a function of their stellar mass, as well as the constant specific SFR (sSFR) lines. We overplot several main sequence relations for different redshifts \citep{Leslie20,Elbaz07}, with the fits derived in \cite{DuartePuertas17} and \cite{Vilella-Rojo21} as local comparisons. It is clear that our sample of EELGs lies well above the local main sequence relation, with sSFR values more comparable to those of typical high redshift star forming galaxies. We compute the difference between the SFR values measured in our sample and those expected if the galaxies followed exactly the \cite{Vilella-Rojo21} relation. The typical difference is very high, with our EELG sample showing SFR values $1.31\substack{+0.55\\ -0.46}$ dex higher. In particular, the differences are higher at lower stellar masses, with $M_\star < 10^{7.5}\ M_{\odot}$ galaxies having $1.71\substack{+0.54\\ -0.48}$ dex difference and those with $M_\star > 10^{8.5}\ M_{\odot}$ presenting $1.08\substack{+0.34\\ -0.32}$ offset. This is partially due to selection effects, given that the detection of galaxies with low stellar masses and low sSFR is less likely. Nevertheless, the detection of high sSFR low mass galaxies and the lack of high sSFR high mass galaxies are not due to biases. This result is consistent with many works that have discovered how low mass galaxies undergo bursts of star formation more intense and more often than their higher mass counterparts, and therefore are more likely to be observed in a starburst phase \citep{Sparre17,Guo16,Khostovan21}. This effect is actually already present (albeit softer) in typical star-forming galaxies, hence the slope of the main sequence being lower than one.

This differences put the vast majority of our sample above the threshold for galaxies undergoing starbursts, which is usually set around 0.48-0.6 dex above the main sequence \citep{Rodighiero11,Elbaz18,Bluck20}. They are in fact more similar to main sequence galaxies at very high redshift, with some reaching values that are typical at ${\rm z} \sim 5$, considering the extrapolation to low mass galaxies of the main sequence at that redshift. 

\subsubsection{Comparison with typical low redshift star-forming galaxies}
Some properties derived from the SED fitting process can be complex to evaluate, given the uncertainties in the models, the assumptions taken (most notably the SFH), and the relatively sparse wavelength coverage of our data. In order to provide a fair comparison with the literature, we performed the same SED fitting analysis, adopting the same assumptions on a different sample of galaxies. The absolute values obtained for ages, metallicities, etc. may differ from others due to our choices in the SED fitting process and not due to physical differences in the galaxies, but the relative difference in those magnitudes when applying the same method provides more accurate insights. 

\begin{figure}
   \centering
   \includegraphics[width=0.48\textwidth,keepaspectratio]{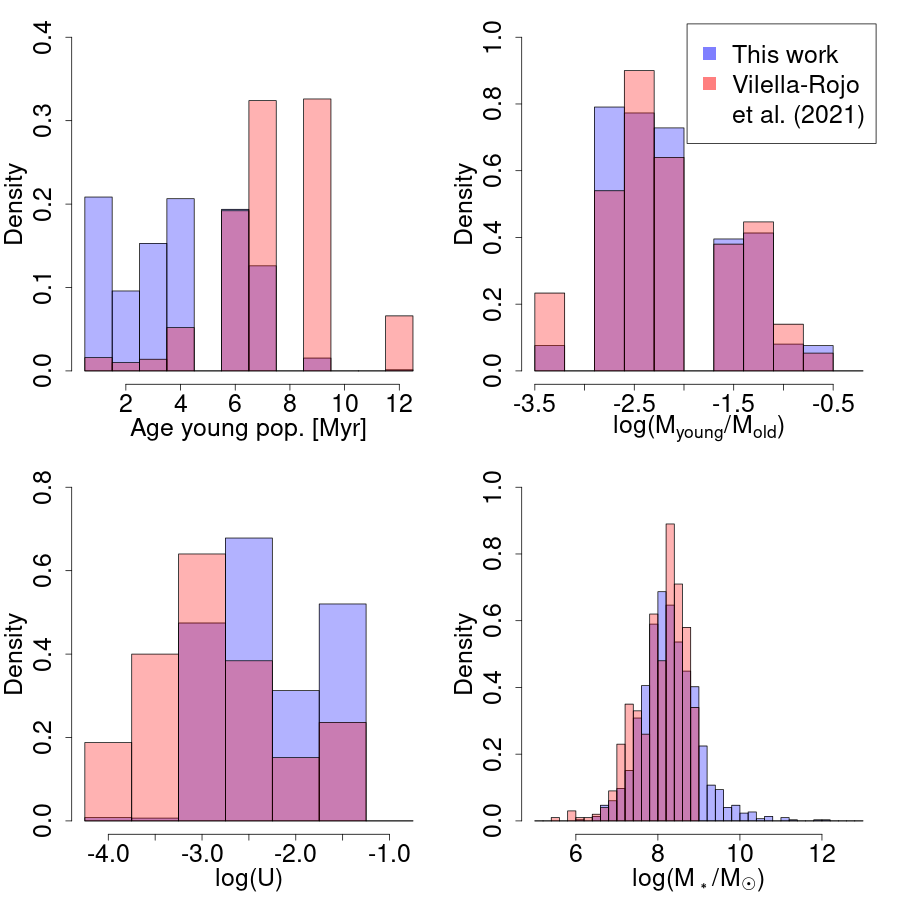}
      \caption{Histograms comparing the properties of our EELG sample (blue) with a subsample of typical star forming galaxies from \cite{Vilella-Rojo21} (red). We show the best fitting values of several magnitudes, obtained fitting J-PLUS photometry to SED models using \texttt{CIGALE}. From left to right and top to bottom, they are the age of the young stellar population, the mass ratio between the young and the old populations, the ionization parameter, and the stellar masses.}
         \label{fig:gon}
   \end{figure}

In order to do this, we choose the sample from \cite{Vilella-Rojo21} as a comparison. In that work, the authors used the J-PLUS first data release (DR1; \citealt{Cenarro19}) to identify $805$ local (${\rm z} < 0.017$) star-forming galaxies selected as H$\alpha$ emitters. The comparison is therefore straightforward, given that the filters used are the same. \texttt{CIGALE} is run over the \cite{Vilella-Rojo21} sample of H$\alpha$ emitters, with the same parameters used for our analysis (Sect.~\ref{sec:cigale}). Even if their galaxies are similar to ours, they selected brighter ($r<18$ mag) objects, at a slightly lower redshift. Therefore, their typical masses are higher than ours, with median and 1$\sigma$ limits of $\log\,(M_{\star}/M_{\odot}) = 8.64\substack{+1.12 \\ -0.88}$ compared to $\log\,(M_{\star}/M_{\odot}) = 8.15\substack{+0.55 \\ -0.59}$. In order to correct for this difference, we limit our analysis to their galaxies showing stellar masses below 10$^{8.9} M_{\odot}$, which lowers their typical mass values to $\log\,(M_{\star}/M_{\odot}) = 8.21\substack{+ 0.44\\ -0.80}$, in agreement with our sample (see Panel d) in Fig. \ref{fig:gon}).

The results of the comparison between the other SED parameters is shown in the rest of the panels in Fig. \ref{fig:gon} (we only plot the most relevant histograms for this analysis). The most striking difference appears in the age of the young stellar populations, with extremely low values for our extreme [OIII] emitters sample ($3.0\substack{+2.7\\ -1.0}$ Myr) compared to more typical values for the \cite{Vilella-Rojo21} sample (7$\substack{+2\\ -1}$ Myr). This result was expected, considering the difference in the sample selection in their work (with a minimum EW in H$\alpha$ of around 12 \text{\AA}) and ours (with a minimum EW in [OIII] of 300 \text{\AA}), resulting in 95\% of our galaxies showing H$\alpha$ EWs higher than 56 \text{\AA}, and 90\% above 135 \text{\AA}. It is clear in all models of star formation that the EW of the emission lines decreases rapidly after the initial burst, in a few Myrs. Therefore, galaxies with higher EW values will tend to have younger bursts of star formation. Other parameters (fraction of young population, metallicity, ionization parameters) will also play a role, but the very high EW in our work necessarily implies very young ages. 

In line with the EW differences stated, we see that the mass fraction of the young population is slightly higher in our sample than in \cite{Vilella-Rojo21} with $\log\,(M_{\rm young}/M_{\rm old}) = -2.0\substack{+0.7\\ -0.3}$ and $\log\,(M_{\rm young}/M_{\rm old}) = -2.3\substack{+1.0\\ -0.3}$, respectively. The metallicities (not shown in the figure) are indistinguishable between the samples, with a p-value of 0.27 in the Kolmogorov-Smirnoff test. The ionization parameter shows a difference going from $\log\,U=-3.0\substack{+1.0\\ -0.5}$ in their sample to $\log\,U=-2.5\substack{+1.0\\ -0.5}$ in ours. However, it must be kept in mind that mass fraction and ionization parameter are subject to degeneracies and are hard to accurately measure just with the photometric data we have.

While the results presented in this work are insufficient to explain why the EELGs are undergoing such extreme events, some inferences can be made. The very young ages of the bursts indicate that it is likely that these galaxies only show these properties for briefs periods of time, which explains in part why they are so uncommon. Many more galaxies may have undergone similar starburst phases in other moments of their SFH \citep[see e.g.][]{SanchezAlmeida08,SanchezAlmeida18}. The trigger of the extreme star formation events remains an open question, yet some hypothesis have been presented in the literature. For decades, mergers have been considered likely to be causing a fraction of the events \citep{Barnes91}, and more recently their importance has also been shown in dwarf galaxy starbursts both from simulations \citep{Bekki08} and observations \citep{Stierwalt15}. Disrupted morphologies have also been observed with increased frequency in EELGs \citep[e.g.][]{Calabro17}. While in the present work most galaxies are compact, there is a significant fraction that show extended, often complex and clumpy morphologies. In addition, while some may appear undisturbed, traces of a recent merger may be hidden due to their low surface brightness \citep{Martinez-Delgado12}. Another key driver for starbursts is the infall of cold gas into the galaxy \citep{Ceverino10,SanchezAlmeida15}, which is inline with the typically low metallicity of these systems, especially if the gas is near pristine and falling from the cosmic web. This process also explains why the EELGs are more common at higher redshift, where cold gas was much more abundant. Another physical process that could trigger the starbursts is the star formation feedback \cite{Sparre17}. It could be due to the removal of gas from the galaxy, that later "rains down" on it, causing he starburst, or by compressing the gas inside the galaxy, triggering further star formation \citep{Tenorio-Tagle05}. While disentangling the causes of the extreme events of star formation is beyond the scope of this paper, the sample presented here it can be useful in that regard in future analyses.

\section{Summary and conclusions}
\label{sec:conclusions}
We have used the J-PLUS DR2 to select a sample of $466$ EELGs in the local Universe over $2\,176$ square degrees, with EW[OIII] above $300$ \text{\AA}. Out of theme, $411$ ($89$\%) were previously unclassified as EELGs.

To build the sample, we have identified objects with excess of emission in the unique $\sim$ 200\text{\AA} J-PLUS $J0515$ filter compared to the $r$ band. To remove contaminants (stars and high redshift systems), we rely in color-color and color-magnitude diagrams using infrared WISE observations, reaching 96\% purity and 92\% completeness considering the available SDSS spectra.

We perform SED fitting on the J-PLUS data using the \texttt{CIGALE} software, finding the following key properties of the sample of galaxies:

\begin{itemize}
    \item Low stellar masses, with typical values of $\log\,(M_{\star}/M_{\odot}) = 8.13\substack{+0.61\\ -0.58}$ .
    \item Very young ages, $3\substack{+2.7\\ -2}$ Myr, the main driver of the very high EW.
    \item Moderately low metallicities, Z=$0.008\substack{+0.012\\ -0.004}$, and high ionization parameters, $\log\,U=-2.50\substack{+1.00\\ -0.35}$, contributing as well to the exceptionally high EW.
    \item Low extinction, $E(B-V)=0.1\substack{+0.2\\ -0.1}$, typical of low mass galaxies with high star formation activity.
\end{itemize}

We have compared our sample with several results in the literature. We find a very good agreement for those galaxies that have spectroscopic data, comparing both line fluxes and EW. We measure negligible offsets in the comparisons between spectroscopic and our photometric J-PLUS measurements in the H$\alpha$ and [OIII]5007 lines, with very little scatter in fluxes ($\sim$ 0.18 dex) and EWs ($\sim$ 0.1 dex). 

We analyze as well the efficiency of our selection process compared to other recent results. We find that the number density of EELGs we calculate is in agreement with magnitude-limited spectroscopic surveys such as the main SDSS legacy survey and GAMA, while reaching fainter galaxies than both and much wider areas than GAMA. We measure comparable densities as narrow-band surveys, with caveats due to selection criteria. We are, in contrast, much more efficient than searches made using broadband surveys, such as the blueberry galaxies \citep{Yang17} or the green peas \citep{Cardamone09}. We find 20 - 50 times more EELGs per unit of volume, when controlling for EW and magnitude or stellar mass. It is also likely that we are able to access lower metallicity systems than broadband surveys, since we are not directly biased against high H$\alpha$/[OIII]5007 systems, which is a property of extremely metal poor galaxies \citep{Senchyna19}. The [OIII]5007 EW and SFR as a function of mass diagrams place our sample of EELGs in the typical values for the high redshift Universe (${\rm z}\sim2-5$) while being located at ${\rm z}<0.06$. 

In conclusion, we have presented a sample of mostly previously unclassified EELGs, which despite residing in the local Universe, share characteristics (mass, sSFR, EW, metallicity) that make them similar to those at high redshift. Therefore, spectroscopic follow-up of this sample may shed light into the properties of the galaxies forming in the very early Universe.

\begin{acknowledgements}

Based on observations made with the JAST80 telescope at the Observatorio Astrof\'{\i}sico de Javalambre (OAJ), in Teruel, owned, managed, and operated by the Centro de Estudios de F\'{\i}sica del  Cosmos de Arag\'on. We acknowledge the OAJ Data Processing and Archiving Unit (UPAD, \citealt{upad}) for reducing and calibrating the OAJ data used in this work.

Funding for the J-PLUS Project has been provided by the Governments of Spain and Arag\'on through the Fondo de Inversiones de Teruel; the Arag\'on Government through the Reseach Groups E96, E103, and E16\_17R; the Spanish Ministry of Science, Innovation and Universities (MCIU/AEI/FEDER, UE) with grants PGC2018-097585-B-C21 and PGC2018-097585-B-C22; the Spanish Ministry of Economy and Competitiveness (MINECO) under AYA2015-66211-C2-1-P, AYA2015-66211-C2-2, AYA2012-30789, and ICTS-2009-14; and European FEDER funding (FCDD10-4E-867, FCDD13-4E-2685). The Brazilian agencies FINEP, FAPESP, and the National Observatory of Brazil have also contributed to this project.

L.A.D.G.~thanks for financial support from the State Agency for Research of the Spanish MCIU through the `Center of Excellence Severo Ochoa' award to the Instituto de Astrof\'isica de Andaluc\'ia (SEV-2017-0709).

This project uses data from the Legacy Surveys. The Legacy Surveys consist of three individual and complementary projects: the Dark Energy Camera Legacy Survey (DECaLS; Proposal ID 2014B-0404; PIs: David Schlegel and Arjun Dey), the Beijing-Arizona Sky Survey (BASS; NOAO Prop. ID 2015A-0801; PIs: Zhou Xu and Xiaohui Fan), and the Mayall z-band Legacy Survey (MzLS; Prop. ID 2016A-0453; PI: Arjun Dey). DECaLS, BASS and MzLS together include data obtained, respectively, at the Blanco telescope, Cerro Tololo Inter-American Observatory, NSF’s NOIRLab; the Bok telescope, Steward Observatory, University of Arizona; and the Mayall telescope, Kitt Peak National Observatory, NOIRLab. The Legacy Surveys project is honored to be permitted to conduct astronomical research on Iolkam Du’ag (Kitt Peak), a mountain with particular significance to the Tohono O’odham Nation. 
BASS is a key project of the Telescope Access Program (TAP), which has been funded by the National Astronomical Observatories of China, the Chinese Academy of Sciences (the Strategic Priority Research Program “The Emergence of Cosmological Structures” Grant XDB09000000), and the Special Fund for Astronomy from the Ministry of Finance. The BASS is also supported by the External Cooperation Program of Chinese Academy of Sciences (Grant 114A11KYSB20160057), and Chinese National Natural Science Foundation (Grant 11433005).

\end{acknowledgements}

%
%

\bibliographystyle{aa}
\bibliography{bibo3}

\begin{thebibliography}{130}
\expandafter\ifx\csname natexlab\endcsname\relax\def\natexlab#1{#1}\fi

\bibitem[{{Ahumada} {et~al.}(2020){Ahumada}, {Prieto}, {Almeida}, {Anders},
  {Anderson}, {Andrews}, {Anguiano}, {Arcodia}, {Armengaud}, {Aubert}, {Avila},
  {Avila-Reese}, {Badenes}, {Balland}, {Barger}, {Barrera-Ballesteros}, {Basu},
  {Bautista}, {Beaton}, {Beers}, {Benavides}, {Bender}, {Bernardi}, {Bershady},
  {Beutler}, {Bidin}, {Bird}, {Bizyaev}, {Blanc}, {Blanton}, {Boquien},
  {Borissova}, {Bovy}, {Brandt}, {Brinkmann}, {Brownstein}, {Bundy}, {Bureau},
  {Burgasser}, {Burtin}, {Cano-D{\'\i}az}, {Capasso}, {Cappellari}, {Carrera},
  {Chabanier}, {Chaplin}, {Chapman}, {Cherinka}, {Chiappini}, {Doohyun Choi},
  {Chojnowski}, {Chung}, {Clerc}, {Coffey}, {Comerford}, {Comparat}, {da
  Costa}, {Cousinou}, {Covey}, {Crane}, {Cunha}, {Ilha}, {Dai}, {Damsted},
  {Darling}, {Davidson}, {Davies}, {Dawson}, {De}, {de la Macorra}, {De Lee},
  {Queiroz}, {Deconto Machado}, {de la Torre}, {Dell'Agli}, {du Mas des
  Bourboux}, {Diamond-Stanic}, {Dillon}, {Donor}, {Drory}, {Duckworth},
  {Dwelly}, {Ebelke}, {Eftekharzadeh}, {Davis Eigenbrot}, {Elsworth},
  {Eracleous}, {Erfanianfar}, {Escoffier}, {Fan}, {Farr},
  {Fern{\'a}ndez-Trincado}, {Feuillet}, {Finoguenov}, {Fofie},
  {Fraser-McKelvie}, {Frinchaboy}, {Fromenteau}, {Fu}, {Galbany}, {Garcia},
  {Garc{\'\i}a-Hern{\'a}ndez}, {Oehmichen}, {Ge}, {Maia}, {Geisler}, {Gelfand},
  {Goddy}, {Gonzalez-Perez}, {Grabowski}, {Green}, {Grier}, {Guo}, {Guy},
  {Harding}, {Hasselquist}, {Hawken}, {Hayes}, {Hearty}, {Hekker}, {Hogg},
  {Holtzman}, {Horta}, {Hou}, {Hsieh}, {Huber}, {Hunt}, {Chitham}, {Imig},
  {Jaber}, {Angel}, {Johnson}, {Jones}, {J{\"o}nsson}, {Jullo}, {Kim},
  {Kinemuchi}, {Kirkpatrick}, {Kite}, {Klaene}, {Kneib}, {Kollmeier}, {Kong},
  {Kounkel}, {Krishnarao}, {Lacerna}, {Lan}, {Lane}, {Law}, {Le Goff}, {Leung},
  {Lewis}, {Li}, {Lian}, {Lin}, {Long}, {Longa-Pe{\~n}a}, {Lundgren}, {Lyke},
  {Ted Mackereth}, {MacLeod}, {Majewski}, {Manchado}, {Maraston}, {Martini},
  {Masseron}, {Masters}, {Mathur}, {McDermid}, {Merloni}, {Merrifield},
  {M{\'e}sz{\'a}ros}, {Miglio}, {Minniti}, {Minsley}, {Miyaji}, {Mohammad},
  {Mosser}, {Mueller}, {Muna}, {Mu{\~n}oz-Guti{\'e}rrez}, {Myers}, {Nadathur},
  {Nair}, {Nandra}, {do Nascimento}, {Nevin}, {Newman}, {Nidever}, {Nitschelm},
  {Noterdaeme}, {O'Connell}, {Olmstead}, {Oravetz}, {Oravetz}, {Osorio},
  {Pace}, {Padilla}, {Palanque-Delabrouille}, {Palicio}, {Pan}, {Pan},
  {Parker}, {Paviot}, {Peirani}, {Ram{\'r}ez}, {Penny}, {Percival},
  {Perez-Fournon}, {P{\'e}rez-R{\`a}fols}, {Petitjean}, {Pieri},
  {Pinsonneault}, {Poovelil}, {Povick}, {Prakash}, {Price-Whelan}, {Raddick},
  {Raichoor}, {Ray}, {Rembold}, {Rezaie}, {Riffel}, {Riffel}, {Rix}, {Robin},
  {Roman-Lopes}, {Rom{\'a}n-Z{\'u}{\~n}iga}, {Rose}, {Ross}, {Rossi},
  {Rowlands}, {Rubin}, {Salvato}, {S{\'a}nchez}, {S{\'a}nchez-Menguiano},
  {S{\'a}nchez-Gallego}, {Sayres}, {Schaefer}, {Schiavon}, {Schimoia},
  {Schlafly}, {Schlegel}, {Schneider}, {Schultheis}, {Schwope}, {Seo},
  {Serenelli}, {Shafieloo}, {Shamsi}, {Shao}, {Shen}, {Shetrone}, {Shirley},
  {Aguirre}, {Simon}, {Skrutskie}, {Slosar}, {Smethurst}, {Sobeck}, {Sodi},
  {Souto}, {Stark}, {Stassun}, {Steinmetz}, {Stello}, {Stermer},
  {Storchi-Bergmann}, {Streblyanska}, {Stringfellow}, {Stutz}, {Su{\'a}rez},
  {Sun}, {Taghizadeh-Popp}, {Talbot}, {Tayar}, {Thakar}, {Theriault}, {Thomas},
  {Thomas}, {Tinker}, {Tojeiro}, {Toledo}, {Tremonti}, {Troup}, {Tuttle},
  {Unda-Sanzana}, {Valentini}, {Vargas-Gonz{\'a}lez}, {Vargas-Maga{\~n}a},
  {V{\'a}zquez-Mata}, {Vivek}, {Wake}, {Wang}, {Weaver}, {Weijmans}, {Wild},
  {Wilson}, {Wilson}, {Wolthuis}, {Wood-Vasey}, {Yan}, {Yang}, {Y{\`e}che},
  {Zamora}, {Zarrouk}, {Zasowski}, {Zhang}, {Zhao}, {Zhao}, {Zheng}, {Zheng},
  {Zhu}, \& {Zou}}]{Ahumada20}
{Ahumada}, R., {Prieto}, C.~A., {Almeida}, A., {et~al.} 2020, \apjs, 249, 3

\bibitem[{{Aihara} {et~al.}(2011){Aihara}, {Allende Prieto}, {An}, {Anderson},
  {Aubourg}, {Balbinot}, {Beers}, {Berlind}, {Bickerton}, {Bizyaev}, {Blanton},
  {Bochanski}, {Bolton}, {Bovy}, {Brandt}, {Brinkmann}, {Brown}, {Brownstein},
  {Busca}, {Campbell}, {Carr}, {Chen}, {Chiappini}, {Comparat}, {Connolly},
  {Cortes}, {Croft}, {Cuesta}, {da Costa}, {Davenport}, {Dawson}, {Dhital},
  {Ealet}, {Ebelke}, {Edmondson}, {Eisenstein}, {Escoffier}, {Esposito},
  {Evans}, {Fan}, {Femen{\'\i}a Castell{\'a}}, {Font-Ribera}, {Frinchaboy},
  {Ge}, {Gillespie}, {Gilmore}, {Gonz{\'a}lez Hern{\'a}ndez}, {Gott}, {Gould},
  {Grebel}, {Gunn}, {Hamilton}, {Harding}, {Harris}, {Hawley}, {Hearty}, {Ho},
  {Hogg}, {Holtzman}, {Honscheid}, {Inada}, {Ivans}, {Jiang}, {Johnson},
  {Jordan}, {Jordan}, {Kazin}, {Kirkby}, {Klaene}, {Knapp}, {Kneib},
  {Kochanek}, {Koesterke}, {Kollmeier}, {Kron}, {Lampeitl}, {Lang}, {Le Goff},
  {Lee}, {Lin}, {Long}, {Loomis}, {Lucatello}, {Lundgren}, {Lupton}, {Ma},
  {MacDonald}, {Mahadevan}, {Maia}, {Makler}, {Malanushenko}, {Malanushenko},
  {Mandelbaum}, {Maraston}, {Margala}, {Masters}, {McBride}, {McGehee},
  {McGreer}, {M{\'e}nard}, {Miralda-Escud{\'e}}, {Morrison}, {Mullally},
  {Muna}, {Munn}, {Murayama}, {Myers}, {Naugle}, {Neto}, {Nguyen}, {Nichol},
  {O'Connell}, {Ogando}, {Olmstead}, {Oravetz}, {Padmanabhan},
  {Palanque-Delabrouille}, {Pan}, {Pandey}, {P{\^a}ris}, {Percival},
  {Petitjean}, {Pfaffenberger}, {Pforr}, {Phleps}, {Pichon}, {Pieri}, {Prada},
  {Price-Whelan}, {Raddick}, {Ramos}, {Reyl{\'e}}, {Rich}, {Richards}, {Rix},
  {Robin}, {Rocha-Pinto}, {Rockosi}, {Roe}, {Rollinde}, {Ross}, {Ross},
  {Rossetto}, {S{\'a}nchez}, {Sayres}, {Schlegel}, {Schlesinger}, {Schmidt},
  {Schneider}, {Sheldon}, {Shu}, {Simmerer}, {Simmons}, {Sivarani}, {Snedden},
  {Sobeck}, {Steinmetz}, {Strauss}, {Szalay}, {Tanaka}, {Thakar}, {Thomas},
  {Tinker}, {Tofflemire}, {Tojeiro}, {Tremonti}, {Vandenberg}, {Vargas
  Maga{\~n}a}, {Verde}, {Vogt}, {Wake}, {Wang}, {Weaver}, {Weinberg}, {White},
  {White}, {Yanny}, {Yasuda}, {Yeche}, \& {Zehavi}}]{Aihara11}
{Aihara}, H., {Allende Prieto}, C., {An}, D., {et~al.} 2011, \apjs, 193, 29

\bibitem[{{Amor{\'{\i}}n} {et~al.}(2009){Amor{\'{\i}}n}, {Aguerri},
  {Mu{\~n}oz-Tu{\~n}{\'o}n}, \& {Cair{\'o}s}}]{Amorin09}
{Amor{\'{\i}}n}, R., {Aguerri}, J.~A.~L., {Mu{\~n}oz-Tu{\~n}{\'o}n}, C., \&
  {Cair{\'o}s}, L.~M. 2009, \aap, 501, 75

\bibitem[{{Amor{\'\i}n} {et~al.}(2015){Amor{\'\i}n}, {P{\'e}rez-Montero},
  {Contini}, {V{\'\i}lchez}, {Bolzonella}, {Tasca}, {Lamareille}, {Zamorani},
  {Maier}, {Carollo}, {Kneib}, {Le F{\`e}vre}, {Lilly}, {Mainieri}, {Renzini},
  {Scodeggio}, {Bardelli}, {Bongiorno}, {Caputi}, {Cucciati}, {de la Torre},
  {de Ravel}, {Franzetti}, {Garilli}, {Iovino}, {Kampczyk}, {Knobel},
  {Kova{\v{c}}}, {Le Borgne}, {Le Brun}, {Mignoli}, {Pell{\`o}}, {Peng},
  {Presotto}, {Ricciardelli}, {Silverman}, {Tanaka}, {Tresse}, {Vergani}, \&
  {Zucca}}]{Amorin15}
{Amor{\'\i}n}, R., {P{\'e}rez-Montero}, E., {Contini}, T., {et~al.} 2015, \aap,
  578, A105

\bibitem[{{Amor{\'\i}n} {et~al.}(2007){Amor{\'\i}n}, {Mu{\~n}oz-Tu{\~n}{\'o}n},
  {Aguerri}, {Cair{\'o}s}, \& {Caon}}]{Amorin07}
{Amor{\'\i}n}, R.~O., {Mu{\~n}oz-Tu{\~n}{\'o}n}, C., {Aguerri}, J.~A.~L.,
  {Cair{\'o}s}, L.~M., \& {Caon}, N. 2007, \aap, 467, 541

\bibitem[{{Arrabal Haro} {et~al.}(2020){Arrabal Haro}, {Rodr{\'\i}guez
  Espinosa}, {Mu{\~n}oz-Tu{\~n}{\'o}n}, {Sobral}, {Lumbreras-Calle}, {Boquien},
  {Hern{\'a}n-Caballero}, {Rodr{\'\i}guez-Mu{\~n}oz}, \& {Alcalde
  Pampliega}}]{ArrabalHaro20}
{Arrabal Haro}, P., {Rodr{\'\i}guez Espinosa}, J.~M.,
  {Mu{\~n}oz-Tu{\~n}{\'o}n}, C., {et~al.} 2020, \mnras, 495, 1807

\bibitem[{{Atek} {et~al.}(2011){Atek}, {Siana}, {Scarlata}, {Malkan},
  {McCarthy}, {Teplitz}, {Henry}, {Colbert}, {Bridge}, {Bunker}, {Dressler},
  {Fosbury}, {Hathi}, {Martin}, {Ross}, \& {Shim}}]{Atek11}
{Atek}, H., {Siana}, B., {Scarlata}, C., {et~al.} 2011, \apj, 743, 121

\bibitem[{{Bacon} {et~al.}(2017){Bacon}, {Conseil}, {Mary}, {Brinchmann},
  {Shepherd}, {Akhlaghi}, {Weilbacher}, {Piqueras}, {Wisotzki}, {Lagattuta},
  {Epinat}, {Guerou}, {Inami}, {Cantalupo}, {Courbot}, {Contini}, {Richard},
  {Maseda}, {Bouwens}, {Bouch{\'e}}, {Kollatschny}, {Schaye}, {Marino},
  {Pello}, {Herenz}, {Guiderdoni}, \& {Carollo}}]{Bacon17}
{Bacon}, R., {Conseil}, S., {Mary}, D., {et~al.} 2017, \aap, 608, A1

\bibitem[{{Barnes} \& {Hernquist}(1991)}]{Barnes91}
{Barnes}, J.~E. \& {Hernquist}, L.~E. 1991, \apjl, 370, L65

\bibitem[{{Bekki}(2008)}]{Bekki08}
{Bekki}, K. 2008, \mnras, 388, L10

\bibitem[{{Benitez} {et~al.}(2014){Benitez}, {Dupke}, {Moles}, {Sodre},
  {Cenarro}, {Marin-Franch}, {Taylor}, {Cristobal}, {Fernandez-Soto}, {Mendes
  de Oliveira}, {Cepa-Nogue}, {Abramo}, {Alcaniz}, {Overzier},
  {Hernandez-Monteagudo}, {Alfaro}, {Kanaan}, {Carvano}, {Reis}, {Martinez
  Gonzalez}, {Ascaso}, {Ballesteros}, {Xavier}, {Varela}, {Ederoclite},
  {Vazquez Ramio}, {Broadhurst}, {Cypriano}, {Angulo}, {Diego}, {Zandivarez},
  {Diaz}, {Melchior}, {Umetsu}, {Spinelli}, {Zitrin}, {Coe}, {Yepes}, {Vielva},
  {Sahni}, {Marcos-Caballero}, {Shu Kitaura}, {Maroto}, {Masip}, {Tsujikawa},
  {Carneiro}, {Gonzalez Nuevo}, {Carvalho}, {Reboucas}, {Carvalho}, {Abdalla},
  {Bernui}, {Pigozzo}, {Ferreira}, {Chandrachani Devi}, {Bengaly}, {Campista},
  {Amorim}, {Asari}, {Bongiovanni}, {Bonoli}, {Bruzual}, {Cardiel}, {Cava},
  {Cid Fernandes}, {Coelho}, {Cortesi}, {Delgado}, {Diaz Garcia}, {Espinosa},
  {Galliano}, {Gonzalez-Serrano}, {Falcon-Barroso}, {Fritz}, {Fernandes},
  {Gorgas}, {Hoyos}, {Jimenez-Teja}, {Lopez-Aguerri}, {Lopez-San Juan},
  {Mateus}, {Molino}, {Novais}, {OMill}, {Oteo}, {Perez-Gonzalez}, {Poggianti},
  {Proctor}, {Ricciardelli}, {Sanchez-Blazquez}, {Storchi-Bergmann}, {Telles},
  {Schoennell}, {Trujillo}, {Vazdekis}, {Viironen}, {Daflon},
  {Aparicio-Villegas}, {Rocha}, {Ribeiro}, {Borges}, {Martins}, {Marcolino},
  {Martinez-Delgado}, {Perez-Torres}, {Siffert}, {Calvao}, {Sako}, {Kessler},
  {Alvarez-Candal}, {De Pra}, {Roig}, {Lazzaro}, {Gorosabel}, {Lopes de
  Oliveira}, {Lima-Neto}, {Irwin}, {Liu}, {Alvarez}, {Balmes}, {Chueca},
  {Costa-Duarte}, {da Costa}, {Dantas}, {Diaz}, {Fabregat}, {Ferrari},
  {Gavela}, {Gracia}, {Gruel}, {Gutierrez}, {Guzman}, {Hernandez-Fernandez},
  {Herranz}, {Hurtado-Gil}, {Jablonsky}, {Laporte}, {Le Tiran}, {Licandro},
  {Lima}, {Martin}, {Martinez}, {Montero}, {Penteado}, {Pereira}, {Peris},
  {Quilis}, {Sanchez-Portal}, {Soja}, {Solano}, {Torra}, \&
  {Valdivielso}}]{Benitez14}
{Benitez}, N., {Dupke}, R., {Moles}, M., {et~al.} 2014, arXiv e-prints
  [\eprint[arXiv]{1403.5237}]

\bibitem[{{Ben{\'{\i}}tez} {et~al.}(2009){Ben{\'{\i}}tez}, {Gazta{\~n}aga},
  {Miquel}, {Castander}, {Moles}, {Crocce}, {Fern{\'a}ndez-Soto}, {Fosalba},
  {Ballesteros}, {Campa}, {Cardiel-Sas}, {Castilla}, {Crist{\'o}bal-Hornillos},
  {Delfino}, {Fern{\'a}ndez}, {Fern{\'a}ndez-Sopuerta},
  {Garc{\'{\i}}a-Bellido}, {Lobo}, {Mart{\'{\i}}nez}, {Ortiz}, {Pacheco},
  {Paredes}, {Pons-Border{\'{\i}}a}, {S{\'a}nchez}, {S{\'a}nchez}, {Varela}, \&
  {de Vicente}}]{Benitez09}
{Ben{\'{\i}}tez}, N., {Gazta{\~n}aga}, E., {Miquel}, R., {et~al.} 2009, \apj,
  691, 241

\bibitem[{{Bertin} \& {Arnouts}(1996)}]{Bertin96}
{Bertin}, E. \& {Arnouts}, S. 1996, \aaps, 117, 393

\bibitem[{{Bluck} {et~al.}(2020){Bluck}, {Maiolino}, {Piotrowska}, {Trussler},
  {Ellison}, {S{\'a}nchez}, {Thorp}, {Teimoorinia}, {Moreno}, \&
  {Conselice}}]{Bluck20}
{Bluck}, A. F.~L., {Maiolino}, R., {Piotrowska}, J.~M., {et~al.} 2020, \mnras,
  499, 230

\bibitem[{{Boquien} {et~al.}(2019){Boquien}, {Burgarella}, {Roehlly}, {Buat},
  {Ciesla}, {Corre}, {Inoue}, \& {Salas}}]{Boquien19}
{Boquien}, M., {Burgarella}, D., {Roehlly}, Y., {et~al.} 2019, \aap, 622, A103

\bibitem[{{Boquien} {et~al.}(2016){Boquien}, {Kennicutt}, {Calzetti}, {Dale},
  {Galametz}, {Sauvage}, {Croxall}, {Draine}, {Kirkpatrick}, {Kumari}, {Hunt},
  {De Looze}, {Pellegrini}, {Rela{\~n}o}, {Smith}, \& {Tabatabaei}}]{Boquien16}
{Boquien}, M., {Kennicutt}, R., {Calzetti}, D., {et~al.} 2016, \aap, 591, A6

\bibitem[{{Bouwens} {et~al.}(2015){Bouwens}, {Illingworth}, {Oesch}, {Trenti},
  {Labb{\'e}}, {Bradley}, {Carollo}, {van Dokkum}, {Gonzalez}, {Holwerda},
  {Franx}, {Spitler}, {Smit}, \& {Magee}}]{Bouwens15}
{Bouwens}, R.~J., {Illingworth}, G.~D., {Oesch}, P.~A., {et~al.} 2015, \apj,
  803, 34

\bibitem[{{Brinchmann} {et~al.}(2004){Brinchmann}, {Charlot}, {White},
  {Tremonti}, {Kauffmann}, {Heckman}, \& {Brinkmann}}]{Brinchmann04}
{Brinchmann}, J., {Charlot}, S., {White}, S.~D.~M., {et~al.} 2004, \mnras, 351,
  1151

\bibitem[{{Bruzual} \& {Charlot}(2003)}]{BC03}
{Bruzual}, G. \& {Charlot}, S. 2003, \mnras, 344, 1000

\bibitem[{{Calabr{\`o}} {et~al.}(2017){Calabr{\`o}}, {Amor{\'\i}n}, {Fontana},
  {P{\'e}rez-Montero}, {Lemaux}, {Ribeiro}, {Bardelli}, {Castellano},
  {Contini}, {De Barros}, {Garilli}, {Grazian}, {Guaita}, {Hathi}, {Koekemoer},
  {Le F{\`e}vre}, {Maccagni}, {Pentericci}, {Schaerer}, {Talia}, {Tasca}, \&
  {Zucca}}]{Calabro17}
{Calabr{\`o}}, A., {Amor{\'\i}n}, R., {Fontana}, A., {et~al.} 2017, \aap, 601,
  A95

\bibitem[{{Calzetti} {et~al.}(2000){Calzetti}, {Armus}, {Bohlin}, {Kinney},
  {Koornneef}, \& {Storchi-Bergmann}}]{Calzetti00}
{Calzetti}, D., {Armus}, L., {Bohlin}, R.~C., {et~al.} 2000, \apj, 533, 682

\bibitem[{{Cardamone} {et~al.}(2009){Cardamone}, {Schawinski}, {Sarzi},
  {Bamford}, {Bennert}, {Urry}, {Lintott}, {Keel}, {Parejko}, {Nichol},
  {Thomas}, {Andreescu}, {Murray}, {Raddick}, {Slosar}, {Szalay}, \&
  {Vandenberg}}]{Cardamone09}
{Cardamone}, C., {Schawinski}, K., {Sarzi}, M., {et~al.} 2009, \mnras, 399,
  1191

\bibitem[{{Cardamone} {et~al.}(2010){Cardamone}, {van Dokkum}, {Urry},
  {Taniguchi}, {Gawiser}, {Brammer}, {Taylor}, {Damen}, {Treister}, {Cobb},
  {Bond}, {Schawinski}, {Lira}, {Murayama}, {Saito}, \&
  {Sumikawa}}]{Cardamone11}
{Cardamone}, C.~N., {van Dokkum}, P.~G., {Urry}, C.~M., {et~al.} 2010, \apjs,
  189, 270

\bibitem[{{Catal{\'a}n-Torrecilla} {et~al.}(2015){Catal{\'a}n-Torrecilla}, {Gil
  de Paz}, {Castillo-Morales}, {Iglesias-P{\'a}ramo}, {S{\'a}nchez},
  {Kennicutt}, {P{\'e}rez-Gonz{\'a}lez}, {Marino}, {Walcher}, {Husemann},
  {Garc{\'\i}a-Benito}, {Mast}, {Gonz{\'a}lez Delgado}, {Mu{\~n}oz-Mateos},
  {Bland-Hawthorn}, {Bomans}, {Del Olmo}, {Galbany}, {Gomes}, {Kehrig},
  {L{\'o}pez-S{\'a}nchez}, {Mendoza}, {Monreal-Ibero}, {P{\'e}rez-Torres},
  {S{\'a}nchez-Bl{\'a}zquez}, {Vilchez}, \& {Califa
  Collaboration}}]{Catalan-Torrecilla15}
{Catal{\'a}n-Torrecilla}, C., {Gil de Paz}, A., {Castillo-Morales}, A.,
  {et~al.} 2015, \aap, 584, A87

\bibitem[{{Cava} {et~al.}(2015){Cava}, {P{\'e}rez-Gonz{\'a}lez},
  {Eliche-Moral}, {Ricciardelli}, {Vidal-Garc{\'\i}a}, {Alcalde Pampliega},
  {Alonso-Herrero}, {Barro}, {Cardiel}, {Cenarro}, {Charlot}, {Daddi},
  {Dessauges-Zavadsky}, {Dom{\'\i}nguez S{\'a}nchez}, {Espino-Briones},
  {Esquej}, {Gallego}, {Hern{\'a}n-Caballero}, {Huertas-Company}, {Koekemoer},
  {Mu{\~n}oz-Tunon}, {Rodriguez-Espinosa}, {Rodr{\'\i}guez-Mu{\~n}oz},
  {Tresse}, \& {Villar}}]{Cava15}
{Cava}, A., {P{\'e}rez-Gonz{\'a}lez}, P.~G., {Eliche-Moral}, M.~C., {et~al.}
  2015, \apj, 812, 155

\bibitem[{{Cenarro} {et~al.}(2019){Cenarro}, {Moles},
  {Crist{\'o}bal-Hornillos}, {Mar{\'\i}n-Franch}, {Ederoclite}, {Varela},
  {L{\'o}pez-Sanjuan}, {Hern{\'a}ndez-Monteagudo}, {Angulo}, {V{\'a}zquez
  Rami{\'o}}, {Viironen}, {Bonoli}, {Orsi}, {Hurier}, {San Roman}, {Greisel},
  {Vilella-Rojo}, {D{\'\i}az-Garc{\'\i}a}, {Logro{\~n}o-Garc{\'\i}a},
  {Gurung-L{\'o}pez}, {Spinoso}, {Izquierdo-Villalba}, {Aguerri}, {Allende
  Prieto}, {Bonatto}, {Carvano}, {Chies-Santos}, {Daflon}, {Dupke},
  {Falc{\'o}n-Barroso}, {Gon{\c{c}}alves}, {Jim{\'e}nez-Teja}, {Molino},
  {Placco}, {Solano}, {Whitten}, {Abril}, {Ant{\'o}n}, {Bello}, {Bielsa de
  Toledo}, {Castillo-Ram{\'\i}rez}, {Chueca}, {Civera},
  {D{\'\i}az-Mart{\'\i}n}, {Dom{\'\i}nguez-Mart{\'\i}nez},
  {Garzar{\'a}n-Calderaro}, {Hern{\'a}ndez-Fuertes}, {Iglesias-Marzoa},
  {I{\~n}iguez}, {Jim{\'e}nez Ruiz}, {Kruuse}, {Lamadrid}, {Lasso-Cabrera},
  {L{\'o}pez-Alegre}, {L{\'o}pez-Sainz}, {Ma{\'\i}cas}, {Moreno-Signes},
  {Muniesa}, {Rodr{\'\i}guez-Llano}, {Rueda-Teruel}, {Rueda-Teruel},
  {Soriano-Lagu{\'\i}a}, {Tilve}, {Valdivielso}, {Yanes-D{\'\i}az}, {Alcaniz},
  {Mendes de Oliveira}, {Sodr{\'e}}, {Coelho}, {Lopes de Oliveira}, {Tamm},
  {Xavier}, {Abramo}, {Akras}, {Alfaro}, {Alvarez-Candal}, {Ascaso}, {Beasley},
  {Beers}, {Borges Fernandes}, {Bruzual}, {Buzzo}, {Carrasco}, {Cepa},
  {Cortesi}, {Costa-Duarte}, {De Pr{\'a}}, {Favole}, {Galarza}, {Galbany},
  {Garcia}, {Gonz{\'a}lez Delgado}, {Gonz{\'a}lez-Serrano},
  {Guti{\'e}rrez-Soto}, {Hernandez-Jimenez}, {Kanaan}, {Kuncarayakti},
  {Landim}, {Laur}, {Licandro}, {Lima Neto}, {Lyman}, {Ma{\'\i}z
  Apell{\'a}niz}, {Miralda-Escud{\'e}}, {Morate}, {Nogueira-Cavalcante},
  {Novais}, {Oncins}, {Oteo}, {Overzier}, {Pereira}, {Rebassa-Mansergas},
  {Reis}, {Roig}, {Sako}, {Salvador-Rusi{\~n}ol}, {Sampedro},
  {S{\'a}nchez-Bl{\'a}zquez}, {Santos}, {Schmidtobreick}, {Siffert}, {Telles},
  \& {Vilchez}}]{Cenarro19}
{Cenarro}, A.~J., {Moles}, M., {Crist{\'o}bal-Hornillos}, D., {et~al.} 2019,
  \aap, 622, A176

\bibitem[{{Cenarro} {et~al.}(2014){Cenarro}, {Moles}, {Mar{\'{\i}}n-Franch},
  {Crist{\'o}bal-Hornillos}, {Yanes D{\'{\i}}az}, {Ederoclite}, {Varela},
  {V{\'a}zquez-Rami{\'o}}, {Valdivielso}, {Ben{\'{\i}}tez}, {Cepa}, {Dupke},
  {Fern{\'a}ndez-Soto}, {Mendes de Oliveira}, {Sodr{\'e}}, {Taylor},
  {Rueda-Teruel}, {Rueda-Teruel}, {Luis-Simoes}, {Chueca}, {Ant{\'o}n},
  {Bello}, {D{\'{\i}}az-Mart{\'{\i}}n}, {Guill{\'e}n-Civera},
  {Hern{\'a}ndez-Fuertes}, {Iglesias-Marzoa}, {Jim{\'e}nez-Mej{\'{\i}}as},
  {Lasso-Cabrera}, {L{\'o}pez-Alegre}, {L{\'o}pez-Sainz},
  {Rodr{\'{\i}}guez-Hern{\'a}ndez}, {Su{\'a}rez}, {Lamadrid}, {Ma{\'{\i}}cas},
  {Abril-Iba{\~n}ez}, {Tilve}, \& {Rodr{\'{\i}}guez-Llano}}]{oaj}
{Cenarro}, A.~J., {Moles}, M., {Mar{\'{\i}}n-Franch}, A., {et~al.} 2014, in
  \procspie, Vol. 9149, Observatory Operations: Strategies, Processes, and
  Systems V, 91491I

\bibitem[{{Ceverino} {et~al.}(2010){Ceverino}, {Dekel}, \&
  {Bournaud}}]{Ceverino10}
{Ceverino}, D., {Dekel}, A., \& {Bournaud}, F. 2010, \mnras, 404, 2151

\bibitem[{{Chabrier}(2003)}]{Chabrier03}
{Chabrier}, G. 2003, \pasp, 115, 763

\bibitem[{{Cook} {et~al.}(2019){Cook}, {Kasliwal}, {Van Sistine}, {Kaplan},
  {Sutter}, {Kupfer}, {Shupe}, {Laher}, {Masci}, {Dale}, {Sesar}, {Brady},
  {Yan}, {Ofek}, {Reitze}, \& {Kulkarni}}]{Cook19}
{Cook}, D.~O., {Kasliwal}, M.~M., {Van Sistine}, A., {et~al.} 2019, \apj, 880,
  7

\bibitem[{{Crist{\'o}bal-Hornillos} {et~al.}(2012){Crist{\'o}bal-Hornillos},
  {Gruel}, {Varela}, {L{\'o}pez-Sainz}, {Ederoclite}, {Moles}, {Cenarro},
  {Mar{\'{\i}}n-Franch}, {Hern{\'a}ndez-Fuertes}, {Yanes-D{\'{\i}}az},
  {Chueca}, {Rueda-Teruel}, {Rueda-Teruel}, \& {Luis-Simoes}}]{upad}
{Crist{\'o}bal-Hornillos}, D., {Gruel}, N., {Varela}, J., {et~al.} 2012, in
  SPIE CS, Vol. 8451

\bibitem[{{De Barros} {et~al.}(2019){De Barros}, {Oesch}, {Labb{\'e}},
  {Stefanon}, {Gonz{\'a}lez}, {Smit}, {Bouwens}, \& {Illingworth}}]{deBarros19}
{De Barros}, S., {Oesch}, P.~A., {Labb{\'e}}, I., {et~al.} 2019, \mnras, 489,
  2355

\bibitem[{{Dey} {et~al.}(2019){Dey}, {Schlegel}, {Lang}, {Blum}, {Burleigh},
  {Fan}, {Findlay}, {Finkbeiner}, {Herrera}, {Juneau}, {Landriau}, {Levi},
  {McGreer}, {Meisner}, {Myers}, {Moustakas}, {Nugent}, {Patej}, {Schlafly},
  {Walker}, {Valdes}, {Weaver}, {Y{\`e}che}, {Zou}, {Zhou}, {Abareshi},
  {Abbott}, {Abolfathi}, {Aguilera}, {Alam}, {Allen}, {Alvarez}, {Annis},
  {Ansarinejad}, {Aubert}, {Beechert}, {Bell}, {BenZvi}, {Beutler}, {Bielby},
  {Bolton}, {Brice{\~n}o}, {Buckley-Geer}, {Butler}, {Calamida}, {Carlberg},
  {Carter}, {Casas}, {Castander}, {Choi}, {Comparat}, {Cukanovaite}, {Delubac},
  {DeVries}, {Dey}, {Dhungana}, {Dickinson}, {Ding}, {Donaldson}, {Duan},
  {Duckworth}, {Eftekharzadeh}, {Eisenstein}, {Etourneau}, {Fagrelius},
  {Farihi}, {Fitzpatrick}, {Font-Ribera}, {Fulmer}, {G{\"a}nsicke},
  {Gaztanaga}, {George}, {Gerdes}, {Gontcho}, {Gorgoni}, {Green}, {Guy},
  {Harmer}, {Hernandez}, {Honscheid}, {Huang}, {James}, {Jannuzi}, {Jiang},
  {Joyce}, {Karcher}, {Karkar}, {Kehoe}, {Kneib}, {Kueter-Young}, {Lan},
  {Lauer}, {Le Guillou}, {Le Van Suu}, {Lee}, {Lesser}, {Perreault Levasseur},
  {Li}, {Mann}, {Marshall}, {Mart{\'\i}nez-V{\'a}zquez}, {Martini}, {du Mas des
  Bourboux}, {McManus}, {Meier}, {M{\'e}nard}, {Metcalfe},
  {Mu{\~n}oz-Guti{\'e}rrez}, {Najita}, {Napier}, {Narayan}, {Newman}, {Nie},
  {Nord}, {Norman}, {Olsen}, {Paat}, {Palanque-Delabrouille}, {Peng},
  {Poppett}, {Poremba}, {Prakash}, {Rabinowitz}, {Raichoor}, {Rezaie},
  {Robertson}, {Roe}, {Ross}, {Ross}, {Rudnick}, {Safonova}, {Saha},
  {S{\'a}nchez}, {Savary}, {Schweiker}, {Scott}, {Seo}, {Shan}, {Silva},
  {Slepian}, {Soto}, {Sprayberry}, {Staten}, {Stillman}, {Stupak}, {Summers},
  {Sien Tie}, {Tirado}, {Vargas-Maga{\~n}a}, {Vivas}, {Wechsler}, {Williams},
  {Yang}, {Yang}, {Yapici}, {Zaritsky}, {Zenteno}, {Zhang}, {Zhang}, {Zhou}, \&
  {Zhou}}]{Dey19}
{Dey}, A., {Schlegel}, D.~J., {Lang}, D., {et~al.} 2019, \aj, 157, 168

\bibitem[{{Driver} \& {Robotham}(2010)}]{Driver10}
{Driver}, S.~P. \& {Robotham}, A. S.~G. 2010, \mnras, 407, 2131

\bibitem[{{Duarte Puertas} {et~al.}(2017){Duarte Puertas}, {Vilchez},
  {Iglesias-P{\'a}ramo}, {Kehrig}, {P{\'e}rez-Montero}, \&
  {Rosales-Ortega}}]{DuartePuertas17}
{Duarte Puertas}, S., {Vilchez}, J.~M., {Iglesias-P{\'a}ramo}, J., {et~al.}
  2017, \aap, 599, A71

\bibitem[{{Elbaz} {et~al.}(2007){Elbaz}, {Daddi}, {Le Borgne}, {Dickinson},
  {Alexander}, {Chary}, {Starck}, {Brandt}, {Kitzbichler}, {MacDonald},
  {Nonino}, {Popesso}, {Stern}, \& {Vanzella}}]{Elbaz07}
{Elbaz}, D., {Daddi}, E., {Le Borgne}, D., {et~al.} 2007, \aap, 468, 33

\bibitem[{{Elbaz} {et~al.}(2018){Elbaz}, {Leiton}, {Nagar}, {Okumura},
  {Franco}, {Schreiber}, {Pannella}, {Wang}, {Dickinson}, {D{\'\i}az-Santos},
  {Ciesla}, {Daddi}, {Bournaud}, {Magdis}, {Zhou}, \& {Rujopakarn}}]{Elbaz18}
{Elbaz}, D., {Leiton}, R., {Nagar}, N., {et~al.} 2018, \aap, 616, A110

\bibitem[{{Fumagalli} {et~al.}(2012){Fumagalli}, {Patel}, {Franx}, {Brammer},
  {van Dokkum}, {da Cunha}, {Kriek}, {Lundgren}, {Momcheva}, {Rix}, {Schmidt},
  {Skelton}, {Whitaker}, {Labbe}, \& {Nelson}}]{Fumagalli12}
{Fumagalli}, M., {Patel}, S.~G., {Franx}, M., {et~al.} 2012, \apjl, 757, L22

\bibitem[{{Garn} \& {Best}(2010)}]{Garn10}
{Garn}, T. \& {Best}, P.~N. 2010, \mnras, 409, 421

\bibitem[{{Geach} {et~al.}(2008){Geach}, {Smail}, {Best}, {Kurk}, {Casali},
  {Ivison}, \& {Coppin}}]{Geach08}
{Geach}, J.~E., {Smail}, I., {Best}, P.~N., {et~al.} 2008, \mnras, 388, 1473

\bibitem[{{Gonz{\'a}lez Delgado} {et~al.}(2021){Gonz{\'a}lez Delgado},
  {D{\'\i}az-Garc{\'\i}a}, {de Amorim}, {Bruzual}, {Cid Fernandes},
  {P{\'e}rez}, {Bonoli}, {Cenarro}, {Coelho}, {Cortesi}, {Garc{\'\i}a-Benito},
  {L{\'o}pez Fern{\'a}ndez}, {Mart{\'\i}nez-Solaeche},
  {Rodr{\'\i}guez-Mart{\'\i}n}, {Magris}, {Mej{\'\i}a-Narvaez}, {Brito-Silva},
  {Abramo}, {Diego}, {Dupke}, {Hern{\'a}n-Caballero},
  {Hern{\'a}ndez-Monteagudo}, {L{\'o}pez-Sanjuan}, {Mar{\'\i}n-Franch},
  {Marra}, {Moles}, {Montero-Dorta}, {Queiroz}, {Sodr{\'e}}, {Varela},
  {V{\'a}zquez Rami{\'o}}, {V{\'\i}lchez}, {Baqui}, {Ben{\'\i}tez},
  {Crist{\'o}bal-Hornillos}, {Ederoclite}, {Mendes de Oliveira}, {Civera},
  {Muniesa}, {Taylor}, {Tempel}, \& {J-PAS Collaboration}}]{GonzalezDelgado21}
{Gonz{\'a}lez Delgado}, R.~M., {D{\'\i}az-Garc{\'\i}a}, L.~A., {de Amorim}, A.,
  {et~al.} 2021, \aap, 649, A79

\bibitem[{{Gronwall} {et~al.}(2007){Gronwall}, {Ciardullo}, {Hickey},
  {Gawiser}, {Feldmeier}, {van Dokkum}, {Urry}, {Herrera}, {Lehmer}, {Infante},
  {Orsi}, {Marchesini}, {Blanc}, {Francke}, {Lira}, \& {Treister}}]{Gronwall07}
{Gronwall}, C., {Ciardullo}, R., {Hickey}, T., {et~al.} 2007, \apj, 667, 79

\bibitem[{{Gronwall} {et~al.}(2004){Gronwall}, {Jangren}, {Salzer}, {Werk}, \&
  {Ciardullo}}]{Gronwall04}
{Gronwall}, C., {Jangren}, A., {Salzer}, J.~J., {Werk}, J.~K., \& {Ciardullo},
  R. 2004, \aj, 128, 644

\bibitem[{{Guo} {et~al.}(2016){Guo}, {Koo}, {Lu}, {Forbes}, {Rafelski},
  {Trump}, {Amor{\'\i}n}, {Barro}, {Dav{\'e}}, {Faber}, {Hathi}, {Yesuf},
  {Cooper}, {Dekel}, {Guhathakurta}, {Kirby}, {Koekemoer},
  {P{\'e}rez-Gonz{\'a}lez}, {Lin}, {Newman}, {Primack}, {Rosario}, {Willmer},
  \& {Yan}}]{Guo16}
{Guo}, Y., {Koo}, D.~C., {Lu}, Y., {et~al.} 2016, \apj, 822, 103

\bibitem[{{Guseva} {et~al.}(2017){Guseva}, {Izotov}, {Fricke}, \&
  {Henkel}}]{Guseva17}
{Guseva}, N.~G., {Izotov}, Y.~I., {Fricke}, K.~J., \& {Henkel}, C. 2017, \aap,
  599, A65

\bibitem[{{Haro}(1956)}]{Haro56}
{Haro}, G. 1956, Boletin de los Observatorios Tonantzintla y Tacubaya, 2, 8

\bibitem[{{Hern{\'a}n-Caballero} {et~al.}(2016){Hern{\'a}n-Caballero},
  {Hatziminaoglou}, {Alonso-Herrero}, \& {Mateos}}]{Hernan-Caballero16}
{Hern{\'a}n-Caballero}, A., {Hatziminaoglou}, E., {Alonso-Herrero}, A., \&
  {Mateos}, S. 2016, \mnras, 463, 2064

\bibitem[{{Hern{\'a}n-Caballero} {et~al.}(2021){Hern{\'a}n-Caballero},
  {Varela}, {L{\'o}pez-Sanjuan}, {Muniesa}, {Civera}, {Chaves-Montero},
  {D{\'\i}az-Garc{\'\i}a}, {Laur}, {Hern{\'a}ndez-Monteagudo}, {Abramo},
  {Angulo}, {Crist{\'o}bal-Hornillos}, {Gonz{\'a}lez-Delgado}, {Greisel},
  {Orsi}, {Queiroz}, {Sobral}, {Tamm}, {Tempel}, {V{\'a}zquez-Rami{\'o}},
  {Alcaniz}, {Ben{\'\i}tez}, {Bonoli}, {Carneiro}, {Cenarro}, {Dupke},
  {Ederoclite}, {Mar{\'\i}n-Frach}, {Mendes de Oliveira}, {Moles}, {Sodr},
  {Taylor}, \& {Cypriano}}]{Hernan-Caballero21}
{Hern{\'a}n-Caballero}, A., {Varela}, J., {L{\'o}pez-Sanjuan}, C., {et~al.}
  2021, arXiv e-prints, arXiv:2108.03271

\bibitem[{{Hinojosa-Go{\~n}i} {et~al.}(2016){Hinojosa-Go{\~n}i},
  {Mu{\~n}oz-Tu{\~n}{\'o}n}, \& {M{\'e}ndez-Abreu}}]{Hinojosa-Goni16}
{Hinojosa-Go{\~n}i}, R., {Mu{\~n}oz-Tu{\~n}{\'o}n}, C., \& {M{\'e}ndez-Abreu},
  J. 2016, \aap, 592, A122

\bibitem[{{Hippelein} {et~al.}(2003){Hippelein}, {Maier}, {Meisenheimer},
  {Wolf}, {Fried}, {von Kuhlmann}, {K{\"u}mmel}, {Phleps}, \&
  {R{\"o}ser}}]{Hippelein03}
{Hippelein}, H., {Maier}, C., {Meisenheimer}, K., {et~al.} 2003, \aap, 402, 65

\bibitem[{{Hood} {et~al.}(2018){Hood}, {Kannappan}, {Stark}, {Dell'Antonio},
  {Moffett}, {Eckert}, {Norris}, \& {Hendel}}]{Hood18}
{Hood}, C.~E., {Kannappan}, S.~J., {Stark}, D.~V., {et~al.} 2018, \apj, 857,
  144

\bibitem[{{Hsyu} {et~al.}(2018){Hsyu}, {Cooke}, {Prochaska}, \&
  {Bolte}}]{Hsyu18}
{Hsyu}, T., {Cooke}, R.~J., {Prochaska}, J.~X., \& {Bolte}, M. 2018, \apj, 863,
  134

\bibitem[{{Inoue}(2011)}]{Inoue11}
{Inoue}, A.~K. 2011, \mnras, 415, 2920

\bibitem[{{Izotov} {et~al.}(2021){Izotov}, {Guseva}, {Fricke}, {Henkel},
  {Schaerer}, \& {Thuan}}]{Izotov21}
{Izotov}, Y.~I., {Guseva}, N.~G., {Fricke}, K.~J., {et~al.} 2021, \aap, 646,
  A138

\bibitem[{{Izotov} {et~al.}(2011){Izotov}, {Guseva}, \& {Thuan}}]{Izotov11}
{Izotov}, Y.~I., {Guseva}, N.~G., \& {Thuan}, T.~X. 2011, \apj, 728, 161

\bibitem[{{Jangren} {et~al.}(2005){Jangren}, {Wegner}, {Salzer}, {Werk}, \&
  {Gronwall}}]{Jangren05}
{Jangren}, A., {Wegner}, G., {Salzer}, J.~J., {Werk}, J.~K., \& {Gronwall}, C.
  2005, \aj, 130, 496

\bibitem[{{Jiang} {et~al.}(2019){Jiang}, {Malhotra}, {Rhoads}, \&
  {Yang}}]{Jiang19}
{Jiang}, T., {Malhotra}, S., {Rhoads}, J.~E., \& {Yang}, H. 2019, \apj, 872,
  145

\bibitem[{{Kakazu} {et~al.}(2007){Kakazu}, {Cowie}, \& {Hu}}]{Kakazu07}
{Kakazu}, Y., {Cowie}, L.~L., \& {Hu}, E.~M. 2007, \apj, 668, 853

\bibitem[{{Kauffmann} {et~al.}(2003){Kauffmann}, {Heckman}, {White}, {Charlot},
  {Tremonti}, {Brinchmann}, {Bruzual}, {Peng}, {Seibert}, {Bernardi},
  {Blanton}, {Brinkmann}, {Castander}, {Cs{\'a}bai}, {Fukugita}, {Ivezic},
  {Munn}, {Nichol}, {Padmanabhan}, {Thakar}, {Weinberg}, \&
  {York}}]{Kauffmann03}
{Kauffmann}, G., {Heckman}, T.~M., {White}, S. D.~M., {et~al.} 2003, \mnras,
  341, 33

\bibitem[{{Kellar} {et~al.}(2012){Kellar}, {Salzer}, {Wegner}, {Gronwall}, \&
  {Williams}}]{Kellar12}
{Kellar}, J.~A., {Salzer}, J.~J., {Wegner}, G., {Gronwall}, C., \& {Williams},
  A. 2012, \aj, 143, 145

\bibitem[{{Kennicutt}(1998)}]{Kennicutt98}
{Kennicutt}, Robert~C., J. 1998, \apj, 498, 541

\bibitem[{{Khostovan} {et~al.}(2021){Khostovan}, {Malhotra}, {Rhoads},
  {Harish}, {Jiang}, {Wang}, {Wold}, {Zheng}, {Barrientos}, {Coughlin}, {Hu},
  {Infante}, {Perez}, {Pharo}, {Valdes}, \& {Walker}}]{Khostovan21}
{Khostovan}, A.~A., {Malhotra}, S., {Rhoads}, J.~E., {et~al.} 2021, \mnras,
  503, 5115

\bibitem[{{Khostovan} {et~al.}(2015){Khostovan}, {Sobral}, {Mobasher}, {Best},
  {Smail}, {Stott}, {Hemmati}, \& {Nayyeri}}]{Khostovan15}
{Khostovan}, A.~A., {Sobral}, D., {Mobasher}, B., {et~al.} 2015, \mnras, 452,
  3948

\bibitem[{{Khostovan} {et~al.}(2016){Khostovan}, {Sobral}, {Mobasher}, {Smail},
  {Darvish}, {Nayyeri}, {Hemmati}, \& {Stott}}]{Khostovan16}
{Khostovan}, A.~A., {Sobral}, D., {Mobasher}, B., {et~al.} 2016, \mnras, 463,
  2363

\bibitem[{{Kojima} {et~al.}(2020){Kojima}, {Ouchi}, {Rauch}, {Ono}, {Nakajima},
  {Isobe}, {Fujimoto}, {Harikane}, {Hashimoto}, {Hayashi}, {Komiyama},
  {Kusakabe}, {Kim}, {Lee}, {Mukae}, {Nagao}, {Onodera}, {Shibuya}, {Sugahara},
  {Umemura}, \& {Yabe}}]{Kojima20}
{Kojima}, T., {Ouchi}, M., {Rauch}, M., {et~al.} 2020, \apj, 898, 142

\bibitem[{{Kron}(1980)}]{Kron80}
{Kron}, R.~G. 1980, \apjs, 43, 305

\bibitem[{{Leitherer} {et~al.}(1999){Leitherer}, {Schaerer}, {Goldader},
  {Delgado}, {Robert}, {Kune}, {de Mello}, {Devost}, \&
  {Heckman}}]{Leitherer99}
{Leitherer}, C., {Schaerer}, D., {Goldader}, J.~D., {et~al.} 1999, \apjs, 123,
  3

\bibitem[{{Leslie} {et~al.}(2020){Leslie}, {Schinnerer}, {Liu}, {Magnelli},
  {Algera}, {Karim}, {Davidzon}, {Gozaliasl}, {Jim{\'e}nez-Andrade}, {Lang},
  {Sargent}, {Novak}, {Groves}, {Smol{\v{c}}i{\'c}}, {Zamorani}, {Vaccari},
  {Battisti}, {Vardoulaki}, {Peng}, \& {Kartaltepe}}]{Leslie20}
{Leslie}, S.~K., {Schinnerer}, E., {Liu}, D., {et~al.} 2020, \apj, 899, 58

\bibitem[{{Logro{\~n}o-Garc{\'\i}a} {et~al.}(2019){Logro{\~n}o-Garc{\'\i}a},
  {Vilella-Rojo}, {L{\'o}pez-Sanjuan}, {Varela}, {Viironen}, {Muniesa},
  {Cenarro}, {Crist{\'o}bal-Hornillos}, {Ederoclite}, {Mar{\'\i}n-Franch},
  {Moles}, {V{\'a}zquez Rami{\'o}}, {Bonoli}, {D{\'\i}az-Garc{\'\i}a}, {Orsi},
  {San Roman}, {Akras}, {Chies-Santos}, {Coelho}, {Daflon}, {Costa-Duarte},
  {Dupke}, {Galbany}, {Gonz{\'a}lez Delgado}, {Hernandez-Jimenez}, {Lopes de
  Oliveira}, {Mendes de Oliveira}, {Oteo}, {Gon{\c{c}}alves},
  {S{\'a}nchez-Portal}, {Schmidtobreick}, \& {Sodr{\'e}}}]{Logrono-Garcia19}
{Logro{\~n}o-Garc{\'\i}a}, R., {Vilella-Rojo}, G., {L{\'o}pez-Sanjuan}, C.,
  {et~al.} 2019, \aap, 622, A180

\bibitem[{{L{\'o}pez-Sanjuan} {et~al.}(2017){L{\'o}pez-Sanjuan}, {Tempel},
  {Ben{\'\i}tez}, {Molino}, {Viironen}, {D{\'\i}az-Garc{\'\i}a},
  {Fern{\'a}ndez-Soto}, {Santos}, {Varela}, {Cenarro}, {Moles}, {Arnalte-Mur},
  {Ascaso}, {Montero-Dorta}, {Povi{\'c}}, {Mart{\'\i}nez}, {Nieves-Seoane},
  {Stefanon}, {Hurtado-Gil}, {M{\'a}rquez}, {Perea}, {Aguerri}, {Alfaro},
  {Aparicio-Villegas}, {Broadhurst}, {Cabrera-Ca{\~n}o}, {Castander}, {Cepa},
  {Cervi{\~n}o}, {Crist{\'o}bal-Hornillos}, {Gonz{\'a}lez Delgado}, {Husillos},
  {Infante}, {Masegosa}, {del Olmo}, {Prada}, \& {Quintana}}]{Lopez-Sanjuan17}
{L{\'o}pez-Sanjuan}, C., {Tempel}, E., {Ben{\'\i}tez}, N., {et~al.} 2017, \aap,
  599, A62

\bibitem[{{L{\'o}pez-Sanjuan} {et~al.}(2019){L{\'o}pez-Sanjuan}, {Varela},
  {Crist{\'o}bal-Hornillos}, {V{\'a}zquez Rami{\'o}}, {Carrasco}, {Tremblay},
  {Whitten}, {Placco}, {Mar{\'\i}n-Franch}, {Cenarro}, {Ederoclite}, {Alfaro},
  {Coelho}, {Civera}, {Hern{\'a}ndez-Fuertes}, {Jim{\'e}nez-Esteban},
  {Jim{\'e}nez-Teja}, {Ma{\'\i}z Apell{\'a}niz}, {Sobral}, {V{\'\i}lchez},
  {Alcaniz}, {Angulo}, {Dupke}, {Hern{\'a}ndez-Monteagudo}, {Mendes de
  Oliveira}, {Moles}, \& {Sodr{\'e}}}]{clsj19jcal}
{L{\'o}pez-Sanjuan}, C., {Varela}, J., {Crist{\'o}bal-Hornillos}, D., {et~al.}
  2019, \aap, 631, A119

\bibitem[{{L{\'o}pez-Sanjuan} {et~al.}(2021){L{\'o}pez-Sanjuan}, {Yuan},
  {V{\'a}zquez Rami{\'o}}, {Varela}, {Crist{\'o}bal-Hornillos}, {Tremblay},
  {Mar{\'\i}n-Franch}, {Cenarro}, {Ederoclite}, {Alfaro}, {Alvarez-Candal},
  {Daflon}, {Hern{\'a}n-Caballero}, {Hern{\'a}ndez-Monteagudo},
  {Jim{\'e}nez-Esteban}, {Placco}, {Tempel}, {Alcaniz}, {Angulo}, {Dupke},
  {Moles}, \& {Sodr{\'e}}}]{clsj21zsl}
{L{\'o}pez-Sanjuan}, C., {Yuan}, H., {V{\'a}zquez Rami{\'o}}, H., {et~al.}
  2021, \aap, in press [\eprint{2101.12407}]

\bibitem[{{Lumbreras-Calle} {et~al.}(2019a){Lumbreras-Calle},
  {Mu{\~n}oz-Tu{\~n}{\'o}n}, {M{\'e}ndez-Abreu}, {Mas-Hesse},
  {P{\'e}rez-Gonz{\'a}lez}, {Alcalde Pampliega}, {Arrabal Haro}, {Cava},
  {Dom{\'{\i}}nguez S{\'a}nchez}, {Eliche-Moral}, {Alonso-Herrero}, {Borlaff},
  {Gallego}, {Hern{\'a}n-Caballero}, {Koekemoer}, \&
  {Rodr{\'{\i}}guez-Mu{\~n}oz}}]{Lumbreras-Calle19a}
{Lumbreras-Calle}, A., {Mu{\~n}oz-Tu{\~n}{\'o}n}, C., {M{\'e}ndez-Abreu}, J.,
  {et~al.} 2019a, \aap, 621, A52

\bibitem[{{Madau} \& {Dickinson}(2014)}]{Madau14}
{Madau}, P. \& {Dickinson}, M. 2014, \araa, 52, 415

\bibitem[{{Maier} {et~al.}(2003){Maier}, {Meisenheimer}, {Thommes},
  {Hippelein}, {R{\"o}ser}, {Fried}, {von Kuhlmann}, {Phleps}, \&
  {Wolf}}]{Maier03}
{Maier}, C., {Meisenheimer}, K., {Thommes}, E., {et~al.} 2003, \aap, 402, 79

\bibitem[{{Mar{\'{\i}}n-Franch} {et~al.}(2015){Mar{\'{\i}}n-Franch}, {Taylor},
  {Cenarro}, {Cristobal-Hornillos}, \& {Moles}}]{t80cam}
{Mar{\'{\i}}n-Franch}, A., {Taylor}, K., {Cenarro}, J., {Cristobal-Hornillos},
  D., \& {Moles}, M. 2015, in IAU General Assembly, Vol.~29, 2257381

\bibitem[{{Markarian}(1967)}]{Markarian67}
{Markarian}, B.~E. 1967, Astrofizika, 3, 24

\bibitem[{{Mart{\'\i}nez-Delgado} {et~al.}(2012){Mart{\'\i}nez-Delgado},
  {Romanowsky}, {Gabany}, {Annibali}, {Arnold}, {Fliri}, {Zibetti}, {van der
  Marel}, {Rix}, {Chonis}, {Carballo-Bello}, {Aloisi}, {Macci{\`o}},
  {Gallego-Laborda}, {Brodie}, \& {Merrifield}}]{Martinez-Delgado12}
{Mart{\'\i}nez-Delgado}, D., {Romanowsky}, A.~J., {Gabany}, R.~J., {et~al.}
  2012, \apjl, 748, L24

\bibitem[{{Maseda} {et~al.}(2018){Maseda}, {van der Wel}, {Rix}, {Momcheva},
  {Brammer}, {Franx}, {Lundgren}, {Skelton}, \& {Whitaker}}]{Maseda18}
{Maseda}, M.~V., {van der Wel}, A., {Rix}, H.-W., {et~al.} 2018, \apj, 854, 29

\bibitem[{{Matthee} {et~al.}(2021){Matthee}, {Sobral}, {Hayes}, {Pezzulli},
  {Gronke}, {Schaerer}, {Naidu}, {R{\"o}ttgering}, {Calhau}, {Paulino-Afonso},
  {Santos}, \& {Amor{\'\i}n}}]{Matthee21}
{Matthee}, J., {Sobral}, D., {Hayes}, M., {et~al.} 2021, \mnras, 505, 1382

\bibitem[{{Meisenheimer} {et~al.}(1998){Meisenheimer}, {Beckwith},
  {Fockenbrock}, {Fried}, {Hippelein}, {Huang}, {Leinert}, {Phleps}, {Roser},
  {Thommes}, {Thompson}, {Wolf}, \& {Chaffee}}]{Meisenheimer98}
{Meisenheimer}, K., {Beckwith}, S., {Fockenbrock}, H., {et~al.} 1998, in
  Astronomical Society of the Pacific Conference Series, Vol. 146, The Young
  Universe: Galaxy Formation and Evolution at Intermediate and High Redshift,
  ed. S.~{D'Odorico}, A.~{Fontana}, \& E.~{Giallongo}, 134

\bibitem[{{Melnick} {et~al.}(1985){Melnick}, {Terlevich}, \&
  {Eggleton}}]{Melnick85}
{Melnick}, J., {Terlevich}, R., \& {Eggleton}, P.~P. 1985, \mnras, 216, 255

\bibitem[{{Mendes de Oliveira} {et~al.}(2019){Mendes de Oliveira}, {Ribeiro},
  {Schoenell}, {Kanaan}, {Overzier}, {Molino}, {Sampedro}, {Coelho}, {Barbosa},
  {Cortesi}, {Costa-Duarte}, {Herpich}, {Hernandez-Jimenez}, {Placco},
  {Xavier}, {Abramo}, {Saito}, {Chies-Santos}, {Ederoclite}, {Lopes de
  Oliveira}, {Gon{\c{c}}alves}, {Akras}, {Almeida}, {Almeida-Fernandes},
  {Beers}, {Bonatto}, {Bonoli}, {Cypriano}, {Vinicius-Lima}, {de Souza},
  {Fabiano de Souza}, {Ferrari}, {Gon{\c{c}}alves}, {Gonzalez},
  {Guti{\'e}rrez-Soto}, {Hartmann}, {Jaffe}, {Kerber}, {Lima-Dias}, {Lopes},
  {Menendez-Delmestre}, {Nakazono}, {Novais}, {Ortega-Minakata}, {Pereira},
  {Perottoni}, {Queiroz}, {Reis}, {Santos}, {Santos-Silva}, {Santucci},
  {Barbosa}, {Siffert}, {Sodr{\'e}}, {Torres-Flores}, {Westera}, {Whitten},
  {Alcaniz}, {Alonso-Garc{\'\i}a}, {Alencar}, {Alvarez-Candal}, {Amram},
  {Azanha}, {Barb{\'a}}, {Bernardinelli}, {Borges Fernandes}, {Branco},
  {Brito-Silva}, {Buzzo}, {Caffer}, {Campillay}, {Cano}, {Carvano}, {Castejon},
  {Cid Fernandes}, {Dantas}, {Daflon}, {Damke}, {de la Reza}, {de Melo de
  Azevedo}, {De Paula}, {Diem}, {Donnerstein}, {Dors}, {Dupke}, {Eikenberry},
  {Escudero}, {Faifer}, {Far{\'\i}as}, {Fernandes}, {Fernandes}, {Fontes},
  {Galarza}, {Hirata}, {Katena}, {Gregorio-Hetem},
  {Hern{\'a}ndez-Fern{\'a}ndez}, {Izzo}, {Jaque Arancibia}, {Jatenco-Pereira},
  {Jim{\'e}nez-Teja}, {Kann}, {Krabbe}, {Labayru}, {Lazzaro}, {Lima Neto},
  {Lopes}, {Magalh{\~a}es}, {Makler}, {de Menezes}, {Miralda-Escud{\'e}},
  {Monteiro-Oliveira}, {Montero-Dorta}, {Mu{\~n}oz-Elgueta}, {Nemmen}, {Nilo
  Castell{\'o}n}, {Oliveira}, {Ort{\'\i}z}, {Pattaro}, {Pereira}, {Quint},
  {Riguccini}, {Rocha Pinto}, {Rodrigues}, {Roig}, {Rossi}, {Saha}, {Santos},
  {Schnorr M{\"u}ller}, {Sesto}, {Silva}, {Smith Castelli}, {Teixeira},
  {Telles}, {Thom de Souza}, {Th{\"o}ne}, {Trevisan}, {de Ugarte Postigo},
  {Urrutia-Viscarra}, {Veiga}, {Vika}, {Vitorelli}, {Werle}, {Werner}, \&
  {Zaritsky}}]{MendesdeOliveira19}
{Mendes de Oliveira}, C., {Ribeiro}, T., {Schoenell}, W., {et~al.} 2019,
  \mnras, 489, 241

\bibitem[{{Moles} {et~al.}(2008){Moles}, {Ben{\'{\i}}tez}, {Aguerri}, {Alfaro},
  {Broadhurst}, {Cabrera-Ca{\~n}o}, {Castander}, {Cepa}, {Cervi{\~n}o},
  {Crist{\'o}bal-Hornillos}, {Fern{\'a}ndez-Soto}, {Gonz{\'a}lez Delgado},
  {Infante}, {M{\'a}rquez}, {Mart{\'{\i}}nez}, {Masegosa}, {del Olmo}, {Perea},
  {Prada}, {Quintana}, \& {S{\'a}nchez}}]{Moles08}
{Moles}, M., {Ben{\'{\i}}tez}, N., {Aguerri}, J.~A.~L., {et~al.} 2008, \aj,
  136, 1325

\bibitem[{{Molino} {et~al.}(2019){Molino}, {Costa-Duarte}, {Mendes de
  Oliveira}, {Cenarro}, {Lima Neto}, {Cypriano}, {Sodr{\'e}}, {Coelho},
  {Chow-Mart{\'\i}nez}, {Monteiro-Oliveira}, {Sampedro}, {Cristobal-Hornillos},
  {Varela}, {Ederoclite}, {Chies-Santos}, {Schoenell}, {Ribeiro},
  {Mar{\'\i}n-Franch}, {L{\'o}pez-Sanjuan}, {Hern{\'a}ndez-Fern{\'a}ndez},
  {Cortesi}, {V{\'a}zquez Rami{\'o}}, {Santos}, {Cibirka}, {Novais}, {Pereira},
  {Hern{\'a}ndez-Jimenez}, {Jimenez-Teja}, {Moles}, {Ben{\'\i}tez}, \&
  {Dupke}}]{Molino19}
{Molino}, A., {Costa-Duarte}, M.~V., {Mendes de Oliveira}, C., {et~al.} 2019,
  \aap, 622, A178

\bibitem[{{Nilsson} {et~al.}(2011){Nilsson}, {{\"O}stlin}, {M{\o}ller},
  {M{\"o}ller-Nilsson}, {Tapken}, {Freudling}, \& {Fynbo}}]{Nilsson11}
{Nilsson}, K.~K., {{\"O}stlin}, G., {M{\o}ller}, P., {et~al.} 2011, \aap, 529,
  A9

\bibitem[{{Noeske} {et~al.}(2007){Noeske}, {Weiner}, {Faber}, {Papovich},
  {Koo}, {Somerville}, {Bundy}, {Conselice}, {Newman}, {Schiminovich}, {Le
  Floc'h}, {Coil}, {Rieke}, {Lotz}, {Primack}, {Barmby}, {Cooper}, {Davis},
  {Ellis}, {Fazio}, {Guhathakurta}, {Huang}, {Kassin}, {Martin}, {Phillips},
  {Rich}, {Small}, {Willmer}, \& {Wilson}}]{Noeske07}
{Noeske}, K.~G., {Weiner}, B.~J., {Faber}, S.~M., {et~al.} 2007, \apjl, 660,
  L43

\bibitem[{{Noll} {et~al.}(2009){Noll}, {Burgarella}, {Giovannoli}, {Buat},
  {Marcillac}, \& {Mu{\~n}oz-Mateos}}]{Noll09}
{Noll}, S., {Burgarella}, D., {Giovannoli}, E., {et~al.} 2009, \aap, 507, 1793

\bibitem[{{Onodera} {et~al.}(2020){Onodera}, {Shimakawa}, {Suzuki}, {Tanaka},
  {Harikane}, {Hayashi}, {Kodama}, {Koyama}, {Nakajima}, \&
  {Shibuya}}]{Onodera20}
{Onodera}, M., {Shimakawa}, R., {Suzuki}, T.~L., {et~al.} 2020, \apj, 904, 180

\bibitem[{{Ouchi} {et~al.}(2009){Ouchi}, {Mobasher}, {Shimasaku}, {Ferguson},
  {Fall}, {Ono}, {Kashikawa}, {Morokuma}, {Nakajima}, {Okamura}, {Dickinson},
  {Giavalisco}, \& {Ohta}}]{Ouchi09}
{Ouchi}, M., {Mobasher}, B., {Shimasaku}, K., {et~al.} 2009, \apj, 706, 1136

\bibitem[{{P{\'e}rez-Gonz{\'a}lez} {et~al.}(2013){P{\'e}rez-Gonz{\'a}lez},
  {Cava}, {Barro}, {Villar}, {Cardiel}, {Ferreras},
  {Rodr{\'{\i}}guez-Espinosa}, {Alonso-Herrero}, {Balcells}, {Cenarro}, {Cepa},
  {Charlot}, {Cimatti}, {Conselice}, {Daddi}, {Donley}, {Elbaz}, {Espino},
  {Gallego}, {Gobat}, {Gonz{\'a}lez-Mart{\'{\i}}n}, {Guzm{\'a}n},
  {Hern{\'a}n-Caballero}, {Mu{\~n}oz-Tu{\~n}{\'o}n}, {Renzini},
  {Rodr{\'{\i}}guez-Zaur{\'{\i}}n}, {Tresse}, {Trujillo}, \&
  {Zamorano}}]{Perez-Gonzalez13}
{P{\'e}rez-Gonz{\'a}lez}, P.~G., {Cava}, A., {Barro}, G., {et~al.} 2013, \apj,
  762, 46

\bibitem[{{P{\'e}rez-Montero} {et~al.}(2020){P{\'e}rez-Montero}, {Kehrig},
  {V{\'\i}lchez}, {Garc{\'\i}a-Benito}, {Duarte Puertas}, \&
  {Iglesias-P{\'a}ramo}}]{Perez-Montero20}
{P{\'e}rez-Montero}, E., {Kehrig}, C., {V{\'\i}lchez}, J.~M., {et~al.} 2020,
  \aap, 643, A80

\bibitem[{{Pirzkal} {et~al.}(2013){Pirzkal}, {Rothberg}, {Ly}, {Malhotra},
  {Rhoads}, {Grogin}, {Dahlen}, {Noeske}, {Meurer}, {Walsh}, {Hathi}, {Cohen},
  {Bellini}, {Holwerda}, {Straughn}, {Mechtley}, \& {Windhorst}}]{Pirzkal13}
{Pirzkal}, N., {Rothberg}, B., {Ly}, C., {et~al.} 2013, \apj, 772, 48

\bibitem[{{Reddy} {et~al.}(2018){Reddy}, {Shapley}, {Sanders}, {Kriek}, {Coil},
  {Shivaei}, {Freeman}, {Mobasher}, {Siana}, {Azadi}, {Fetherolf}, {Fornasini},
  {Leung}, {Price}, {Zick}, \& {Barro}}]{Reddy18}
{Reddy}, N.~A., {Shapley}, A.~E., {Sanders}, R.~L., {et~al.} 2018, \apj, 869,
  92

\bibitem[{{Rodighiero} {et~al.}(2011){Rodighiero}, {Daddi}, {Baronchelli},
  {Cimatti}, {Renzini}, {Aussel}, {Popesso}, {Lutz}, {Andreani}, {Berta},
  {Cava}, {Elbaz}, {Feltre}, {Fontana}, {F{\"o}rster Schreiber},
  {Franceschini}, {Genzel}, {Grazian}, {Gruppioni}, {Ilbert}, {Le Floch},
  {Magdis}, {Magliocchetti}, {Magnelli}, {Maiolino}, {McCracken}, {Nordon},
  {Poglitsch}, {Santini}, {Pozzi}, {Riguccini}, {Tacconi}, {Wuyts}, \&
  {Zamorani}}]{Rodighiero11}
{Rodighiero}, G., {Daddi}, E., {Baronchelli}, I., {et~al.} 2011, \apjl, 739,
  L40

\bibitem[{{Salim} {et~al.}(2007){Salim}, {Rich}, {Charlot}, {Brinchmann},
  {Johnson}, {Schiminovich}, {Seibert}, {Mallery}, {Heckman}, {Forster},
  {Friedman}, {Martin}, {Morrissey}, {Neff}, {Small}, {Wyder}, {Bianchi},
  {Donas}, {Lee}, {Madore}, {Milliard}, {Szalay}, {Welsh}, \& {Yi}}]{Salim07}
{Salim}, S., {Rich}, R.~M., {Charlot}, S., {et~al.} 2007, \apjs, 173, 267

\bibitem[{{Salpeter}(1955)}]{Salpeter55}
{Salpeter}, E.~E. 1955, \apj, 121, 161

\bibitem[{{Salzer} {et~al.}(2020){Salzer}, {Feddersen}, {Derloshon},
  {Gronwall}, {Van Sistine}, {Sugden}, {Janowiecki}, {Hirschauer}, \&
  {Kellar}}]{Salzer20}
{Salzer}, J.~J., {Feddersen}, J.~R., {Derloshon}, K., {et~al.} 2020, \aj, 160,
  242

\bibitem[{{Salzer} {et~al.}(2005){Salzer}, {Jangren}, {Gronwall}, {Werk},
  {Chomiuk}, {Caperton}, {Melbourne}, \& {McKinstry}}]{Salzer05}
{Salzer}, J.~J., {Jangren}, A., {Gronwall}, C., {et~al.} 2005, \aj, 130, 2584

\bibitem[{{S{\'a}nchez Almeida} {et~al.}(2015){S{\'a}nchez Almeida},
  {Elmegreen}, {Mu{\~n}oz-Tu{\~n}{\'o}n}, {Elmegreen}, {P{\'e}rez-Montero},
  {Amor{\'{\i}}n}, {Filho}, {Ascasibar}, {Papaderos}, \&
  {V{\'{\i}}lchez}}]{SanchezAlmeida15}
{S{\'a}nchez Almeida}, J., {Elmegreen}, B.~G., {Mu{\~n}oz-Tu{\~n}{\'o}n}, C.,
  {et~al.} 2015, \apjl, 810, L15

\bibitem[{{S{\'a}nchez Almeida} {et~al.}(2008){S{\'a}nchez Almeida},
  {Mu{\~n}oz-Tu{\~n}{\'o}n}, {Amor{\'\i}n}, {Aguerri}, {S{\'a}nchez-Janssen},
  \& {Tenorio-Tagle}}]{SanchezAlmeida08}
{S{\'a}nchez Almeida}, J., {Mu{\~n}oz-Tu{\~n}{\'o}n}, C., {Amor{\'\i}n}, R.,
  {et~al.} 2008, \apj, 685, 194

\bibitem[{{S{\'a}nchez Almeida} {et~al.}(2018){S{\'a}nchez Almeida},
  {Olmo-Garc{\'\i}a}, {Elmegreen}, {Elmegreen}, {Filho},
  {Mu{\~n}oz-Tu{\~n}{\'o}n}, {P{\'e}rez-Montero}, \&
  {Rom{\'a}n}}]{SanchezAlmeida18}
{S{\'a}nchez Almeida}, J., {Olmo-Garc{\'\i}a}, A., {Elmegreen}, B.~G., {et~al.}
  2018, \apj, 869, 40

\bibitem[{{S{\'a}nchez Almeida} {et~al.}(2016){S{\'a}nchez Almeida},
  {P{\'e}rez-Montero}, {Morales-Luis}, {Mu{\~n}oz-Tu{\~n}{\'o}n},
  {Garc{\'{\i}}a-Benito}, {Nuza}, \& {Kitaura}}]{SanchezAlmeida16}
{S{\'a}nchez Almeida}, J., {P{\'e}rez-Montero}, E., {Morales-Luis}, A.~B.,
  {et~al.} 2016, \apj, 819, 110

\bibitem[{{Sargent} \& {Searle}(1970)}]{Sargent70}
{Sargent}, W. L.~W. \& {Searle}, L. 1970, \apjl, 162, L155

\bibitem[{{Schlafly} {et~al.}(2019){Schlafly}, {Meisner}, \&
  {Green}}]{Schlafly19}
{Schlafly}, E.~F., {Meisner}, A.~M., \& {Green}, G.~M. 2019, \apjs, 240, 30

\bibitem[{{Searle} \& {Sargent}(1972)}]{Searle72}
{Searle}, L. \& {Sargent}, W. L.~W. 1972, \apj, 173, 25

\bibitem[{{Senchyna} \& {Stark}(2019)}]{Senchyna19}
{Senchyna}, P. \& {Stark}, D.~P. 2019, \mnras, 484, 1270

\bibitem[{{Sobral} {et~al.}(2014){Sobral}, {Best}, {Smail}, {Mobasher},
  {Stott}, \& {Nisbet}}]{Sobral14}
{Sobral}, D., {Best}, P.~N., {Smail}, I., {et~al.} 2014, \mnras, 437, 3516

\bibitem[{{Sobral} {et~al.}(2015){Sobral}, {Matthee}, {Best}, {Smail},
  {Khostovan}, {Milvang-Jensen}, {Kim}, {Stott}, {Calhau}, {Nayyeri}, \&
  {Mobasher}}]{Sobral15}
{Sobral}, D., {Matthee}, J., {Best}, P.~N., {et~al.} 2015, \mnras, 451, 2303

\bibitem[{{Sobral} {et~al.}(2018){Sobral}, {Santos}, {Matthee},
  {Paulino-Afonso}, {Ribeiro}, {Calhau}, \& {Khostovan}}]{Sobral18}
{Sobral}, D., {Santos}, S., {Matthee}, J., {et~al.} 2018, \mnras, 476, 4725

\bibitem[{{Sobral} {et~al.}(2013){Sobral}, {Smail}, {Best}, {Geach}, {Matsuda},
  {Stott}, {Cirasuolo}, \& {Kurk}}]{Sobral13}
{Sobral}, D., {Smail}, I., {Best}, P.~N., {et~al.} 2013, \mnras, 428, 1128

\bibitem[{{Sparre} {et~al.}(2017){Sparre}, {Hayward}, {Feldmann},
  {Faucher-Gigu{\`e}re}, {Muratov}, {Kere{\v s}}, \& {Hopkins}}]{Sparre17}
{Sparre}, M., {Hayward}, C.~C., {Feldmann}, R., {et~al.} 2017, \mnras, 466, 88

\bibitem[{{Speagle} {et~al.}(2014){Speagle}, {Steinhardt}, {Capak}, \&
  {Silverman}}]{Speagle14}
{Speagle}, J.~S., {Steinhardt}, C.~L., {Capak}, P.~L., \& {Silverman}, J.~D.
  2014, \apjs, 214, 15

\bibitem[{{Spinoso} {et~al.}(2020){Spinoso}, {Orsi}, {L{\'o}pez-Sanjuan},
  {Bonoli}, {Viironen}, {Izquierdo-Villalba}, {Sobral}, {Gurung-L{\'o}pez},
  {Hern{\'a}n-Caballero}, {Ederoclite}, {Varela}, {Overzier},
  {Miralda-Escud{\'e}}, {Muniesa}, {V{\'\i}lchez}, {Alcaniz}, {Angulo},
  {Cenarro}, {Crist{\'o}bal-Hornillos}, {Dupke}, {Hern{\'a}ndez-Monteagudo},
  {Mar{\'\i}n-Franch}, {Moles}, {Sodr{\'e}}, \&
  {V{\'a}zquez-Rami{\'o}}}]{Spinoso20}
{Spinoso}, D., {Orsi}, A., {L{\'o}pez-Sanjuan}, C., {et~al.} 2020, \aap, 643,
  A149

\bibitem[{{Stierwalt} {et~al.}(2015){Stierwalt}, {Besla}, {Patton}, {Johnson},
  {Kallivayalil}, {Putman}, {Privon}, \& {Ross}}]{Stierwalt15}
{Stierwalt}, S., {Besla}, G., {Patton}, D., {et~al.} 2015, \apj, 805, 2

\bibitem[{{Strauss} {et~al.}(2002){Strauss}, {Weinberg}, {Lupton}, {Narayanan},
  {Annis}, {Bernardi}, {Blanton}, {Burles}, {Connolly}, {Dalcanton}, {Doi},
  {Eisenstein}, {Frieman}, {Fukugita}, {Gunn}, {Ivezi{\'c}}, {Kent}, {Kim},
  {Knapp}, {Kron}, {Munn}, {Newberg}, {Nichol}, {Okamura}, {Quinn}, {Richmond},
  {Schlegel}, {Shimasaku}, {SubbaRao}, {Szalay}, {Vanden Berk}, {Vogeley},
  {Yanny}, {Yasuda}, {York}, \& {Zehavi}}]{Strauss02}
{Strauss}, M.~A., {Weinberg}, D.~H., {Lupton}, R.~H., {et~al.} 2002, \aj, 124,
  1810

\bibitem[{{Stroe} {et~al.}(2017{\natexlab{a}}){Stroe}, {Sobral}, {Matthee},
  {Calhau}, \& {Oteo}}]{Stroe17a}
{Stroe}, A., {Sobral}, D., {Matthee}, J., {Calhau}, J., \& {Oteo}, I.
  2017{\natexlab{a}}, \mnras, 471, 2558

\bibitem[{{Stroe} {et~al.}(2017{\natexlab{b}}){Stroe}, {Sobral}, {Matthee},
  {Calhau}, \& {Oteo}}]{Stroe17b}
{Stroe}, A., {Sobral}, D., {Matthee}, J., {Calhau}, J., \& {Oteo}, I.
  2017{\natexlab{b}}, \mnras, 471, 2575

\bibitem[{{Tang} {et~al.}(2021){Tang}, {Stark}, {Chevallard}, {Charlot},
  {Endsley}, \& {Congiu}}]{Tang20}
{Tang}, M., {Stark}, D.~P., {Chevallard}, J., {et~al.} 2021, \mnras, 503, 4105

\bibitem[{{Taniguchi} {et~al.}(2015){Taniguchi}, {Kajisawa}, {Kobayashi},
  {Shioya}, {Nagao}, {Capak}, {Aussel}, {Ichikawa}, {Murayama}, {Scoville},
  {Ilbert}, {Salvato}, {Sanders}, {Mobasher}, {Miyazaki}, {Komiyama}, {Le
  F{\`e}vre}, {Tasca}, {Lilly}, {Carollo}, {Renzini}, {Rich}, {Schinnerer},
  {Kaifu}, {Karoji}, {Arimoto}, {Okamura}, {Ohta}, {Shimasaku}, \&
  {Hayashino}}]{Taniguchi15}
{Taniguchi}, Y., {Kajisawa}, M., {Kobayashi}, M.~A.~R., {et~al.} 2015, \pasj,
  67, 104

\bibitem[{{Tenorio-Tagle} {et~al.}(2005){Tenorio-Tagle}, {Silich},
  {Rodr{\'\i}guez-Gonz{\'a}lez}, \&
  {Mu{\~n}oz-Tu{\~n}{\'o}n}}]{Tenorio-Tagle05}
{Tenorio-Tagle}, G., {Silich}, S., {Rodr{\'\i}guez-Gonz{\'a}lez}, A., \&
  {Mu{\~n}oz-Tu{\~n}{\'o}n}, C. 2005, \apjl, 628, L13

\bibitem[{{Tremonti} {et~al.}(2004){Tremonti}, {Heckman}, {Kauffmann},
  {Brinchmann}, {Charlot}, {White}, {Seibert}, {Peng}, {Schlegel}, {Uomoto},
  {Fukugita}, \& {Brinkmann}}]{Tremonti04}
{Tremonti}, C.~A., {Heckman}, T.~M., {Kauffmann}, G., {et~al.} 2004, \apj, 613,
  898

\bibitem[{{van der Wel} {et~al.}(2011){van der Wel}, {Straughn}, {Rix},
  {Finkelstein}, {Koekemoer}, {Weiner}, {Wuyts}, {Bell}, {Faber}, {Trump},
  {Koo}, {Ferguson}, {Scarlata}, {Hathi}, {Dunlop}, {Newman}, {Dickinson},
  {Jahnke}, {Salmon}, {de Mello}, {Kocevski}, {Lai}, {Grogin}, {Rodney}, {Guo},
  {McGrath}, {Lee}, {Barro}, {Huang}, {Riess}, {Ashby}, \&
  {Willner}}]{vanderWel11}
{van der Wel}, A., {Straughn}, A.~N., {Rix}, H.~W., {et~al.} 2011, \apj, 742,
  111

\bibitem[{{Vilella-Rojo} {et~al.}(2021){Vilella-Rojo},
  {Logro{\~n}o-Garc{\'\i}a}, {L{\'o}pez-Sanjuan}, {Viironen}, {Varela},
  {Moles}, {Cenarro}, {Crist{\'o}bal-Hornillos}, {Ederoclite},
  {Hern{\'a}ndez-Monteagudo}, {Mar{\'\i}n-Franch}, {V{\'a}zquez Rami{\'o}},
  {Galbany}, {Gonz{\'a}lez Delgado}, {Hern{\'a}n-Caballero}, {Lumbreras-Calle},
  {S{\'a}nchez-Bl{\'a}zquez}, {Sobral}, {V{\'\i}lchez}, {Alcaniz}, {Angulo},
  {Dupke}, \& {Sodr{\'e}}}]{Vilella-Rojo21}
{Vilella-Rojo}, G., {Logro{\~n}o-Garc{\'\i}a}, R., {L{\'o}pez-Sanjuan}, C.,
  {et~al.} 2021, \aap, 650, A68

\bibitem[{{Wegner} {et~al.}(2003){Wegner}, {Salzer}, {Jangren}, {Gronwall}, \&
  {Melbourne}}]{Wegner03}
{Wegner}, G., {Salzer}, J.~J., {Jangren}, A., {Gronwall}, C., \& {Melbourne},
  J. 2003, \aj, 125, 2373

\bibitem[{{Wolf} {et~al.}(2003){Wolf}, {Meisenheimer}, {Rix}, {Borch}, {Dye},
  \& {Kleinheinrich}}]{Wolf03}
{Wolf}, C., {Meisenheimer}, K., {Rix}, H.-W., {et~al.} 2003, \aap, 401, 73

\bibitem[{{Wolf} {et~al.}(2001){Wolf}, {Meisenheimer}, \&
  {R{\"o}ser}}]{Wolf01a}
{Wolf}, C., {Meisenheimer}, K., \& {R{\"o}ser}, H.-J. 2001, \aap, 365, 660

\bibitem[{{Yang} {et~al.}(2017){Yang}, {Malhotra}, {Rhoads}, \&
  {Wang}}]{Yang17}
{Yang}, H., {Malhotra}, S., {Rhoads}, J.~E., \& {Wang}, J. 2017, \apj, 847, 38

\bibitem[{{Zou} {et~al.}(2019){Zou}, {Zhou}, {Fan}, {Zhang}, {Zhou}, {Peng},
  {Nie}, {Jiang}, {McGreer}, {Cai}, {Chen}, {Chen}, {Dey}, {Fan}, {Findlay},
  {Gao}, {Gu}, {Guo}, {He}, {Jiang}, {Jin}, {Kong}, {Lang}, {Lei}, {Lesser},
  {Li}, {Li}, {Lin}, {Ma}, {Maxwell}, {Meng}, {Myers}, {Ning}, {Schlegel},
  {Shao}, {Shi}, {Sun}, {Wang}, {Wang}, {Wang}, {Wei}, {Wu}, {Wu}, {Wu},
  {Yang}, {Yang}, {Yuan}, \& {Yue}}]{Zou19}
{Zou}, H., {Zhou}, X., {Fan}, X., {et~al.} 2019, \apjs, 245, 4

\bibitem[{{Zwicky}(1966)}]{Zwicky66}
{Zwicky}, F. 1966, \apj, 143, 192

\end{thebibliography}

\begin{appendix} 
\section{ADQL query}
\label{appendix:query}

In this appendix we reproduce the ADQL query used in Section \ref{sec:databas} to obtain the parent sample for this work. In addition to the main condition (significant emission in $J0515$ compared to $r$), described in Sect. \ref{sec:databas}, we include some other restrictions:
\begin{itemize}
    \item Objects must have been detected with at least 3$\sigma$ significance in $g$, $r$, and $J0515$.
    \item The flux in the $J0515$ and $g$ filters cannot be compatible, considering the uncertainties. This is to ensure that there is significant emission in $J0515$, not just a very steep slope towards $r$.
    \item The $r$ fluxes in both PSFCOR and AUTO photometry must be higher than certain threshold. This is only to ensure appropriate detections and avoid numeric errors, given that the thresholds are much lower than the final one imposed ($r$< 20 mag).
    \item Objects with FLAGS other than 0, 1, or 2 are rejected, only keeping those with either no issues (0), close neighbors (1), deblended (2), or a combination of those.
    \item Objects with any MASK\_FLAG are rejected.
\end{itemize}
\begin{lstlisting}[basicstyle=\ttfamily]

SELECT flu.*, z.*, s.*, e.* 
FROM jplus.FLambdaDualObj flu, 
jplus.PhotoZLephare z, 
jplus.StarGalClass s, jplus.MWExtinction e 
WHERE  flu.TILE_ID = z.TILE_ID 
AND z.TILE_ID = s.TILE_ID 
AND s.TILE_ID = e.TILE_ID 
AND flu.NUMBER = z.NUMBER  
AND z.NUMBER=s.NUMBER 
AND s.NUMBER=e.NUMBER 
AND flu.FLUX_AUTO[jplus::rSDSS]>44.8 
AND flu.FLUX_PSFCOR[jplus::rSDSS]>17.8 
AND flu.FLUX_RELERR_PSFCOR[jplus::rSDSS]
<0.333 
AND flu.FLUX_RELERR_PSFCOR[jplus::gSDSS]
<0.333 
AND flu.FLUX_RELERR_PSFCOR[jplus::J0515]
<0.333 
AND (flu.FLUX_PSFCOR[jplus::J0515] -
flu.FLUX_PSFCOR[jplus::rSDSS])/
flu.FLUX_PSFCOR[jplus::rSDSS] > 1    
AND (flu.FLUX_PSFCOR[jplus::J0515] 
- flu.FLUX_RELERR_PSFCOR[jplus::J0515]
*flu.FLUX_PSFCOR[jplus::J0515])
>(flu.FLUX_PSFCOR[jplus::gSDSS]+
flu.FLUX_RELERR_PSFCOR[jplus::gSDSS]
*flu.FLUX_PSFCOR[jplus::gSDSS]) 
AND flu.FLAGS[jplus::rSDSS] <4   
AND flu.MASK_FLAGS[jplus::rSDSS] = 0 
AND flu.FLAGS[jplus::J0515] < 4   
AND flu.MASK_FLAGS[jplus::J0515] = 0

\end{lstlisting}

\section{Re-calculation of J-PLUS photometry for extended sources with blending flag}
\label{appendix:rephot}

As mentioned in Section \ref{sec:blended}, some of the objects in our sample present photometric flags indicating that they have been deblended from other source. This, in some cases, just means that there was a different object nearby (such a star or a different galaxy), and the photometric data we are using only belongs to the galaxy we are interested in. Nevertheless, in some cases, the nearby object is in fact the main galaxy body, and the object we are detecting is only a specific region of the galaxy. Keeping these regions as "galaxies" would bias our analysis towards lower masses and higher EWs. In order to avoid this, we run SExtractor over the J-PLUS images.

\subsection{Visual classification}

We inspect the 371 objects with deblending flag in the main sample, selecting those where our object of interest is in fact part of a larger galaxy. While can be ultimately a "judgment call", we follow some rules to use our own SExtractor photometry of an object instead of the original J-PLUS one:

\begin{itemize}
    \item If in the Legacy survey image (or in the PANSTARRS, if outside the Legacy footprint) there is a clear connection of blue emission between our object and an extended galaxy that is not covered by the J-PLUS aperture.

\item If there is a large galaxy nearby, with several HII regions showing similar color as our object, at similar distances.

\end{itemize}
Some cases were not considered as improper deblending:
\begin{itemize}

 \item If the color of the object near our target is very red and no clear structure is found between them, we consider that they do not belong to the same physical object and keep the J-PLUS original photometry.

 \item If the object and our target have a similar size and color, and there is no emission between them, we consider them as satellites and keep the original J-PLUS photometry.
\end{itemize}
In some cases, particularly involving mergers, the decision between considering a system one galaxy or several based upon the photometry is, to a certain degree, arbitrary. We have chosen to keep the original photometry if the system is very disturbed and our object is physically separated from the larger galaxy or galaxies. We have as well preferred to keep the original photometry in cases were the SExtractor apertures were contaminated significantly by other nearby objects, that were however masked out in the J-PLUS photometry. After this visual inspection, we identified 175 galaxies where the original J-PLUS photometry was appropriate, and 179 where additional SExtractor runs where necessary to capture all the flux from the galaxy. We identified in addition 20 spurious objects, that were removed from the sample. Most of them were spikes from bright stars, measured over large apertures.

\begin{figure}
   \centering
   \includegraphics[width=0.48\textwidth,keepaspectratio]{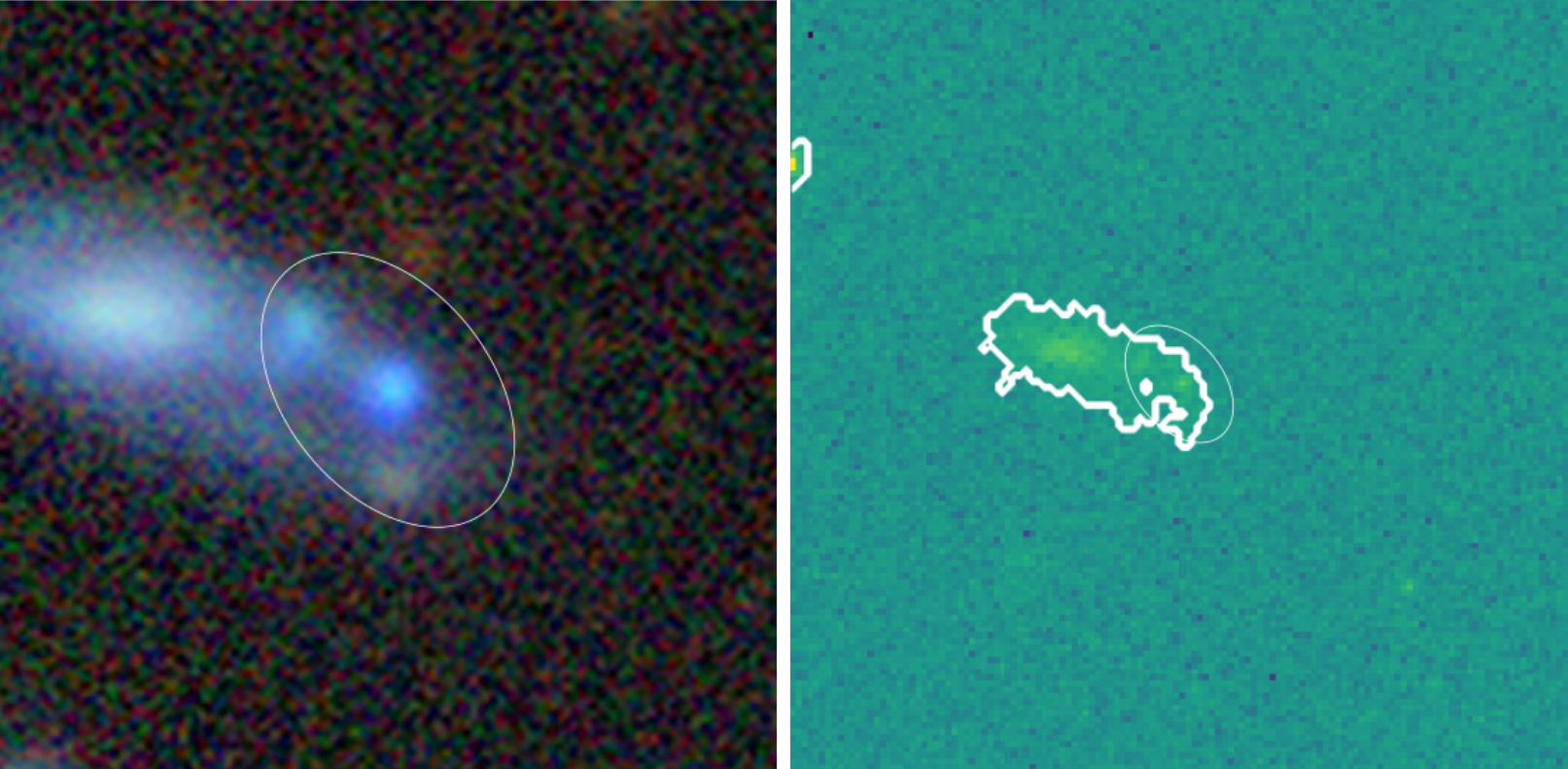}
      \caption{Example of a improperly deblended source, and the SExtractor run that covers the whole galaxy}
         \label{fig:exam_delend}
   \end{figure}

An example of a galaxy with deblending flag where we decided to keep the SExtractor photometry is shown in Fig. \ref{fig:exam_delend}.
   
\subsection{SExtractor runs}
\begin{table}
\caption{Parameters for the SExtractor runs in blended objects}             
\label{tab:sextract}      
\centering                          
\begin{tabular}{c c c}        
\hline\hline                 
 & Set 1 & Set 2 \\    

\hline                        

DETECT\_MINAREA & 20 & 20 \\
DETECT\_THRESH & 1.1 & 0.6 \\
ANALYSIS\_THRESH & 3 & 3 \\
DEBLEND\_NTHRESH & 1 & 1 \\
DEBLEND\_MINCONT & 0.05 & 0.005 \\ 

\end{tabular}
\end{table}

We run SExtractor over the sample of 179 galaxies that were not accurately represented by the original J-PLUS photometric aperture, using the default SExtractor parameter set except for a few parameters shown in Table \ref{tab:sextract} (Set 1). After the first run using Set 1 we inspected visually the result, and considered that for 36 of these galaxies neither the SExtractor run nor the original photometry were appropriate. In most of those cases the our target was a region of a larger galaxy, but both the original photometry and our first SExtractor run failed to identify them as the same object. We chose to run SExtractor with a second set of parameters (Set 2 in Table \ref{tab:sextract}). This was enough to recover accurate photometry for 27 out of the 36 galaxies. The remaining 9 were removed from the sample, since in any case they are so extended that they would not make the EW cut to join the EELG sample.

\section{CIGALE SED fits and parameters}

In this Appendix we will review in more detail the SED fitting process and describe in full the \texttt{CIGALE} input parameters we used. We reproduce here the full set of input parameter in the main SED fit presented in this work.

\begin{lstlisting}[basicstyle=\ttfamily]
 
 [[sfh2exp]]

e-folding time of the main stellar
population model in Myr.
tau_main = 50

e-folding 1 of the late starburst 
population model in Myr.
tau_burst = 1

Mass fraction of the late burst 
population. 
f_burst = 0.0005, 0.0025, 
0.005, 0.01, 0.03, 0.05, 0.075

Age of the main stellar population in the 
galaxy in Myr. The precision is 1 Myr.
age = 200, 500, 1000, 2000, 5000

Age of the late burst in Myr. 
The precision is 1 Myr.
burst_age = 1, 2, 3, 4, 5, 7, 12

Value of SFR at t = 0 in M_sun/yr.
sfr_0 = 1.0

Normalise the SFH to produce one 
solar mass.
normalise = True


  [[bc03]]

Initial mass function: 0 (Salpeter) or 
1 (Chabrier).
imf = 0

Metalicity. Possible values are: 
0.0001, 0.0004, 0.004, 0.008, 0.02, 0.05.
metallicity = 0.0001, 0.0004, 0.004, 
0.008, 0.02


  [[nebular]]

Ionisation parameter
logU = -4.0, -3.5, -3.0, -2.5, -2.0, -1.5

Fraction of Lyman continuum photons 
escaping the galaxy
f_esc = 0.0, 0.2

Fraction of Lyman continuum photons 
     absorbed by dust
f_dust = 0.0

Line width in km/s
lines_width = 300.0

Include nebular emission.
emission = True


  [[dustatt_calzleit]]

E(B-V)*, the colour excess of the stellar 
continuum light for the young population.

E_BVs_young = 0, 0.1, 0.2, 0.3, 0.5

Reduction factor for the E(B-V)* of the old
population compared to the young one (<1).
E_BVs_old_factor = 0.44, 1

Central wavelength of the UV bump in nm.
uv_bump_wavelength = 217.5

Width (FWHM) of the UV bump in nm.
uv_bump_width = 35.0

Amplitude of the UV bump. 
For the Milky Way: 3.
uv_bump_amplitude = 0.0

Slope delta of the power law modifying 
the attenuation curve.
powerlaw_slope = 0.0

\end{lstlisting}

It is important to note that they main focus of the SED fits is to compute accurate values for the continuum and the emission line fluxes, while also providing a stellar mass estimate. The specific results for the stellar population parameters are subject to our choices in the free parameters and several physical degeneracies, thus extracting high level scientific conclusions from them is very challenging. We perform several \texttt{CIGALE} runs on fractions of the main sample, with variations on the input parameters, to test the effect that different choices may have on the results.

The SFH we choose, with two single stellar populations, represents two almost instantaneous bursts of star formation. Other possibilities may as well provide good fits to the data, for instance, including options for a more extended SFH in the old stellar populations ($\tau_{\mathrm{old}}$=50, 500, 1000 Myr). This would provide accurate fits as well, while making almost no impact in the main results (negligible in log(U) or E(B-V), and just 0.1 dex in stellar mass, 0.02-0.04 dex in emission line fluxes, and 0.6 Myr in age of the young population). Even comparing the age of the old stellar population we only see a small scatter (0.15 dex) and no offset.

In our main run we only considered two values (0 and 0.2) for the fraction of Lyman continuum photons escaping the galaxy (\texttt{f\_esc}), and only one value (0) for the fraction of Lyman photons absorbed by dust (\texttt{f\_dust}). We included some variation to provide more flexible fits, but we make no attempt at extracting physical conclusions from these parameters. Increasing the set of possible values to \texttt{f\_esc} = 0, 0.2, 0.5 and \texttt{f\_dust}=0, 0.5 yields very similar results to the original SED fits: the median values of the difference between parameters remain virtually the same, with only small 1$\sigma$ scatters: 0.08 dex in stellar mass, 0.02-0.04 dex in emission line fluxes, 1.5 Myr in age of the young population, 0.05 mag in E(B-V), and negligible in log(U).

Finally, we changed our assumption of a \cite{Salpeter55} IMF to a \cite{Chabrier03} one. In this case, we see the expected 0.23 dex offset in the stellar mass, along with a 0.16 dex scatter. The offsets in the rest of the parameters are negligible, with low scatter, as in the previous tests: 0.04 - 0.07 dex in emission line fluxes, 1.5 Myr in the age of the young population, 0.1 mag in E(B-V), and negligible in log(U).

\label{appendix:cigale}

\end{appendix}

\end{document}